\documentclass{ws-rv9x6}   
\newcommand{\be}{\begin{eqnarray}} 
 
\newcommand{\ee}{\end{eqnarray}}

\begin{document} 
 
\setcounter{chapter}{0} 
 
\chapter{JET QUENCHING AND RADIATIVE ENERGY LOSS \\
 IN DENSE NUCLEAR MATTER}

\author{Miklos Gyulassy$^1$, Ivan Vitev$^2$, 
Xin-Nian Wang$^3$ and Ben-Wei Zhang$^4$} 
 
\address{$^1$Department of Physics, Columbia University, 
538 W. 120th Street,  \\  New York, NY 10027\\ 
%E-mail: gyulassy@mail-cunuke.phys.columbia.edu\\ 
$^2$ Department of Physics and Astronomy, Iowa State University, 
Ames,~IA~50010 \\ 
$^3$Nuclear Science Division, MS 70R0319, \\ 
Lawrence Berkeley National Laboratory, Berkeley, CA 94720\\ 
%E-mail: xnwang@lbl.gov 
$^4$Institute of Particle Physics, Huazhong Normal University, \\
         Wuhan 430079, China \\ 
}

\begin{abstract} 
 We review recent finite opacity  approaches (GLV, WW, WOGZ) 
to the computation of the induced gluon radiative energy loss  
and  their  application to   
the tomographic studies of the density evolution in 
ultra-relativistic  nuclear collisions. 
\end{abstract}

\section{Introduction} 
Since June 2000 measurements of $Au+Au$ reactions at the Relativistic 
Heavy Ion Collider (RHIC) at the Brookhaven National Laboratory (BNL) at 
$\sqrt{s}=56, 130, 200$~AGeV (GeV per nucleon pair) have revealed  a 
variety of novel multiparticle  phenomena  not  observed previously in 
$e^+e^-,\;  ep,\; pp$ collisions at any energy nor in nuclear collisions 
at  lower (SPS/CERN and AGS/BNL) energies ($\sqrt{s}=17,5$~AGeV).  While 
the bulk global observables such as the rapidity dependence of hadron 
multiplicities and transverse energy scale geometrically from 
elementary $p+p$ collisions, the most striking new phenomena  
were discovered in rare  high transverse momentum 
observables\cite{Adcox:2002pe}$^-$\cite{Levai:2001dc}.
The preliminary RHIC data are discussed extensively in the Quark Matter 
2001 and 2002  proceedings\cite{qm01}.

At RHIC collider energies the hard pQCD rate of rare high $p_T$ 
parton scattering becomes sufficiently large  
that jets can be used to probe  the dense quark-gluon plasma 
formed in nuclear collisions. In 1982 Bjorken  proposed that elastic
final state energy loss of partons may ``extinguish'' jets 
in high energy $p+p$ collisions\cite{Bjorken:1982tu}. However, 
the elastic energy loss of partons in a QCD plasma of temperature 
$T\sim 300$ MeV turned out to be too small 
($dE_{el}/dx < 500$ MeV/fm)\cite{Thoma:1990fm,Thoma:1993vs} 
for jet extinction. 
The data on $p+p$ jets  up to Tevatron energies
show no sign of deviations from unquenched factorized pQCD.  
Di-jet acoplanarity was  proposed as another possible manifestation of 
elastic final state 
interactions\cite{Appel:dq,Blaizot:1986ma,Rammerstorfer:js}.
Prior to current RHIC data on nuclear collisions at 200 AGeV,
the data showed no hint of this effect.

\begin{figure} 
\begin{center} 
\psfig{file=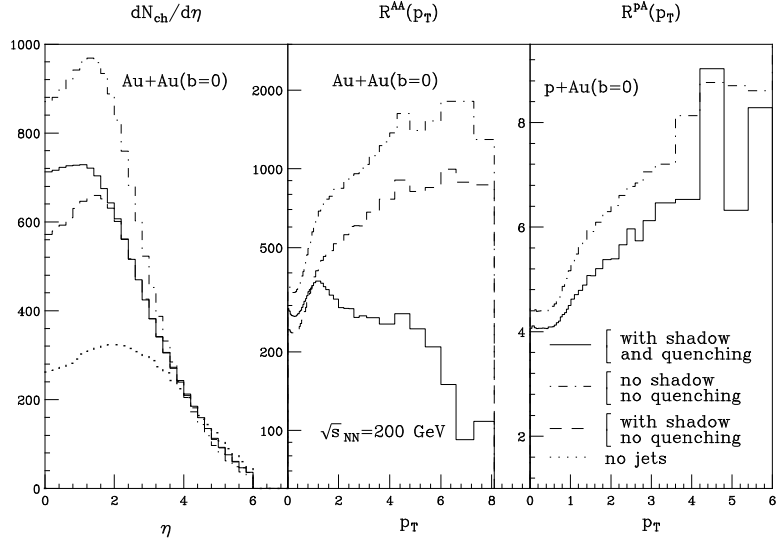,height=4.0in, width=2.8in,clip=5,angle=-90}  
\caption{HIJING predictions\protect\cite{Wang:1992xy}  
  of the inclusive charged hadron spectra in central $Au+Au$ and 
  $p+Au$ collisions at $\sqrt{s}=200$~AGeV.  The competing effects of 
  minijet production (dash-dotted), gluon shadowing (dashed) (assuming 
  that gluon shadowing is identical to that of quarks), and jet 
  quenching (solid) with $dE/dx=$ 2 GeV/fm are shown.  $R^{AB}(p_T)$ is 
  the ratio of the inclusive $p_T$ spectrum of charged hadrons in 
  $A+B$ collisions to that of $p+p$. In contrast to $Au+Au$, no 
  significant quenching is expected in $p+Au$ (or $d+Au$) since only 
  $\sim 20\% $ initial state shadowing and Cronin effects modify the 
  pQCD spectrum at high $p_T$.  } 
\label{hj_92prl} 
\end{center} 
\end{figure}

In some early studies\cite{Gyulassy:1990ye,Gyulassy:1990bh,Gyulassy:1991xb} 
it was suggested that induced radiative energy loss in nuclear 
collisions could be much 
larger than the elastic energy loss and jet quenching should 
become observable at least in collisions of heavy nuclei. 
With the development of the Monte 
Carlo HIJING event generator\cite{Wang:1991ht,Wang:1991us,Gyulassy:ew}, 
predictions for the magnitude 
of the suppression pattern high $p_T$ hadrons  
were made in a first study\cite{Wang:1992xy} as shown   
in Fig.~\ref{hj_92prl}. 
The middle panel shows that 
up to an order of magnitude suppression of charged hadrons 
was expected in the moderate $p_T\sim 5$~GeV  range. The input assumption 
was that  the gluon energy loss in the plasma was $dE/dx=2$ GeV/fm  
due to induced gluon radiative energy loss.  The HIJING  model  is  a 
two component model with semi-hard and hard pQCD jet above $p_T>p_0=2$~GeV 
computed via the PYTHIA code and the ``soft'' beam jet fragments computed 
via a hybrid LUND and Dual Parton model algorithm. See the
original article\cite{Wang:1991ht} 
for a detailed discussion and references.

\begin{figure} 
\vspace{0.2in}
\centerline{\psfig{file=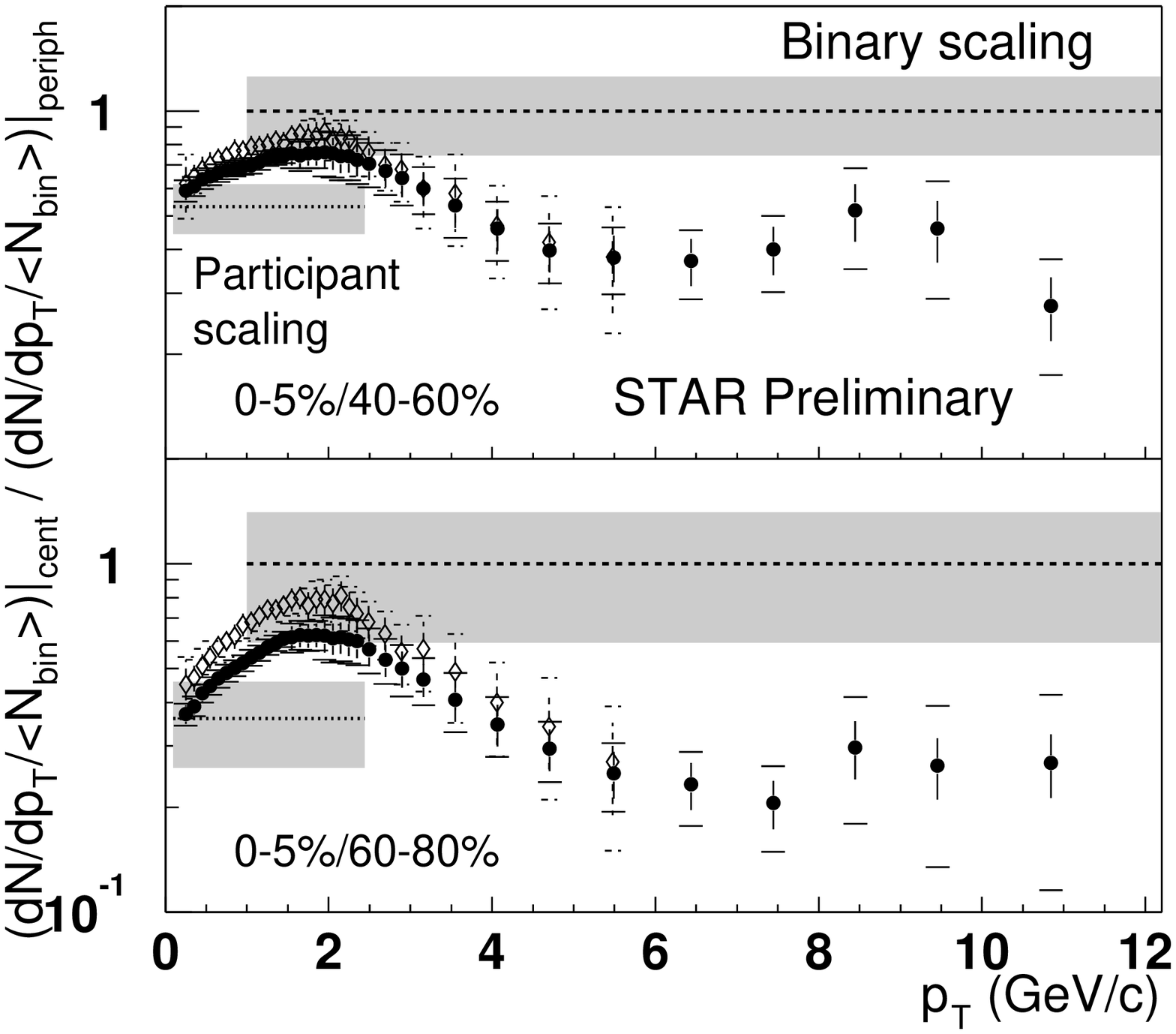,height=2.75in,width=3.in}} 
\caption{Preliminary STAR\protect\cite{Kunde:2002pb}  
data on charged hadron quenching between central and peripheral 
 $Au+Au$ collisions  at   
 $\sqrt{s}=200$ AGeV is shown for peripheral event classes.  
The data are scaled by the Glauber binary collision scaling factor 
for each centrality class. Upper gray band is the expected result 
from pQCD scaling from $p+p$. Lower gray bands correspond to 
scaling with the nucleon participant number instead. Beyond 
4~GeV, charged hadrons are suppressed by 0.2-0.4  (factor 2.5 to 5) 
relative to  naive binary scaling.} 
\label{kunde_fig2} 
\end{figure}

\begin{figure} 
\centerline{\psfig{file=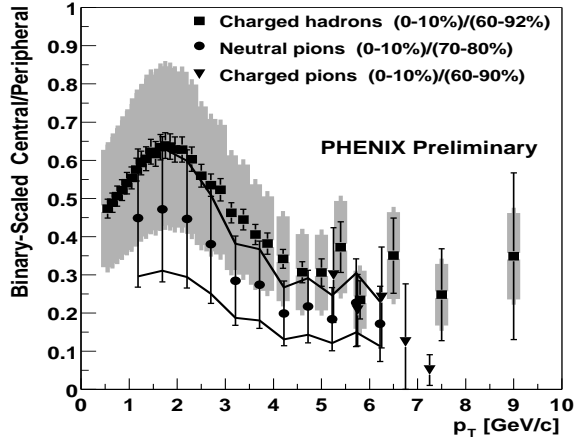,height=2.5in,width=3.25in}}
\vspace{0.2in}
\caption{Preliminary PHENIX\protect\cite{Mioduszewski:2002wt} 
data on charged and $\pi^0$ hadron quenching between central and 
peripheral  $Au+Au$ collisions at $\sqrt{s}=200$ AGeV is shown. 
The pions are generally more quenched than the summed  
charged hadrons ($\pi +K+p$).} 
\label{phenix_fig7} 
\end{figure}

\begin{figure} 
\centerline{\psfig{file=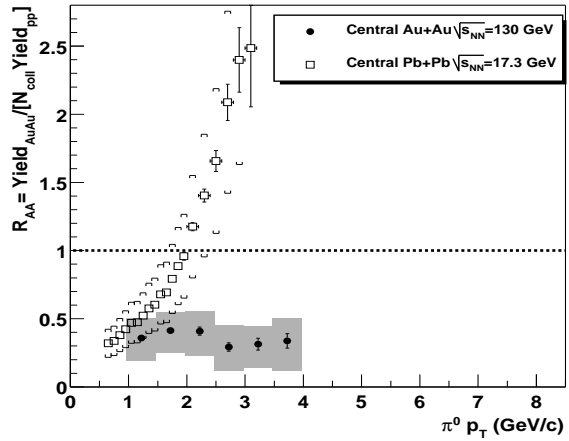,height=2.5in,width=3.in}}  
\caption{The striking contrast between 
PHENIX\protect\cite{Adcox:2002pe} 
data on $\pi^0$ quenching in central  
 $Au+Au$ collisions  at  $\sqrt{s}=130$ AGeV  
compared to the (Cronin) enhancement found 
in $Pb+Pb$ at 17~AGeV from\protect\cite{Aggarwal:1998vh} demonstrates 
that jet quenching is a new nuclear phenomenon first  seen  at RHIC.} 
\label{phenix_fig4} 
\end{figure}

\begin{figure} 
\vspace{0.3in}
\centerline{\psfig{file=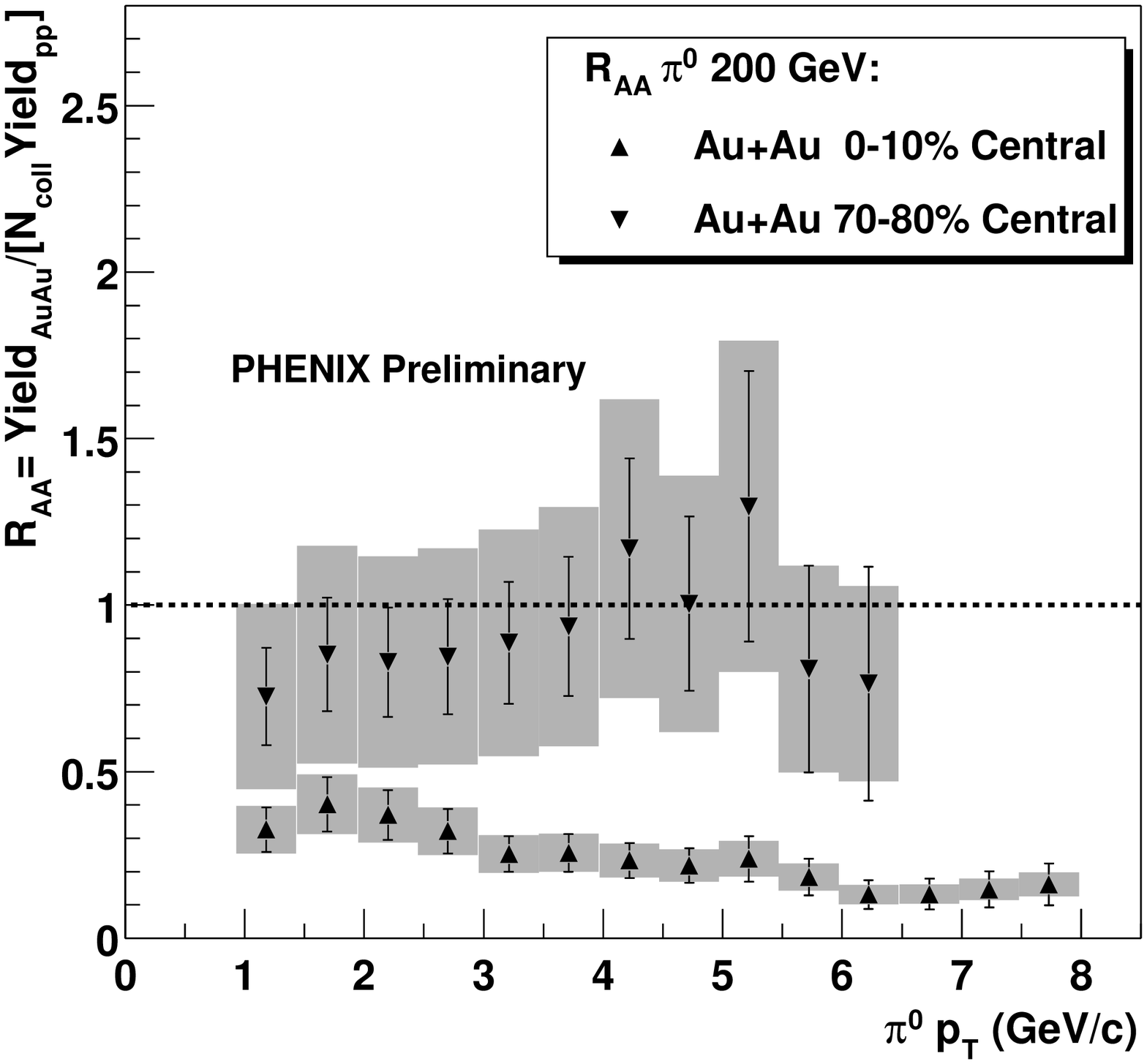,height=2.5in,width=3.25in}} 
\caption{Preliminary $\sqrt{s}=200$ AGeV  
PHENIX\protect\cite{Mioduszewski:2002wt} 
data on $\pi^0$ quenching  on  central and peripheral $Au+Au$ 
collisions  is compared  relative to preliminary PHENIX data on $p+p$. 
Peripheral reactions are consistent with simple binary ($\sim A^{4/3}$) 
scaling from $p+p$  while in central collisions 
substantial quenching is observed.} 
\label{phenix_fig5} 
\end{figure}

A selection of recent data from RHIC that confirm strong  nuclear 
attenuation of  moderate high transverse momentum charged hadron 
and $\pi^0$  production are shown in Fig.~\ref{kunde_fig2} 
through  Fig.~\ref{phenix_fig5}.  
In order to appreciate just how remarkable and different
this experimental discovery is relative to previous lower energy
data from the SPS,
the quenching pattern of $\pi^0$ measured at RHIC is compared 
in Fig.~\ref{phenix_fig4} to the 
strong {\em enhancement} 
of high $p_T$ $\pi^0$ measured in $Pb+Pb$ at 17~AGeV 
at the SPS\cite{Aggarwal:1998vh}.  

The enhancement of high $p_T$ hadrons at SPS is an amplified
version of the well known Cronin effect first observed in $p+A$ collisions.
As we elaborate in later sections, it is due to multiple {\em initial} state
interactions and completely masks any possible energy loss effects 
at the SPS. In contrast, at RHIC the high energy loss overwhelms the
Cronin enhancement in our present interpretation of the data.  
\begin{figure} 
\begin{center} 
\psfig{file=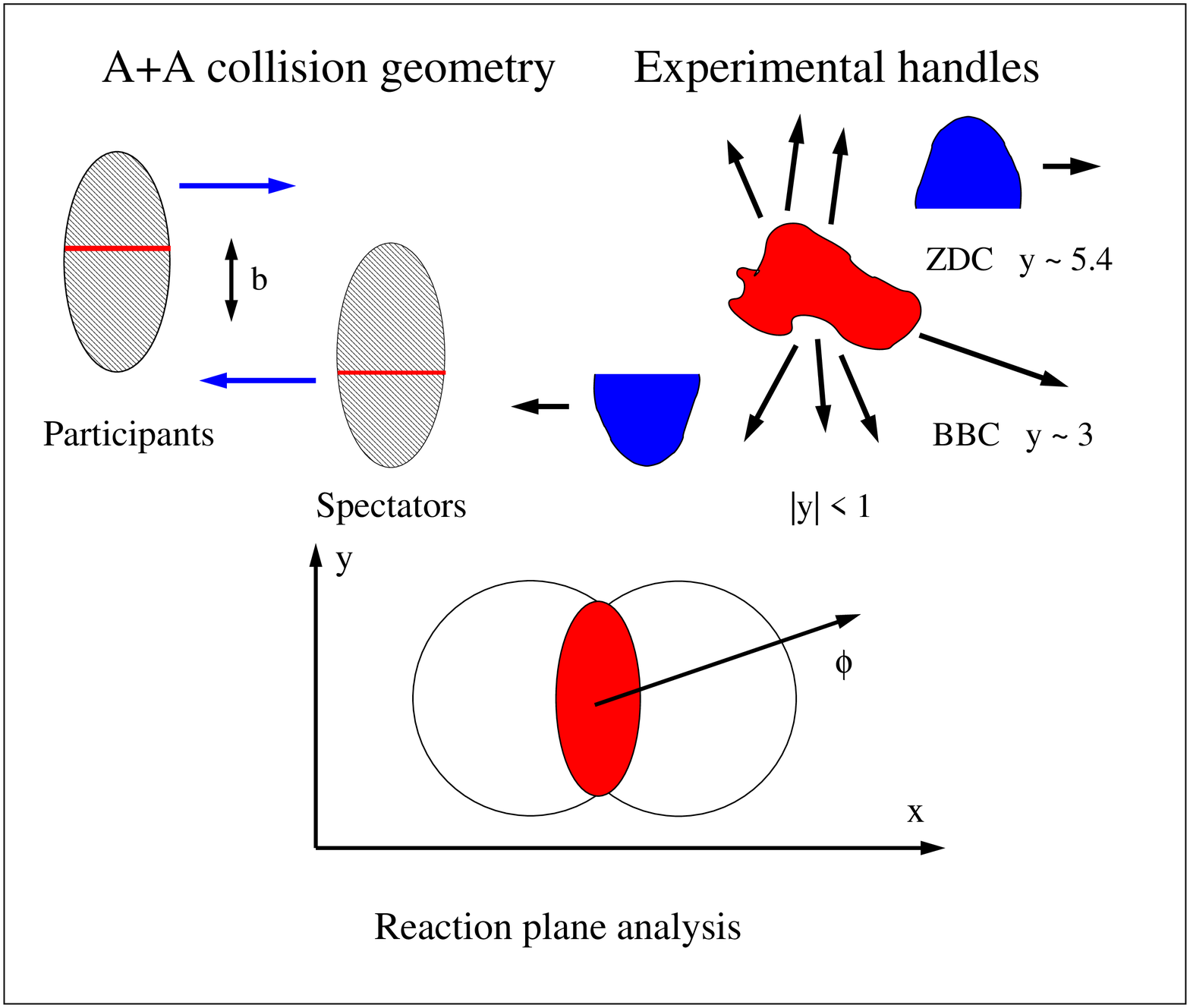,height=3.8in,width=4.2in,clip=5,angle=0}  
\vspace*{.5cm}
\caption{Illustration of key aspects of the relation between 
the geometry  of nuclear collisions and the participant   
$N_{part}(b)=2 \int d^2r \,  T_{A}(b-r)(1-\exp(-\sigma^{pp} T_A(b) )) 
\sim A^1 \leq 2A$ 
 and collision  
$N_{coll}(b)=\sigma^{pp} T_{AA}(b) \sim A^{4/3}$  number at 
a fixed impact parameter. The observables  $dN^{ch}/d \eta$,
$dE_T/d\eta$, $v_2(p_T)$ 
(see Refs.\protect\cite{Adler:qm,Zajc:2001va,Roland:2001me,Bearden:qk}) 
used to constrain the geometry experimentally are also illustrated.} 
\label{aageom} 
\end{center} 
\end{figure} 

As discussed for example in Refs.\cite{Wang:2001bf,Wang:2001cy}, 
it is useful to  
decompose the nuclear geometry dependence of  invariant hadron  
distributions produced in $A+B\rightarrow h+X$ at  impact parameter 
$b$ into a phenomenological ``soft'' and perturbative  
QCD calculable ``hard'' components as 
\begin{eqnarray} 
E\frac{dN_{AB}({b})  }{d^3p} &=&\;  
\frac{ N_{part}({b})}{2}\, 
\frac{dN_{soft}({b}) }{dyd^2{ \bf p}_{T}} 
+ N_{coll}({b})\, \frac{1}{\sigma_{in}^{pp}} 
\frac{d\sigma_{hard}({b})}{dyd^2{\bf p}_{T}} \; , 
\label{decomp} 
\end{eqnarray} 
where $N_{part}({b})$ is the number of nucleon participants 
and $N_{coll}({b}) = \sigma_{in}^{pp} T_{AB}({b})$ 
is the number of binary $NN$ collisions at impact parameter ${b}$. 
The nuclear geometry of hard collisions is expressed  in terms 
of the Glauber profile density per unity area 
$T_{AB}({b})=\int  d^2{\bf r} \;T_A({\bf r})T_B({\bf r}-{\bf b})$ 
where  $T_A({r})=\int dz \;\rho_A({\bf r},z)$ (see Fig.~6).  
The hard part scales with the number of binary collisions $\propto A^{4/3}$ 
because  the probability of high $p_T$ processes 
is small  and built up from all possible independent 
parton scattering processes. The soft part scales with 
only $N_{part}\propto A$ 
because the  probability of low transverse momenta processes 
is bounded by unitarity. This is sometimes referred to as Glauber 
shadowing or saturation depending on the calculational frame. 
The ``soft'' part is actually the overwhelming bulk of 
the produced hadron distribution. It is expected to reflect the 
collective hydrodynamic properties of the produced plasma in $A+A$ 
collisions. Relativistic hydrodynamics predicts specific flow patterns, 
such as hadron mass dependent transverse elliptic  
flow\cite{Kolb:2000fh,Huovinen:2001cy,Kolb:2001qz,Teaney:2001gc}, 
which may provide direct constraints on the QCD equation of state. 
Unfortunately, low transverse momentum processes are not directly 
computable via QCD and many competing phenomenological models 
can be adjusted to fit the data not only at RHIC but SPS and AGS 
as well.  

\begin{figure} 
\centerline{\psfig{file=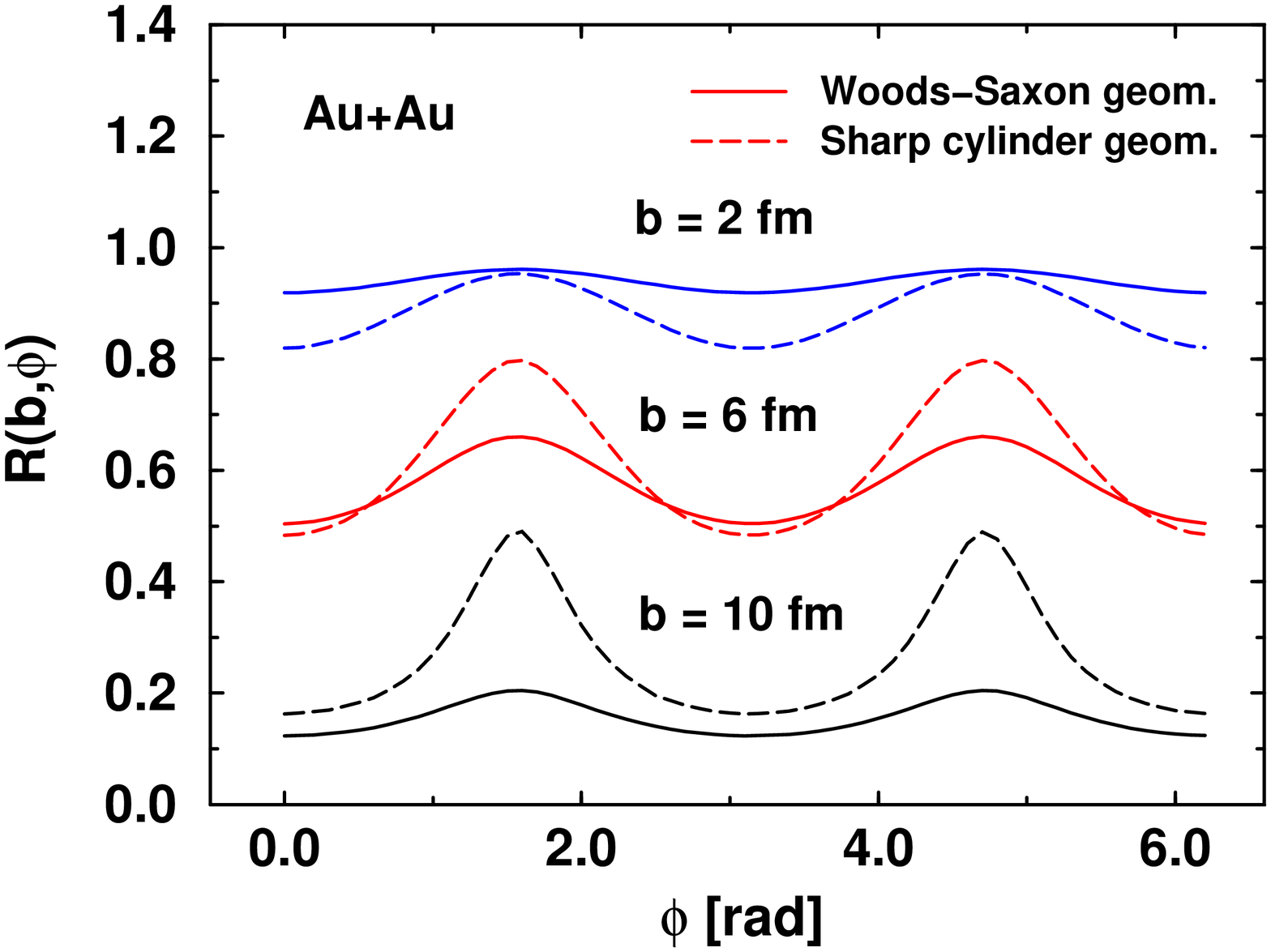,height=3.5in,width=2.75in,angle=-90}} 
\caption{The average normalized 
optical depth ($R$ from Fig.~2 of Ref.\protect\cite{Gyulassy:2000gk}) seen 
by a parton propagating through a plasma in azimuthal direction $\phi$ 
for different impact parameters and different density profiles.} 
\label{azimuth_gvw} 
\end{figure}

\begin{figure} 
\vspace*{.6cm}
\centerline{\psfig{file=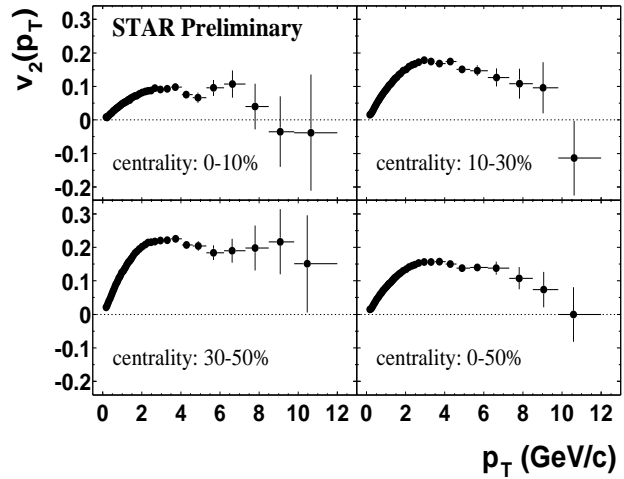,height=2.5in,width=3.25in}} 
\caption{Preliminary $\sqrt{s}=200$ AGeV  
STAR\protect\cite{Kunde:2002pb} 
on high $p_T$ elliptic flow is shown. 
A value of $v_2=0.2$ corresponds to a 2-to-1 azimuthal asymmetry 
of hadrons.} 
\label{kunde_qm2_v2} 
\end{figure}

\begin{figure} 
\centerline{\epsfig{file=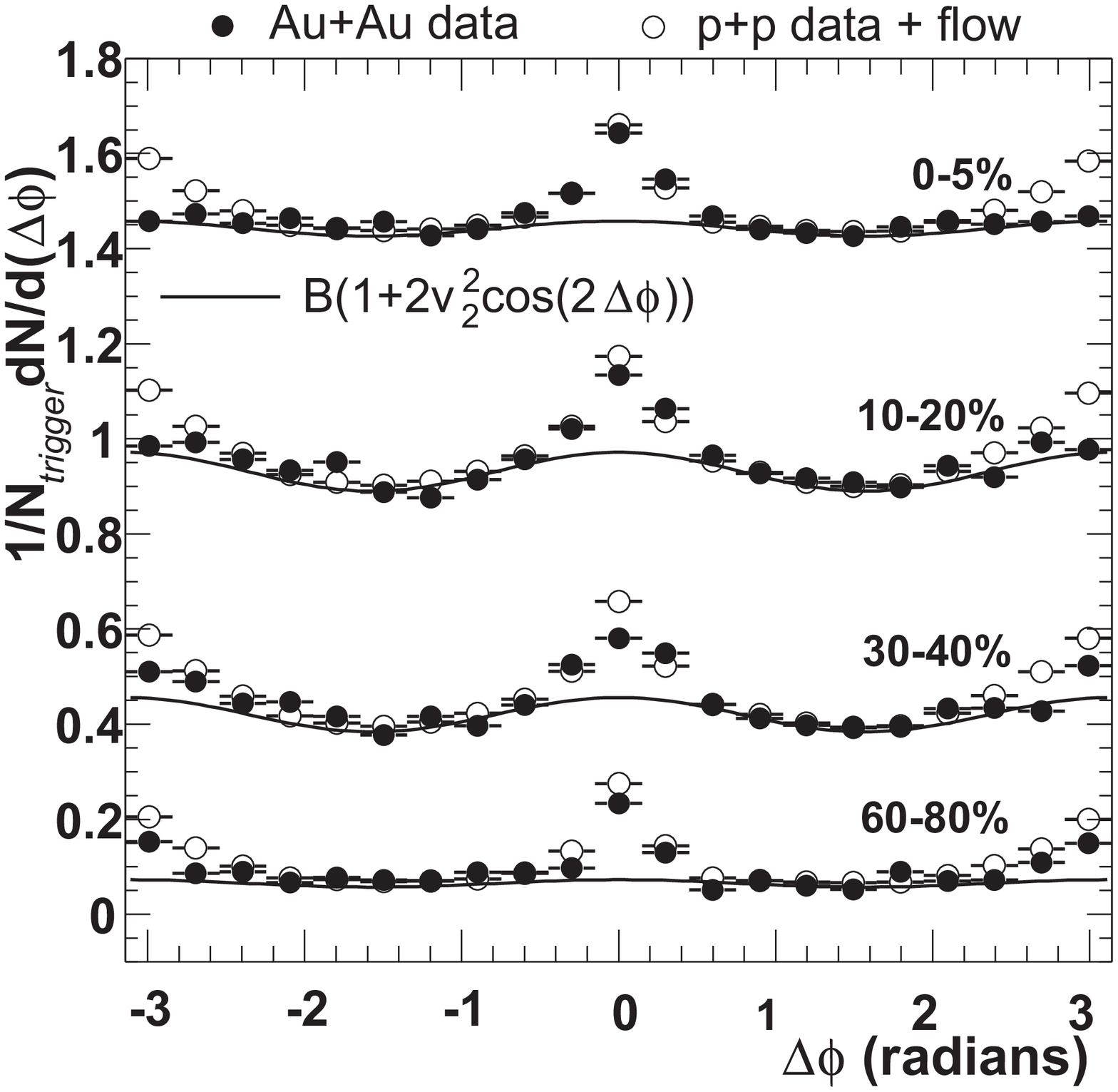,height=3.5in,width=3.0in}}  
\caption{STAR data on high $p_T$ azimuthal correlations  
of charged hadrons at $200$ AGeV 
from Ref.\protect\cite{Adler:2002tq} showing clear evidence 
near side and away side correlations characteristic of two 
jet production in elementary $p+p$. However, in $Au+Au$ 
reactions the away side jet correlations  are reduced in central 
collisions. The Back-to-Back jet correlations 
are attenuated by an order of magnitude  in the most central 
collisions as expected if the plasma produced 
is opaque\protect\cite{Gyulassy:1990ye,Gyulassy:1990bh,Pluemer:dp}.} 
\label{hardke_b2b} 
\end{figure}

\begin{figure} 
\centerline{\epsfig{file=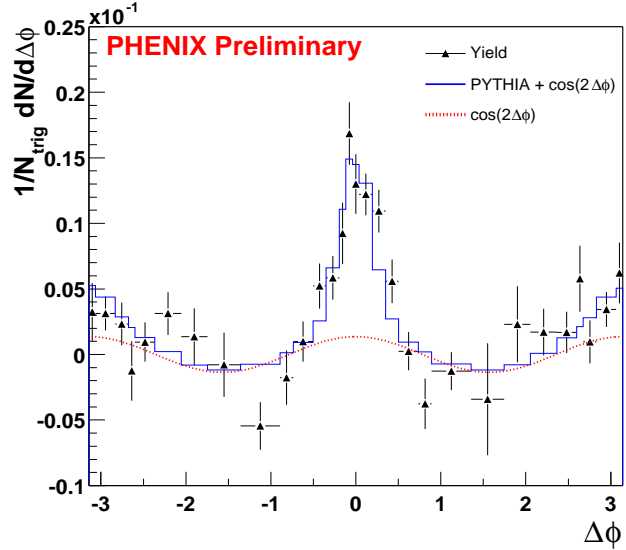,height=3.0in,width=3.5in}}  
\caption{Preliminary PHENIX data\protect\cite{d'Enterria:2002bw} on 
high $p_T$  azimuthal\protect\cite{Gyulassy:2000gk} correlations 
of {\em neutral} pions also show clear jet structure amid an elliptic
flow component.} 
\label{Chui_b2b} 
\end{figure}

The great advantage of RHIC over the previous AGS and SPS explorations 
of nuclear collisions is that the computable high $p_T$ pQCD processes 
are sufficiently abundant, and that they can be used as effective ``external 
probes'' of the quark-gluon plasma that is produced. 
High $p_T$ and heavy mass partons are produced first, 
at time $\delta \tau\sim 1/m_T$~fm/$c$, 
while  the most of the partons of the 
plasma with temperature $T$ form and equilibrate  
at later times $\sim 1/gT\sim 0.5 $ fm/$c$. 
Hard jets propagate along approximate straight
eikonal lines through the plasma until 
$\tau \sim R \sim 5 $ fm/$c$. 
The energy loss and transverse momentum broadening 
suffered by the jet  prior to hadronization provides the tomographic handle 
to probe the opacity of the evolving plasma. Since non-central collisions 
have a well known  geometrical asymmetry, the azimuthal distribution 
of final high $p_T$ hadrons provides even more 
information\cite{Wang:2000fq,Gyulassy:2000gk}. 
This is illustrated in Fig.~\ref{azimuth_gvw} by the normalized 
average thickness of nuclear matter as a function of the angle 
with respect to  the reaction plane that a fast parton sees on its way
out.  The azimuthal distribution can be Fourier decomposed into 
\begin{eqnarray} 
\frac{dN({ b})}{dyd^2{\bf p}_{T}} &=& 
 \frac{1}{\pi} \frac{dn(b)}{dy d p_T^2} 
\left(1+2 v_{1}(y,b,p_{T}) \cos(\phi-\phi_b) \right. \nonumber \\[.5ex]
 & & + \left. v_{2}(y,b,p_{T})\cos(2(\phi-\phi_b)) +  \cdots \right)  \; , 
\end{eqnarray} 
where $\phi_b$ is the azimuthal angle of the impact parameter ${ b}$. 
In Fig.~\ref{kunde_qm2_v2} the large elliptic flow component $v_2$ 
is seen to extend up to the highest $p_T$ in non-central collisions
at mid-rapidity. 
Another  critical new experimental test for jettiness 
at  moderate $p_T\sim 5$~GeV is provided by two particle correlation. 
The data from both STAR and PHENIX show clear evidence for 
back-to-back correlations predicted by pQCD (and checked via 
the PYTHIA generator). 
However, there appears to be a difference between the correlation pattern 
of charged and neutral hadrons as seen in Figs.~\ref{hardke_b2b} 
and~\ref{Chui_b2b}. The STAR data are in accord with predictions 
in Refs.\cite{Gyulassy:1990ye,Gyulassy:1990bh,Pluemer:dp} that the two jet 
structure should be  quenched by an order of magnitude relative 
to jet correlations in $p+p$ due to the high opacity of the plasma. 
Similar jet structure in high $p_T$ neutral hadron correlations is 
also found in the preliminary PHENIX  data\cite{d'Enterria:2002bw}.
Whether the suppression of away-side jet seen in STAR data also 
manifests in the PHENIX experiment has yet to be addressed.


For our review, the most important point 
about both data sets is that narrow azimuthal near side and away 
side correlations do exist. This is a necessary (though not sufficient) 
conditions to enable us to interpret 
moderate $p_T$ hadronic quenching patterns as due to jet energy loss.

In this review we  focus on the rare hard pQCD tail  
of the hadron yields where all these interesting new nuclear phenomena 
have been discovered at RHIC.  
The central element of the theory that will be elaborated in the following
sections is the predicted dependence of
the induced QCD radiative energy loss on the jet energy, the plasma density, 
and the expansion properties of the plasma. We restrict the discussion
to  two  complementary approaches, the GLV and WW/WOGZ  formalisms,
that are best suited in our opinion to applications to nuclear collision
problems  where the opacity is  necessarily  finite and the energy range 
of minijets accessible experimentally is typically $3-15$ GeV.

In the absence of final state interactions  we recall  
the well-known lowest order invariant pQCD differential cross section for  
inclusive $p+p\rightarrow h+X$   given by 
\begin{eqnarray} 
E_{h}\frac{d\sigma_{hard}^{pp\rightarrow h}}{d^3p} &=&\! 
K   \sum_{abcd}\! \int\! \!dx_a  
dx_b f_{a/p}(x_a,Q^2_a) f_{b/p}(x_b,Q^2_b) \nonumber \\ 
&\ &\hspace{1.0in} \times 
\frac{d\sigma}{d{\hat t}}(ab\rightarrow cd) 
\frac{D_{h/c}(z_c,{Q}_c^2)}{\pi z_c}\; , 
\label{hcrossec} 
\end{eqnarray} 
where $x_a=p_a/P_A, x_b=p_b/P_B$ are the initial momentum fractions carried  
by the interacting partons, $z_c=p_h/p_c$ is the momentum fraction carried  
by the final observable hadron, $f_{\alpha/p}(x_\alpha,Q^2_\alpha)$ 
are the parton distribution functions (PDFs), and  
$D_{h/c}(z_c,{Q}_c^2)$ is the fragmentation function (FFs) for 
the parton of flavor $c$ into $h$.

\begin{figure}
\begin{center} 
\psfig{file=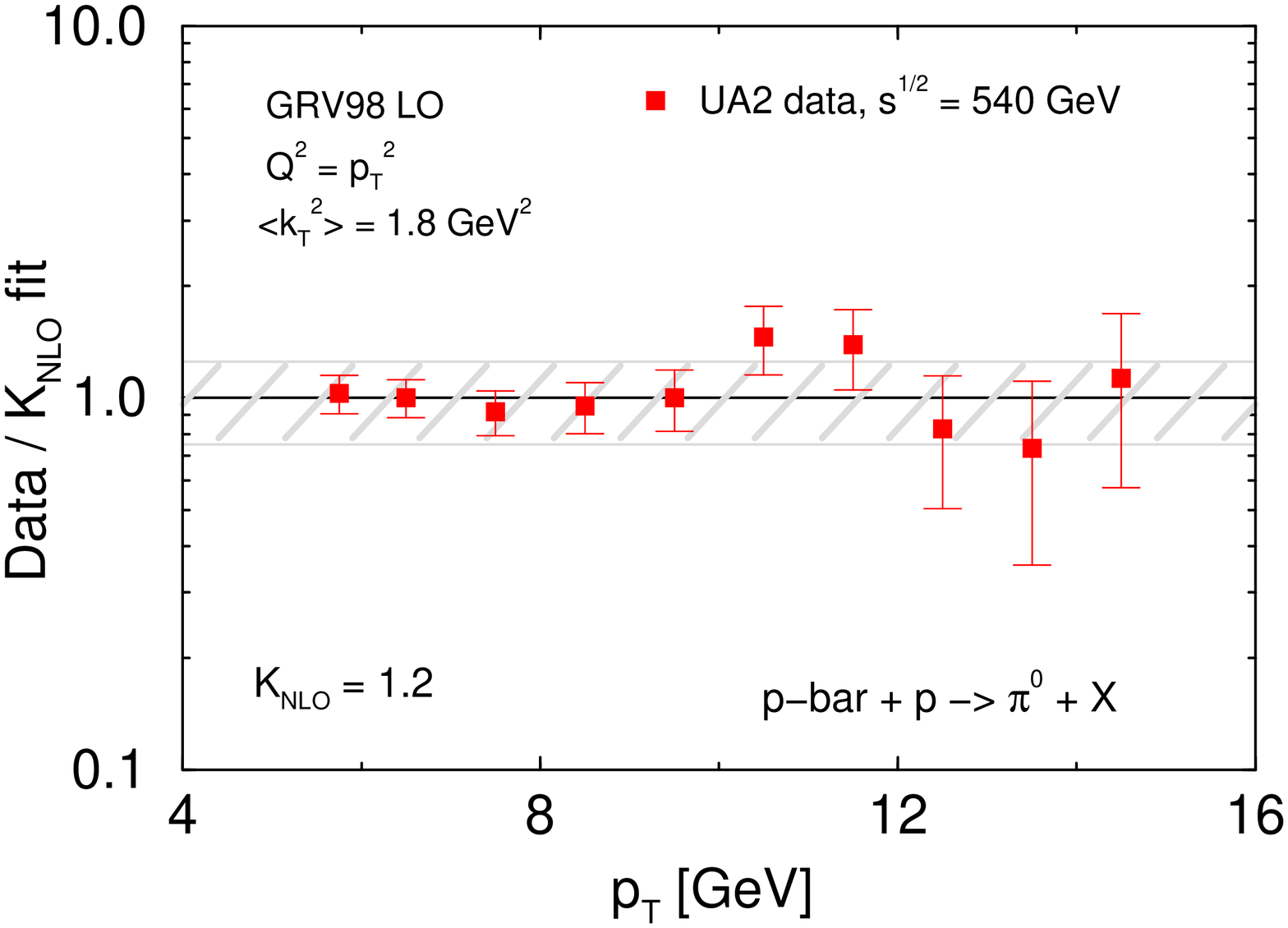,height=3.20in,width=2.3in,clip=5,angle=-90}  
\psfig{file=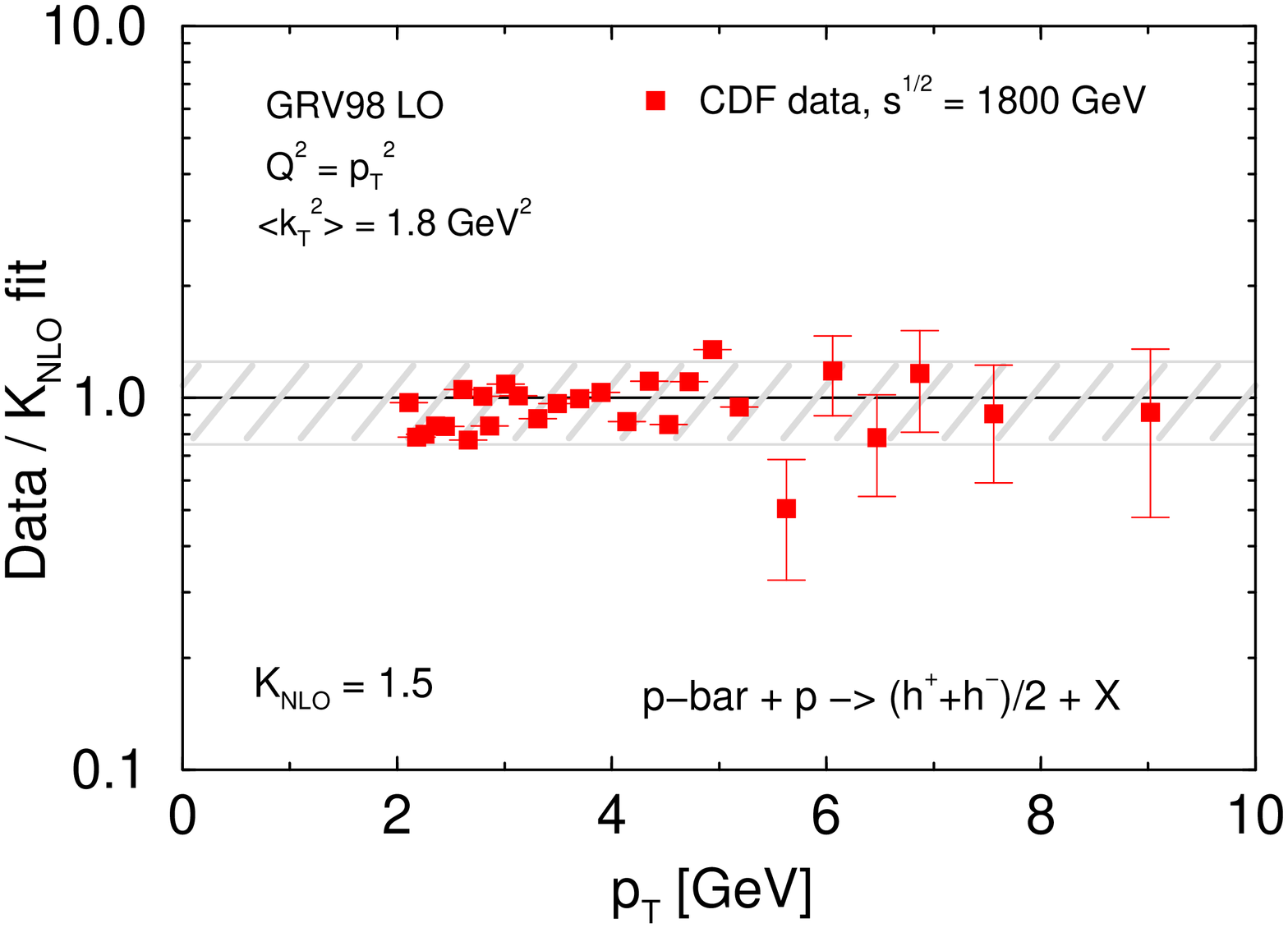,height=3.20in,width=2.3in,clip=5,angle=-90}  
\end{center} 
\caption{Comparison of the pQCD model\protect\cite{Vitev:2002aa} 
 with intrinsic $\langle k_T^2 \rangle=1.8\; 
{\rm GeV}^2$ and GRV~98 PDFs  to $\bar{p}+p$ 
data at $\sqrt{s}=540,1800$ GeV. The normalization $K$ factor simulating 
higher order corrections is fit at each energy. A $\pm25\%$ error band 
is also included.} 
\label{ivan_cern} 
\end{figure}

The phenomenological $K$ factor is introduced  
to mimic next-to-leading order (NLO) corrections.  
One  finds that Eq.(\ref{hcrossec}) tends to over-predict  
the  curvature  of the inclusive hadron spectra  
in the $p_T \leq 4$~GeV range.   This can be partially corrected 
via  the  intrinsic $k_T$-smearing of partons  
associated with vacuum radiation  and  generalized  
parton distributions  $\tilde{f}_\alpha(x,k_T,Q^2)$.   
For the corresponding modification of the kinematics 
in (\ref{hcrossec}) in addition to the $\int d^2 k_T^a  
\int  d^2 k_T^b \,(\cdots) $   see Refs.\cite{Owens:1986mp,Wang:1998ww}. 
The generalized parton distributions are often approximated as 
\begin{equation} 
\tilde{f}_\alpha(x,k_T,Q^2) \approx f_\alpha(x,Q^2) g(k_T), 
\quad  g(k_T) = \frac{e^{-k_T^2/\langle {k}_T^2 \rangle}} 
{\pi\langle {k}_T^2 \rangle } \; , 
\label{gassm} 
\end{equation} 
where the width  $\langle {k}_T^2 \rangle$ of the Gaussian 
is related to initial state vacuum radiation. Discussion 
of some qualitative features of  $\langle {k}_T^2 \rangle_{pp}$
is given in Refs.\cite{Wang:1998ww,Vitev:2003xu}.
For  comparison to experimental data the reader is 
referred to Refs.\cite{Wang:1998ww,Eskola:2002kv,Vitev:2002aa}. 
Fig.~\ref{ivan_cern} shows that this parton model approach 
provides a good description of the high $p_T$ data on hadron 
production in the range above a few GeV at all center of mass 
energies. 

In order to calculate the effects of parton energy loss on the attenuation 
pattern of high $p_T$ partons in nuclear collisions, 
we must modify the free space fragmentation 
functions\cite{Wang:1998ww,Wang:1998bh}. 
Energy loss of the parton prior to hadronization  changes the kinematic 
variables of the effective fragmentation  function. In the first 
approximation, this effect 
can be taken into account by replacing the vacuum fragmentation 
functions in  Eq.(\ref{hcrossec}) by effective quenched 
ones\cite{Wang:1996yh,Wang:1996pe}
\begin{eqnarray} 
&&z_c D_{h/c}^\prime (z_c,Q^2_c) = z_c^\prime D_{h/c} 
(z_c^\prime,Q^2_{c^\prime})  +N_g z_g D_{h/g}(z_g,Q^2_g)\; ; \nonumber \\ 
&&z_c^\prime = \frac{p_h}{p_c-\Delta E_c(p_c,\phi)} \;, 
\quad z_g  =  \frac{p_h}{\Delta E_c(p_c,\phi)/N_g} \; , 
\label{modfrag} 
\end{eqnarray} 
where $z_c^\prime,z_g$ are the rescaled momentum fractions.
The first term is the fragmentation function is of 
the jet $c$ after losing  energy $\Delta E_c(p_c,\phi)$ 
due to {\em medium induced} gluon radiation. 
The second term is the  feedback due to the fragmentation  
of the $N_g(p_c,\phi)$ radiated gluons.  
The modified fragmentation function satisfies the light-cone 
sum rule $\int dz_c\; z_c D_{h/c}^\prime (z_c,Q^2_c)=1$.  
Eq.(\ref{modfrag}) takes into account the dependence of the 
energy loss  on the parent parton energy  and the possibly azimuthally 
asymmetric region of high density nuclear matter. 

In the above first approximation, only the average value of the energy loss 
is used. One can also include multigluon 
fluctuations\cite{Baier:2001yt,Gyulassy:2001nm} 
 of the energy loss 
via an energy loss distribution $P(\epsilon,E)$ where 
$\epsilon=\sum_i \omega_i /E$ 
is the fractional energy loss of a jet of energy $E$ in the rest frame 
of the plasma.  The mean energy loss in the first approximation is related 
to $P$ via 
\begin{eqnarray} 
\int_0^\infty d\epsilon \; P(\epsilon,E) 
\epsilon= {\Delta E}/{E}  \;\; . 
\end{eqnarray} 
 
The invariant hadron distribution attenuated by fluctuating energy loss 
in  $A+A$ collision is then given by 
\begin{eqnarray} 
\label{fullaa} 
E_{h}\frac{dN_{h}^{AA}}{d^3p} &=& T_{AA}(b) \sum_{abcd}   
\int dx_1 dx_2 
\int \!  d^2 {\bf{k}}_a d^2{\bf{k}}_b \;   
g({\bf{k}}_a) g({\bf{k}}_b)  \nonumber \\[.5ex] 
 && \times \, S_A(x_a,Q^2_a)  S_B(x_b,Q^2_b)  
\; f_{a/A}(x_a,Q^2_a) f_{b/B}(x_b,Q^2_b)  \nonumber \\[1.ex] 
&& \times \,  \frac{d\sigma}{d{\hat t}}^{ab\rightarrow cd} 
 \int_0^1 d \epsilon \, P(\epsilon) 
 \frac{z^*_c}{ z_c}  \frac{D_{h/c}(z^*_c,{Q}_c^2)}{\pi z_c} \; , 
\end{eqnarray}  
where $z^*_c = z_c/(1-\epsilon)$. Here, 
$T_{AA}(b)$ is the Glauber binary collision profile density 
at impact parameter $b$. It is found numerically\cite{Gyulassy:2001nm},  
that fluctuations tend to reduce the magnitude of the attenuation 
by a factor about two at RHIC. It is important to include this effect when 
inverting  the attenuation pattern in jet tomography to determine 
the  initial plasma density.  In $B+A$ reactions isospin effects 
can be accounted   for a nucleus with $Z$ protons and  $N$ neutrons  
by $ f_{\alpha/A}(x_\alpha,Q_\alpha^2)=  
(Z/A) \, f_{\alpha/p}(x_\alpha,Q_\alpha^2) 
 +  (N/A) \, f_{\alpha/n} (x_\alpha,Q_\alpha^2)$. 
Nuclear modifications to the PDFs are  
estimated in our applications by using the
shadowing function $S_A(x_\alpha,Q^2_\alpha)$ 
proposed by  EKS'98\cite{Eskola:1998df}.

\section{GLV Theory of Radiative Energy Loss} 
 
In this section we turn  to the theoretical problem of computing 
the energy loss  of a fast parton penetrating a finite, 
expanding quark-gluon plasma. 
High $p_T$ many body pQCD  is a new frontier of research at RHIC 
and eventually LHC. 
The needed new theoretical development is  
the  non-Abelian analogue of the radiative energy loss theory 
familiar from classical E\&M.
An important practical problem in QCD is that 
there exists no external  beam of isolated  
colored quarks or gluons, and that the jets are always 
produced in hard processes
just before the plasma is created. In addition,
the nuclear medium has a small 
dimension compared to the  jet  fragmentation coherence length. 
Therefore, the basic   formation time physics of 
Landau-Pomeranchuk-Migdal\cite{Landau:um,Landau:gr,Migdal:1956tc} (LPM) 
is expected to lead  to strong  destructive interference effects 
that have to be  carefully  taken into account.

In this section we review one of those 
approaches\cite{Gyulassy:2000er,Gyulassy:2000fs,Gyulassy:1999zd,Gyulassy:1999ig} 
developed by Gyulassy-Levai-Vitev (GLV) 
based on an algebraic reaction  
operator formalism. The expansion parameter is the 
opacity $\chi=L/\lambda=\sigma\rho L$ of the system and the result 
is presented to  all orders in powers of $\chi$ or equivalently 
to all twist parton-parton correlations.
Other approaches relying on asymptotic techniques include 
BDMS\cite{Baier:1996kr,Baier:1996sk,Baier:1998kq,Baier:1999ds}, 
Z\cite{Zakharov:1996fv,Zakharov:1997uu,Zakharov:2000iz,Zakharov:2002ik}, 
and 
SW\cite{Wiedemann:2000tf,Wiedemann:2000za,Wiedemann:2000ez,Salgado:2002cd} 
have been reviewed 
elsewhere\cite{Baier:2000mf}. In the second part of this report the 
twist expansion approach\cite{Wang:2001if,Guo:2000nz,Wang:2002ri} for
parton energy loss in nuclear matter is reviewed.

\subsection{The GW Plasma Model} 

The GLV approach is built around a simple model of multiple 
scattering in a plasma formulated in GW\cite{Gyulassy:1994hr}. 
However, the results of GLV\cite{Gyulassy:2000er,Gyulassy:2000fs} 
are more general since 
even out of thermal equilibrium the effective  in-medium 
interactions are of finite range $R \simeq \mu^{-1}$.
Consider 
the sequential elastic scattering of a high energy (jet) 
parton  in the random color field produced by an ensemble of $m$ 
static partons located at ${\bf x}_i=(z_i,{\bf x}_{\perp i})$ 
such that $z_{i+1}>z_i$ and $(z_{i+1}-z_i)\gg \mu^{-1}$, 
where  $\mu$ is the color screening mass in the medium. 
As a simplified 
model of multiple scattering in  a color neutral quark-gluon plasma, we 
 assume a  static Debye screened potential for each target parton: 
\begin{equation} 
V_i^{a}({\bf q})= g(T^a_i)_{c,c^\prime} 
 \frac{1}{{\bf q}^2  + \mu^2}  \, e^{-i{\bf q}\cdot{\bf x}_i} 
\; \; , 
\label{aia} 
\end{equation} 
where $T^a_i$ 
is a $d_i$-dimensional generator of $SU(N)$ 
corresponding to the representation of the 
target parton  at ${\bf x}_i$. 
The initial and final color indices, 
$c, c^\prime$, which refer to the target  parton, are  averaged 
and summed over when computing  the ensemble averaged cross sections. 
With $V_i^{a}\propto T^a_i$ the  ensemble averaged potential 
vanishes everywhere,  $\langle V_i^a \rangle\propto TrT^a_i= 0$. 
However, since 
\begin{equation}Tr{T}^a_i{T}^b_j=\delta_{ab}\delta_{ij} 
({d}_i/d_A){C}_{2i} 
\; \; ,
\label{tai} \end{equation} 
the diagonal mean square fluctuations and the cross sections are finite. 
Recall that for $SU(N)$  the second order Casimir, 
${C}_{2i}=(N^2-1)/2N\equiv  C_F$ 
for quarks in the fundamental ($d_i=N)$ representation, 
while ${C}_{2i}=N\equiv C_A$ for gluons in the adjoint 
($d_i=N^2-1\equiv d_A$) representation. 
 
In this potential,  each scattering leads on the average to only 
a relatively small momentum transfer 
$q_i^\mu=(q_i^0,q_{zi}, {\bf q}_{\perp i})$ with each component 
being much less than  the incident energy, $E_0$. 
The assumption that the  potentials are  static is approximately 
valid in a high temperature  plasma 
of  massless quarks and gluons in the following sense: 
As  $T\rightarrow \infty$, the effective coupling $g\rightarrow 0$ 
(albeit very slowly). The perturbative 
 Debye screening mass $\mu \sim gT$ limits 
$q_\perp$\raisebox{-.6ex}{$\stackrel{<}{\sim}$}$gT$. The typical thermal energy 
$E_T\sim 3T$ of the plasma constituents is therefore large 
compared to $\mu$. Consequently, the average energy loss per 
elastic collision, $ -q^0 \approx -q^z \approx q_\perp^2/2 E_T 
\propto  g^2 T $, is $\sim g$ times smaller than the average 
transverse momentum transfer.

Because we are interested in relatively low momentum transfer scattering 
($\Lambda_{QCD}\ll q_\perp\sim gT \ll T)$, 
 the  spin of the partons can be neglected. 
The jet parton is allowed, however, 
to be in an arbitrary $d$-dimensional representation of $SU(N)$ 
with generators, $T^a$, satisfying  $T^aT^a=C_2{\bf 1}_d$.

The  Born (color matrix) amplitude  to scatter from an incident 
four momentum $p_{i-1}^\mu$ 
to $p_i^\mu$ in the potential centered at ${\bf x}_i$ is then given by 
\begin{equation}
M_i(p_i,p_{i-1})=2\pi\delta(p^0_i-p^0_{i-1})A_i({\bf q}_i) 
e^{-i{\bf q}_i\cdot{\bf x}_i} \; \; ,  
\label{vi} 
\end{equation} 
where ${\bf q}_i={\bf p}_i-{\bf p}_{i-1}$, and $A_i$ is shorthand for 
\begin{equation}A_i({\bf q}_i)= T^a A_i^a 
({\bf q}_i)=-2igE_0 T^a V_i^{a}({\bf q}_i) \; \; . 
\label{ai} 
\end{equation} 
The differential cross section  averaged over initial and summed over 
final colors of both projectile and target partons reduces to the familiar 
form for  low transverse momentum transfers: 
\begin{equation}
d\sigma_{i}/d q_{\perp i}^2 \approx 
 C_i\frac{4\pi\alpha^2}{(q_{\perp i}^2+\mu^2)^2} \; \; , 
\label{dsigi} 
\end{equation} 
 where the color factor is 
\begin{equation}C_i=\frac{1}{d d_i} 
Tr({T}^a{T}^b)Tr (T^a_i{T}^b_i)= C_2 {C}_{2i}/d_A 
\; \; . \label{ci} \end{equation} 
For $SU(3)$, the number $2C_i$ gives the usual 
color factors 
$4/9,1,9/4$ for $qq,qg,gg$ scattering respectively. 
In our notation, the angular distribution is given by 
\begin{equation}d\sigma_{i}/d\Omega_i=\frac{1}{d d_i}Tr 
|A_i({\bf q}_i)|^2/(4\pi)^2 \; \; .
 \label{dsido} 
\end{equation}

\subsection{GLV Formalism} 
 
\begin{figure} 
\begin{center} 
\epsfig{file=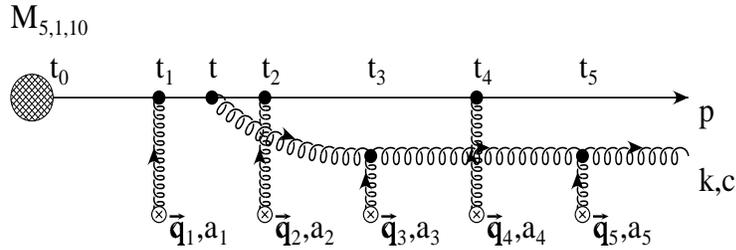,height=5.0cm,  
width=10cm,clip=5,angle=0}  
\vspace*{-.4cm}
\caption{Induced radiation amplitude\protect\cite{Gyulassy:1999zd}
 contributing 
to fifth order and higher orders in the opacity expansion of QCD energy 
loss  in the GW model\protect\cite{Gyulassy:1994hr}. The crosses denote 
color screened Yukawa interactions on a scale $\mu$.  The blob is the 
initial hard jet amplitude. } 
\label{psmq5xfig} 
\end{center} 
\end{figure}

\begin{figure} 
\centerline{\psfig{file=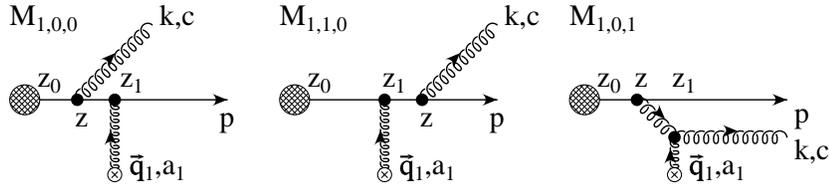,width=11.cm,angle=0}}
\caption{Three first order (singe Born) direct  
amplitudes that 
serve to define the $\hat{D}_n$ component 
of the reaction operator in Eq.(\protect\ref{reacop}) 
are derived in Refs.\protect\cite{Gyulassy:2000er,Gyulassy:2000fs}.} 
\label{dandv} 
\end{figure}

\begin{figure} 
\centerline{\hspace*{.7cm}\psfig{file=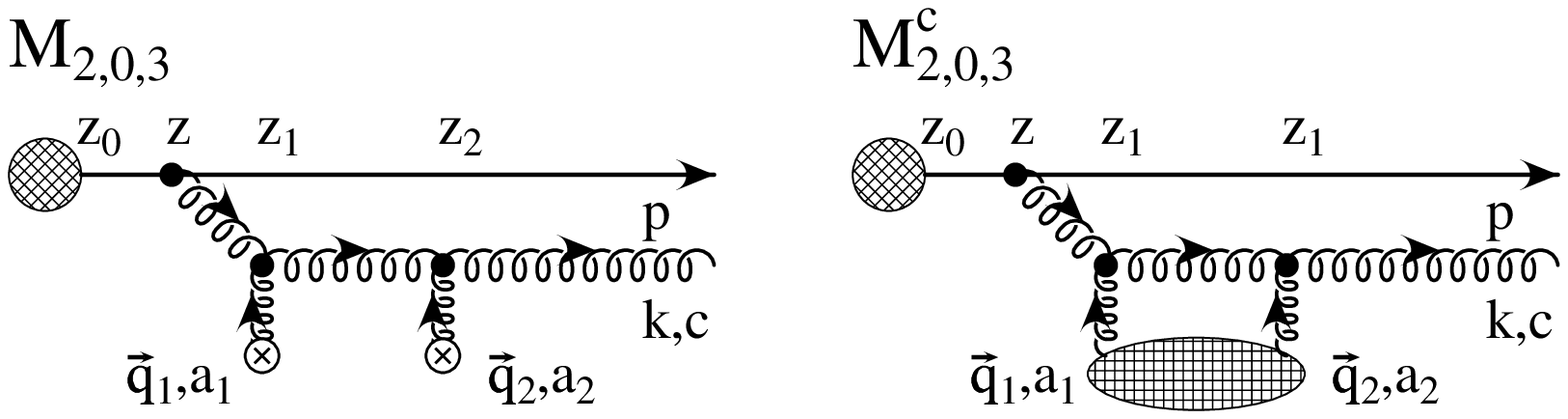,width=9.cm,angle=0} }
\vspace*{.5cm}
\centerline{\hspace*{.7cm}\psfig{file=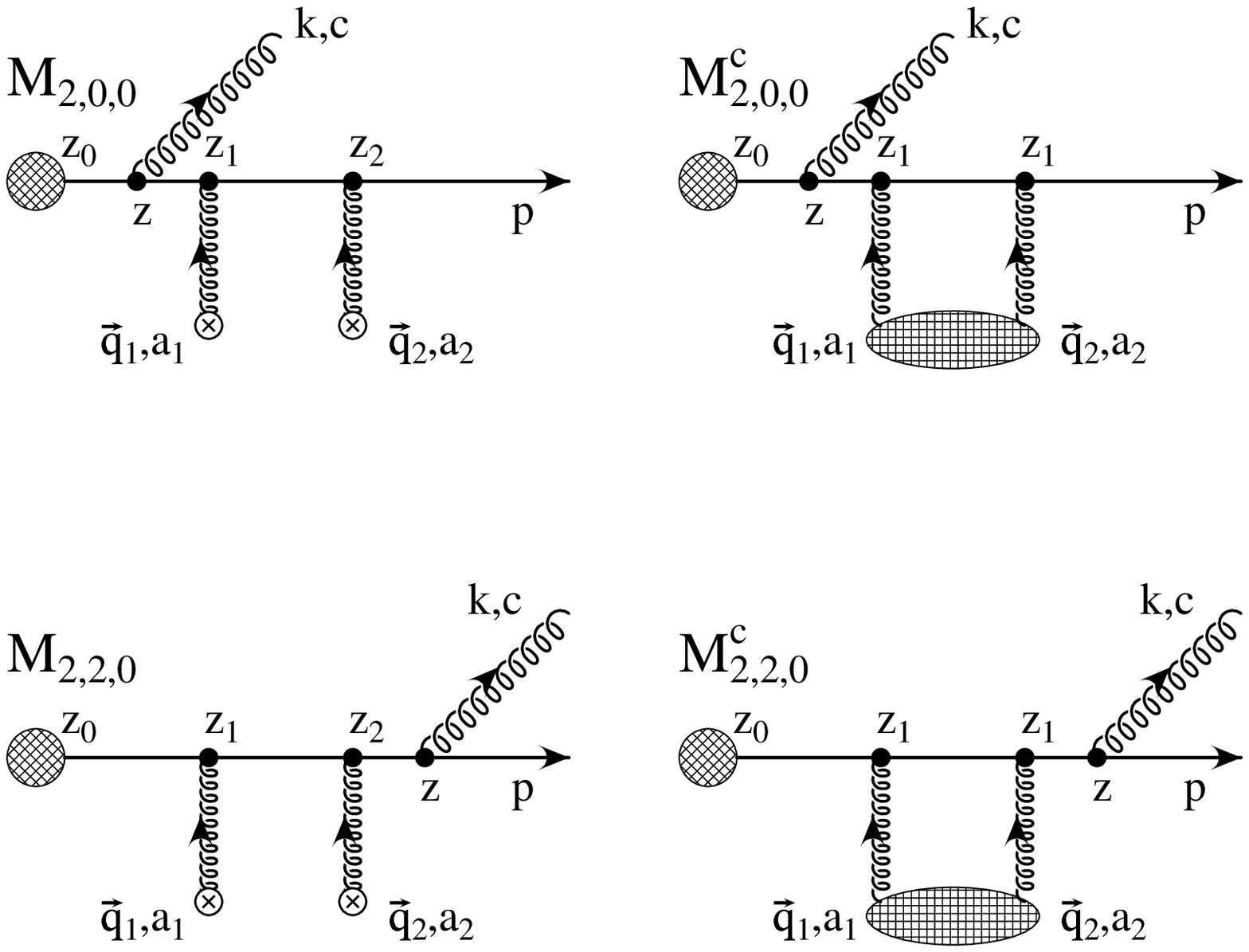,width=9.cm,angle=0} }   
\centerline{\hspace*{.7cm}\psfig{file=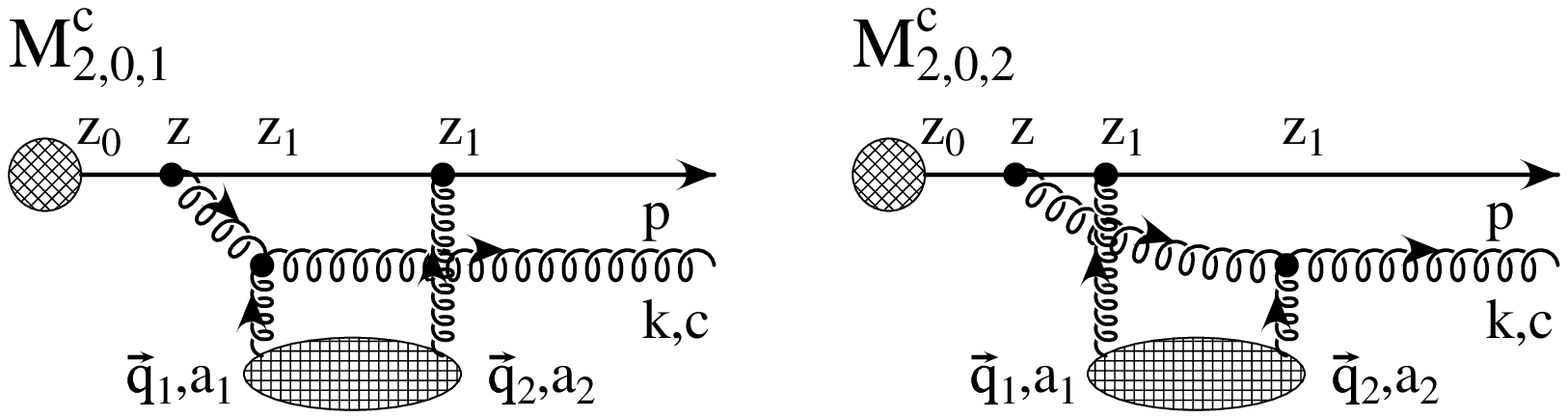,width=9.cm,angle=0} }     
\vspace*{.5cm}
\centerline{\hspace*{.7cm}\psfig{file=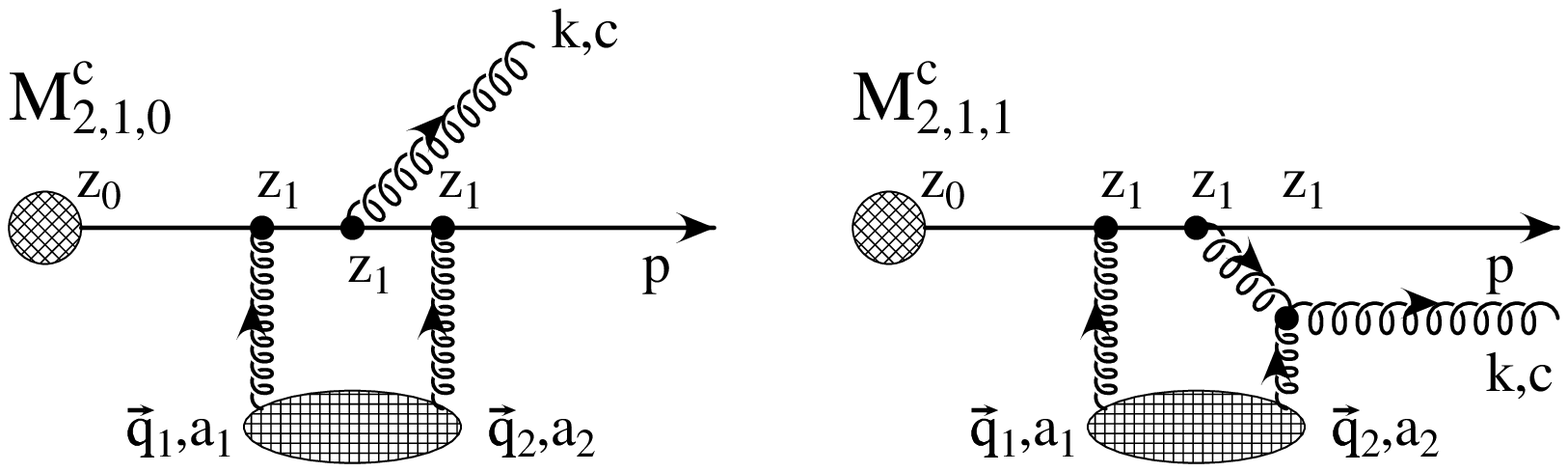,width=9.cm,angle=0} }    
\caption{ Diagrams with two momentum exchanges that contribute 
to first and second orders in opacity. In the contact ($z_2=z_1$)
limit the  (double Born) virtual (contact) amplitudes
define the $\hat{V}_n$ components 
of the reaction operator in 
Eq.(\protect\ref{reacop})\protect\cite{Gyulassy:2000er,Gyulassy:2000fs}. 
Note that two of the diagrams vanish due to zero
measure $\int_{z_1}^{z_1} dz \, \cdots = 0$ of the radiation interval     
and two are topologically 
indistinct\protect\cite{Gyulassy:2000er,Gyulassy:2000fs}. } 
\label{dandv1} 
\end{figure}

In Refs.\cite{Gyulassy:1999zd,Gyulassy:1999ig} a systematic 
recursive graphical technique was developed and translated into an 
algebraic operator method. The goal was to  compute medium induced  
gluon radiation 
amplitudes of the type shown   in Fig.~\ref{psmq5xfig}. 
The exponential growth of the number of graphs with the number of 
interactions  makes it very tedious to go beyond order three in opacity
$\chi$ or twist 8 ($2+3 \times 2$) from the parton-parton 
correlations in the medium. 
In  GLV\cite{Gyulassy:2000er,Gyulassy:2000fs} the combinatorial 
problem is solved by summing  the inclusive radiative gluon 
distributions recursively. The first step in the approach is to 
compute the three direct (single Born) and four surviving
virtual (contact double Born) diagrams shown in 
Figs.~\ref{dandv} and~\ref{dandv1} 
that contribute to the first order in opacity induced radiation.
Detailed instructive calculation of these amplitudes  
is given in\cite{Gyulassy:2000er}.

\begin{figure} 
\begin{center} 
\centerline{\psfig{file=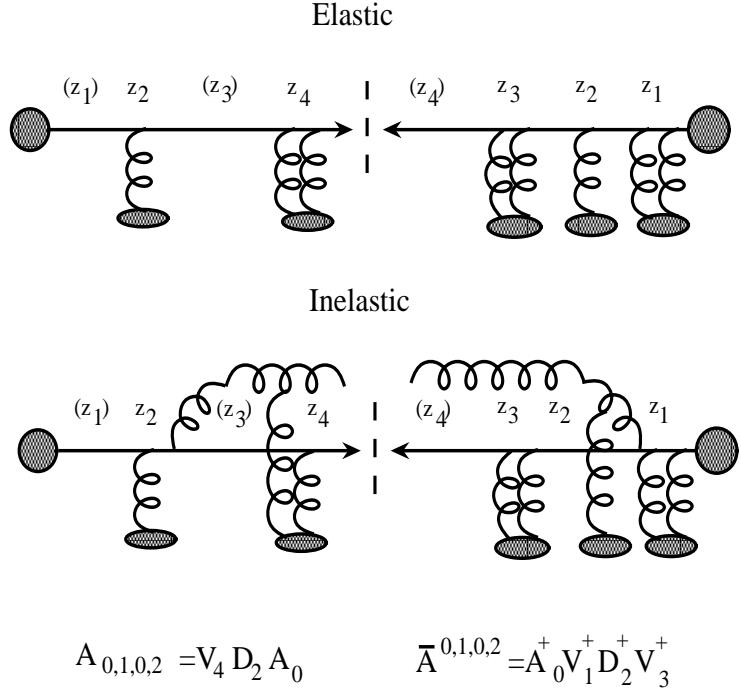,height=4.4in, width=5.5in,angle=0} }
\caption{Example of graphs constructed via $\hat{D}_i,\hat{V}_i$ 
that contribute to the 4-th order in opacity 
(or twist 2+8 in-medium parton-parton correlations) 
in elastic and inclusive inelastic final state interactions. 
The longitudinal depths of active 
scattering centers are denoted by $z_i$ and inactive (created with  
$\hat{1}_i$) by $(z_i)$. The form of $\hat{D}_i,\hat{V}_i$ depend 
on the process type but the tensorial bookkeeping of partial sums 
of amplitudes is the same.} 
\label{4thorder} 
\end{center} 
\end{figure} 
 
%macros 
\newcommand{\beq}{\begin{equation}} 
\newcommand{\eeq}[1]{\label{#1}\end{equation}} 
\newcommand{\ch}{{\rm ch}} 
\renewcommand{\vec}[1]{{\bf #1}} 
\newcommand{\vAi}{{\cal A}_{i_1\cdots i_n}} 
\newcommand{\vAim}{{\cal A}_{i_1\cdots i_{n-1}}} 
\newcommand{\vAbi}{\bar{\cal A}^{i_1\cdots i_n}} 
\newcommand{\vAbim}{\bar{\cal A}^{i_1\cdots i_{n-1}}} 
\newcommand{\htS}{\hat{S}} 
\newcommand{\htR}{\hat{R}} 
\newcommand{\htB}{\hat{B}} 
\newcommand{\htD}{\hat{D}} 
\newcommand{\htV}{\hat{V}} 
\newcommand{\cT}{{\cal T}} 
\newcommand{\cM}{{\cal M}} 
\newcommand{\cMs}{{\cal M}^*} 
\newcommand{\vk}{{\vec k}} 
\newcommand{\vK}{{\vec K}} 
\newcommand{\vb}{{\vec b}} 
\newcommand{{\vq}}{{\vec q}} 
\newcommand{\vQ}{{\vec Q}} 
\newcommand{\tr}{{{\rm Tr}}} 
\newcommand{\half}{\frac{1}{2}} 
For scattering of a jet off  of $n$ scattering centers  
located at depths $z_i$ in a homogeneous medium of large area 
$\mu^2 A_\perp \gg 1$, 
we can write the inclusive radiative gluon ``probability'',
 $P_n({\bf k},c)$, as a sum over products of partial 
sums of amplitudes and complementary  complex conjugate 
amplitudes\cite{Gyulassy:2000er,Gyulassy:2000fs}. Every term 
in the sum  contributes  to the same ${\cal O}(g^{4n+2})$ 
in the strong coupling constant $g$, where 
$g^2$ comes from  the radiation vertex and $(g^4)^n$ is related to 
the $n$ elastic scattering cross sections. The average value 
of $n$ is  the opacity, $\chi = \langle n \rangle=L/\lambda$, 
of the medium. It turns out 
surprisingly\cite{Baier:1996kr,Baier:1996sk} 
that $\lambda$  is the mean free path of radiated gluons rather than 
the jet  itself,  as we shall show. 
The partial sums of diagrams at order $n$ in the opacity expansion 
can be conveniently expressed in a tensor notation  and  constructed 
by repeated operations of $\hat{\bf 1},\htD_i$, or $\htV_i$ 
corresponding to no, direct, or virtual interactions at  
scattering center $i$:
\beq 
\vAi(x,{\bf k},c)=\prod_{m=1}^n 
\left( \delta_{0,i_m} + \delta_{1,i_m} \htD_m + \delta_{2,i_m}  
\htV_m  \right) G_0(x,{\bf k},c)  \; \; . 
\eeq{atens} 
Here $G_0$ is the initial hard $\{q,g\}+g$ color matrix amplitude, 
$x=k^+/p^+$ is the gluon lightcone momentum fraction, ${\bf k}$ is the 
gluon transverse momentum, and $c$ is its color index.

In the inclusive probability each class contracts with 
a unique complementary  class,
\beq  
P_n(x,{\bf k})=\vAbi(x,{\bf k},c)\vAi(x,{\bf k},c).
\eeq{} 
The complementary class is constructed to satisfy 
the fixed opacity power $\propto \chi^n$ requirement  as 
\beq 
\vAbi(x,{\bf k},c) \equiv G_0^\dagger 
(x,{\bf k},c) \prod_{m=1}^n 
\left( \delta_{0,i_m} \hat{V}_m^\dagger +  
\delta_{1,i_m} \hat{D}_m^\dagger  
+ \delta_{2,i_m} \right)  \; \; .
\eeq{} 
Fig.~\ref{4thorder} shows an example of how this  formalism works at 
4th order in opacity for elastic and inelastic inclusive distributions.

Detailed  diagrammatic calculations that illustrate the color and 
kinematic modifications to the amplitudes for both direct and double
Born momentum and color exchanges with the propagating
jet+gluon system  are given in Refs.\cite{Gyulassy:2000er,Gyulassy:2000fs}. 
Direct  single momentum transfer  interactions  
enlarge  rank $n-1$ class elements as follows: 
\begin{eqnarray} 
& &  \htD_n \vAim(x,{\bf k},c) 
\equiv  (a_n + \htS_n + \htB_n)  
\vAim(x,{\bf k},c) \nonumber  \\[1.5ex]  \nonumber 
&& = {a_n} \vAim(x,{\bf k},c) +  
e^{i(\omega_0-\omega_n)z_n }  
\vAim(x,{\bf k}- {{\bf q}_{n}},{[c,a_n]}) 
\nonumber \\[1.ex] 
&&
-{\left(-\half \,\right )^{N_v(\vAim)}} {\bf B}_n \,  
e^{i \omega_0 z_n} {[c,a_n]} \,  T_{el}(\vAim) \;\; ,  
\end{eqnarray} 
where ${\bf B}_n \, = 
{\bf H}-{\bf C}_n=\vk/\vk^2-(\vk-\vq_n)/(\vk-\vq_n)^2)$ 
is the so-called Bertsch-Gunion amplitude 
for producing a gluon with transverse momentum $\vk$ 
in an isolated single collision 
with scattering center $n$. The momentum transfer to the jet is 
$\vq_n$. 
The notation $\omega_n=(\vk-\vq_n)^2/2\omega$ is 
for a gluon with energy $\omega$ 
and $a_n$ is the color matrix in the $d_R$ dimensional 
representation of the jet with color Casimir $C_R$. 
$N_v=\sum^{n-1}_{m=1}\delta_{i_m,2}$  
counts the number of virtual interactions 
in $\vAim$.   $T_{el}(\vAim)$ is the elastic color factor associated with  
all $n-1$ momentum transfers from the medium to the jet line.

Unitarity (virtual forward scattering) corrections  
to the direct processes  involve the sum of four surviving double 
Born contact diagrams in Fig.~\ref{dandv} 
that  enlarge  rank $n-1$ classes via the action of 
\begin{equation}
\htV_n  =  -\half(C_A+C_R) - a_n \htS_n- a_n\htB_n 
= -a_n \htD_n - \half(C_A-C_R) \;\;.
\label{key} 
\end{equation} 
This {\em key} operator relationship between direct and virtual 
insertions discovered  in Refs.\cite{Gyulassy:2000er,Gyulassy:2000fs} 
reduces the problem  to  recursive algebra. 
More specifically:
\begin{eqnarray}
&&\htV_n \vAim(x,{\bf k},c)
 \equiv    \left(  -\half(C_A+C_R) - a_n \htS_n- a_n\htB_n  
\right)  \vAim(x,{\bf k},c)  \nonumber \\[.5ex]
&&= - \frac{C_R+C_A}{2} \vAim(x,{\bf k},c)
 -  e^{i(\omega_0-\omega_n)z_n } a_n
\vAim(x,{\bf k}- {\bf q}_{n}, [c,a_n])
\nonumber \\[.5ex]
&& - \left(-\half\,\right)^{N_v}
\frac{C_A}{2} \, {\bf B}_n \,
e^{i \omega_0 z_n} c \, a_{n-1}^{i_{n-1}}\cdots a_1^{i_1} \;\;.
\label{vidamit}
\end{eqnarray}

The tensorial bookkeeping of classes of diagrams 
makes it possible to construct the distribution of radiated 
gluons in the case of $n$  interactions, 
$P_n(x,{\bf k}) $, recursively  from lower rank (opacity) classes 
via a {``reaction''} operator 
\begin{equation} 
P_n=\bar{\cal A}^{i_1\cdots i_{n-1}}\htR_n 
{\cal A}_{i_1\cdots i_{n-1}} 
\;, \qquad \htR_n  = \hat{D}_n^\dagger 
\hat{D}_n+\hat{V}_n+\hat{V}_n^\dagger \;\;. 
\label{reacop} 
\end{equation} 
Using the key identity (\ref{key}), the reaction matrix simplifies to 
$${\htR_n=  (\htD_{n}-a_{n})^\dagger  (\htD_{n}-a_{n}) - C_A} 
{=(\htS_{n}+\htB_{n})^\dagger  (\htS_{n}+\htB_{n}) -C_A} \;\;. $$ 
 
The next major simplification 
occurs because both $\htS$ and $\htB$ involve the same 
gluon color rotation through  $if^{ca_{n} d}$. This fact reduces the 
color algebra to simple multiplicative Casimir factors in the
adjoint representation 
\begin{eqnarray} 
\vAbim(\htS_{n}^\dagger \htS_{n}-C_A)\vAim  
&=& C_A\left( P_{n-1}({\bf k}-{\bf q}_{n}) 
- P_{n-1}({\bf k})\right)  \nonumber      \\[.5ex] 
&=& C_A\left( e^{i{\bf q}_{n}\cdot\hat{\bf b}}-1 \right) 
P_{n-1}({\bf k}) \;\;, 
\label{need1}
\end{eqnarray} 
where $\hat{\bf b}=i\nabla_{\bf k}$ is the transverse momentum 
shift operator.  We have proved in Refs.\cite{Gyulassy:2000er,Gyulassy:2000fs} 
that 
\beq 
\vAbim\htB_{n}^\dagger \htB_{n}\vAim =0 \; \; .  
\eeq{need2} 
The interference term is also found to be color trivial 
(containing  only powers of the Casimir in the adjoint representation 
$C_A$) and can be  expressed in recursive form: 
\beq 
2 {\bf Re}\, \vAbim\htB_{n}^\dagger \htS_{n}\vAim 
= -2 C_A\, {\bf B}_n\cdot \left({\rm Re}\; e^{-i\omega_nz_n}  
e^{i{\bf q_n}\cdot\hat{\bf b}} {\bf I}_{n-1}\right) \;\; .  
\eeq{need3}
Eqs.(\ref{need1})--(\ref{need3})  present a clear  
proof that only the gluon mean free path which is proportional to 
$C_A$ (or equivalently the gluon elastic scattering cross section) 
appears in the induced non-Abelian bremsstrahlung formulas. 
 
 The transverse vector amplitude ${\bf I}_n$ above obeys 
 a recursion relation  
\beq 
{\bf I }_n =C_A\left(e^{i(\omega_0-\omega_n)z_n} 
 e^{i{\bf q}_{n}\cdot\hat{\bf b}} -1\right) {\bf I}_{n-1} 
-\delta_{n,1} C_A C_R {\bf B}_{1} e^{i\omega_0 z_{1}} \;\; ,  
\eeq{bfin} 
where  ${\bf I}_0= -C_R {\bf H}e^{i\omega_0 z_0}$,   
and the Gunion-Bertsch amplitude can be written as  
\beq 
{\bf B}_{1} e^{i\omega_0 z_{1}}=-\left(e^{i(\omega_0-\omega_1) z_1}  
    e^{i{\bf q}_{1}\cdot \hat{\bf b}} -1\right)  
    {\bf H}e^{i\omega_0 z_1} \;\; .  
\eeq{bg1r} 
For $n\ge 1$ Eq.(\ref{bfin}) can be solved in  closed form   
\beq 
{\bf I}_n= C_R C_A^n\left[\, \prod_{m=1}^n 
\left(e^{i(\omega_0-\omega_m)z_m} 
 e^{i{\bf q}_{m}\cdot\hat{\bf b}} -1\right)\right]  
\; {\bf H}\left(e^{i\omega_0 z_1}-e^{i\omega_0 z_0}\right) \;\;,  
\eeq{reci} 
where the product is understood as ordered for left to right 
in decreasing order in  operators labeled by $m$.

The  inclusive radiation  ``probability''  is then found to obey 
the soluble  recursion  relation 
\begin{eqnarray} 
P_{n}({\bf k})&=& C_A  (P_{n-1}({\bf k}-{\bf q}_{n}) 
  - P_{n-1}({\bf k)})    \nonumber  \\[.5ex] 
&-&2 C_A\, {\bf B}_n\cdot \left( {\bf Re}\; e^{-i\omega_nz_n}  
e^{i{\bf q_n}\cdot\hat{\bf b}} {\bf I}_{n-1} \right) 
+ \delta_{n,1} C_A C_R |{\bf B}_1|^2 \;\; .
\end{eqnarray} 
The initial condition for this recursion relation 
is the initial hard vertex radiation amplitude without final 
state  interactions that is given by  $P_0 = C_R \, 
{\bf H}^2={C_R}/{{\bf k}^2} $.

The  solution to the problem for any order $n$ in mutiparticle
interactions can therefore 
be expressed in closed form as 
\begin{eqnarray} 
P_{n}({\bf k})&=& -2{C_R C_A^n} \, {\bf Re} 
\sum_{i=1}^n  
\left\{\prod_{j=i+1}^n( e^{i{\bf q}_{j}\cdot \hat{\bf b}} - 1)   
\right\}  {\otimes} {\bf B}_{i} \cdot\;  
e^{i{\bf q}_{i}\cdot \hat{\bf b}}   e^{-i\omega_0 z_i} 
\nonumber \\[.5ex]  &\times& \left\{ 
 \prod_{m=1}^{i-1}(e^{i(\omega_0-\omega_m)z_m} 
  e^{i{\bf q}_{m}\cdot \hat{\bf b}} -1)   \right\} 
{\otimes}   {\bf H}(e^{i\omega_0 z_1}-e^{i\omega_0 z_0})  \;\;.
\end{eqnarray} 
This expression  is very general and is suitable for any 
distribution of interaction centers 
as well as any distribution of $z_i$ dependent transverse 
momenta exchanges that can arise due to expansion of the medium.
 Therefore, it can be directly applied to realistic expanding 
plasmas where the distance between adjacent centers (the mean free path) 
varies along the medium as does the local screening scale, $\mu(z)$.  
This form is also well suited for possible future 
 Monte Carlo implementation  for arbitrary ${\bf q}_i, z_i$ 
medium ensemble averages.

The final complete solution to the inclusive induced gluon  
radiation valid to {\em all} orders in opacity 
can be expressed in terms of the following infinite series:
\begin{eqnarray} 
x\frac{dN^g}{dx\, d^2 {\bf k}} &=& 
\frac{C_R \alpha_s}{\pi^2} \sum_{n=1}^\infty 
\frac{1}{n!}  
 \prod_{i=1}^n \left( \int_{z_{i-1}}^\infty dz_i \int d^2{\bf q}_{i}\,  
\left[ \frac{d^2\sigma_g(z_i)}{d^2{\bf q}_{i}}  
 - \sigma_g(z_i) \delta^2({\bf q}_{i}) \right]\right) \nonumber \\ 
&\times& \rho_n(z_1,\cdots,z_n) 
\left( -2\,{\bf C}_{(1, \cdots ,n)} \cdot  
\sum_{m=1}^n {\bf B}_{(m+1, \cdots ,n)(m, \cdots, n)} \right. 
\nonumber \\ 
&\times& \left.\left[ \cos \left ( 
\, \sum_{k=2}^m \omega_{(k,\cdots,n)} \Delta z_k \right) 
-   \cos \left (\, \sum_{k=1}^m \omega_{(k,\cdots,n)} \Delta z_k \right) 
\right]\; \right) \;\;, 
\label{ndifdis}  
\end{eqnarray} 
where $\sum_2^1 \equiv 0$ is understood. 
The notation is defined as follows:
\begin{eqnarray}
{\bf H}&=&{{\bf k} \over {\bf k}^2 }\; , \qquad \qquad
{\bf C}_{(i_1i_2 \cdots i_m)}={({\bf k} - {\bf q}_{i_1} - 
{\bf q}_{i_2}- 
\cdots -{\bf q}_{i_m} ) 
\over ({\bf k} - {\bf q}_{i_1} - 
{\bf q}_{i_2}- 
\cdots -{\bf q}_{i_m}   )^2 } \;, 
\nonumber \\[1.ex]
{\bf B}_i &= &{\bf H} - {\bf C}_i \; , \qquad
{\bf B}_{(i_1 i_2 \cdots i_m )(j_1j_2 \cdots i_n)} = 
{\bf C}_{(i_1 i_2 \cdots j_m)} - {\bf C}_{(j_1 j_2 \cdots j_n)}
\nonumber \\[1.ex]
\omega_{(m,\cdots , n)} &=&
\frac{({\bf k}-{\bf q}_m-\cdots-{\bf q}_n)^2}{2 x E} \;\; .
\label{hbgcdef}
\end{eqnarray}
The infinite  opacity series can be understood as a sum  over 
{\em all  twist}  parton-parton correlations, 
$\langle \cdots (2n \, FF )  \cdots   \rangle$, in the nuclear matter.    
We emphasize that Eq.(\ref{ndifdis}) is not restricted 
to uncorrelated geometries as 
in Refs.\cite{Baier:1996kr}$^-$\cite{Wiedemann:2000ez}. 
It also allows the inclusion of finite kinematic boundaries 
on the ${\bf q}_i$ as well as different functional forms and 
elastic cross sections $\sigma_g(i)$ along the eikonal path. 
The first and second orders in opacity (twists $2+2$, $2+4$) have also 
been  checked\cite{Gyulassy:2000er,Gyulassy:2000fs} through explicit 
calculations of the corresponding  direct and virtual cut diagrams.  
The  $ \sum_i {\bf q}_i  = {\bf k}  $  divergences,   
naively present in the propagators   
${\bf C}_{(i_1i_2 \cdots i_m)}, 
{\bf B}_{(i_1 i_2 \cdots i_m )(j_1j_2 \cdots i_n)}$  in Eq.(\ref{ndifdis}), 
are canceled by  the interferences phases
$\omega_{(m,\cdots , n)}$.  Eq.(\ref{ndifdis}) also provides  
further insight in the LPM effect  in QCD at a 
microscopic/diagrammatic  level by keeping track of the 
destructive interference among all pairs of amplitudes
to all orders in opacity.

For particular models of the target geometry 
one can proceed further analytically. For a  homogeneous rectangular  
geometry, the average over the longitudinal target profile with 
\beq 
\rho_n 
%mg note new notation
(z_1,\cdots,z_n)= n! \, \rho_0^n \,  
\;\theta(L-z_n)\theta(z_n-z_{n-1})  \cdots \theta(z_2-z_{1}) 
\;\; ,\eeq{boxgeom} 
where the mean density is $\rho_0={N_s}/{L A_\perp}$,
leads to an  oscillatory pattern that  
is an artifact of the assumed sharp edges. 
Here $A_\perp$ is the transverse  area of the 
interaction region and $N_s$ is the total number of scattering 
centers. 
A somewhat more realistic model may be 
an exponential 
longitudinal distribution of scattering center separations 
\beq 
{\rho}_n(z_1,\cdots,z_n)=  \prod_{j=1}^n 
\frac{\theta(\Delta z_j)N_s }{L_e(n) A_\perp}e^{-\Delta z_j/L_e(n)} \;\;. 
\eeq{expgeom} 
This converts the oscillating formation phase  factors 
in Eq.(\ref{ndifdis}) into  Lorentzian factors 
\beq 
\int d{\rho}_n 
\cos \left (\, \sum_{k=j}^m \omega_{(k,\cdots,n)} \Delta z_k \right) 
= \frac{N_s^n}{A_\perp^n}
{\rm Re}\;\prod_{k=j}^m \frac{1}{1+i\omega_{(k,\cdots,n)}L_e(n)} 
\; \; . 
\eeq{loren}  
In order to fix $L_e(n)$, we can impose   $\langle z_k-z_0 \rangle 
=k L/(n+1)$. 
This constrains\cite{Gyulassy:2000er,Gyulassy:2000fs,Gyulassy:1999zd} 
$L_e(n)=L/(n+1)$. 
 
\subsection{First Order Non-Abelian Energy Loss}

The simplest and fortunately dominant (a {\em posteriori}) 
application\cite{Gyulassy:2000er,Gyulassy:2000fs} 
of our general solution to the energy 
loss problem was to calculate numerically the total 
radiated energy loss as a function of jet energy $E$, plasma depth 
$L$, and the typical transverse momentum transfer $\mu$. 
In the absence of a medium, the gluon  radiation 
associated with the parton jet (in the small $x$ approximation, 
where we do not distinguish between quark and gluon parents)   
is distributed as 
\begin{equation}   
x\frac{dN^{(0)}}{dx\, d {\bf k}^2}=  
 \frac{C_R \alpha_s}{\pi} \frac{1}{{\bf k}^2} \;\;,  
\label{hdist} 
\end{equation} 
where  $x=k^+/E^+ \approx \omega/E$, and $C_R$ is the Casimir  
of the  jet in the $d_R$ dimensional color representation. 
The differential energy distribution outside a cone defined by 
${\bf k}^2 > \mu^2$ is given by 
\begin{equation}  
\frac{dI^{(0)}}{dx} = \frac{2 C_R \alpha_s}{\pi}  
\, E \, \log \frac{|{\bf k}|_{\rm max}}{\mu} \;,     
\label{di0} 
\end{equation}  
where the upper kinematic limit is 
$\quad {\bf k}^2_{ \max}=\min\, [4E^2x^2,4E^2x(1-x)]\;$.  
The energy loss outside the cone in the vacuum is then given by 
\begin{equation} 
\Delta E^{(0)}=\frac{2C_R\alpha_s}{\pi}\, E \, 
\log \frac{E}{\mu} \; .
\label{de0} 
\end{equation} 
Beyond the small $x$ approximation, the splitting functions
$P_{gq}(x)$ and $P_{gg}(x)$ can give rise to some small corrections.
While  Eq.(\ref{de0})  overestimates the  
radiative  energy loss in the vacuum (self-quenching), it is important to note 
that $\Delta E^{(0)}/E \sim 50\%$ is typically much larger 
than the medium induced energy loss. However, the vacuum energy loss is
 included in the DGLAP evolution of the fragmentation functions
$D_{h/c}(z,Q^2)$.

\begin{figure} 
\centerline{\psfig{file=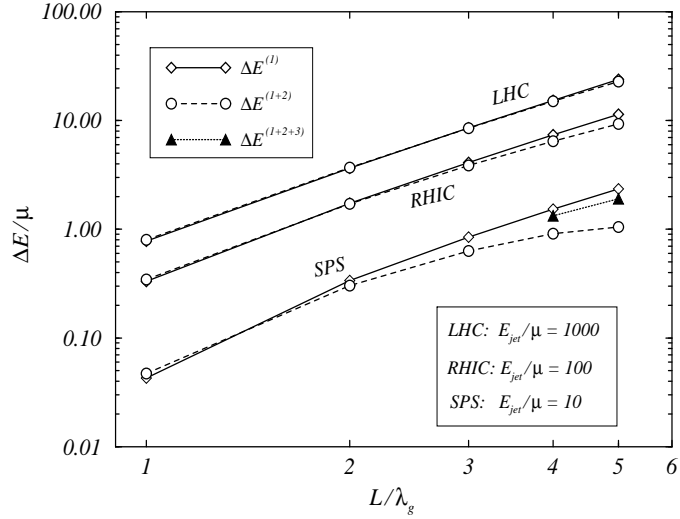,height=3.5in,width=2.750in,angle=-90}}
\caption{ The radiative energy loss of a quark jet
        with energy $E_{jet}=5,50,500$~GeV  (at SPS, RHIC, LHC) 
        is plotted as a function of 
        the opacity $L/\lambda_g$  ($\lambda_g=1$~fm, $\mu=0.5$~GeV)
for a static medium from Ref.\protect\cite{Gyulassy:2000er,Gyulassy:2000fs}.
   Solid curves show the first order in opacity
results. The  dashed curves show
        results up to second order in opacity, and two third order
 results are shown by solid triangles for SPS energies. 
}
\label{elossconv} 
\end{figure}

To compute the medium induced  radiation we focus on the density of 
the $N_s$ scattering centers given by Eq.(\ref{expgeom}).  
Averaging over the momentum transfer ${\bf q}_{1}$ via the color 
Yukawa potential leads to a very simple first order opacity  
result for  the $x\ll 1$ gluon double differential distribution 
\begin{eqnarray} 
x\frac{dN^{(1)}}{dx\, d {\bf k}^2}&=&  
x\frac{dN^{(0)}}{dx\, d {\bf k}^2}  
\, \frac{L}{\lambda_g}  \int_0^{q_{\max}^2} d^2{\bf q}_{1} \,  
\frac{ \mu_{eff}^2 }{\pi ({\bf q}_{1}^2 + \mu^2)^2 } 
\nonumber \\ 
&\ & \hspace{1.0in} \times \, 
\frac{ 2\,{\bf k} \cdot {\bf q}_{1} 
  ({\bf k} - {\bf q}_1)^2  L^2} 
{16x^2E^2 +({\bf k} - {\bf q}_1)^4  L^2 } \;\; ,        
\label{dnx1} 
\end{eqnarray} 
where the opacity factor $L/\lambda_g= N_s \sigma_{el}^{(g)}/A_\perp$  
arises from the sum over the $N_s$ distinct 
targets. Note that the radiated gluon mean free path 
$\lambda_g=(C_A/C_R)\lambda$ appears rather than the jet mean free path.  
The upper kinematic bound on the momentum transfer is 
$q^2_{\rm max}= s/4 \simeq 3 E \mu$ and $1/\mu_{eff}^2=1/\mu^2- 
1/(\mu^2+q_{\max}^2)$. For SPS and RHIC energies
this finite limit cannot be ignored. 
Numerical results comparing the first three orders in opacity corrections 
to the hard distribution Eq.(\ref{hdist}) were presented 
in Refs.\cite{Gyulassy:2000er,Gyulassy:2000fs} 
and the opacity series was shown to 
converge fast is dominated by its first order term for 
realistic nuclear opacities.  Fig.~\ref{elossconv} illustrates  
the convergence properties of Eq.(\ref{ndifdis}) in the example  of
the mean energy loss $\Delta E $, as well as the quadratic dependence
of $\Delta E $ on the size of the nuclear matter for  static media.

In Ref.\cite{Gyulassy:2001kr} we considered various asymptotic limits of 
the first order energy loss in  1+1D and  1+3D expanding plasmas. 
In the case that the mean density decreases  as 
\beq  
\rho(z=\tau)=\rho_0\left(\frac{\tau_0}{\tau}\right)^\alpha \;\; 
\eeq{bjork}  
due to a longitudinal Bjorken expansion,
it is possible to obtain  a  closed analytic 
formula\cite{Gyulassy:2000gk}     
under the strong assumption that no kinematic bounds need to be considered. 
For a hard jet  penetrating the quark-gluon plasma,  
\begin{eqnarray} 
\frac{d \Delta E^{(1)}}{dx} &=&   \frac{2C_R\alpha_s}{\pi} E    
\int^\infty_{\tau_0} \frac{d \tau}{\lambda(\tau)}\; f(Z(x,\tau)),  
\label{de11}   
\end{eqnarray} 
where $x\simeq \omega/E$ is the momentum fraction of the  
radiated gluon and  the formation physics function $f(Z(x,\tau))$ is 
defined\cite{Gyulassy:2000gk} to be  
\beq  
f(x,\tau)= \int_0^\infty \frac{du}{u(1+u)}    
\left[ 1 -   \cos \left (\,u Z(x,\tau) \right) \right]  \;.  
\eeq{fz}  
With $Z(x,\tau)=(\tau-\tau_0)\mu^2(\tau)/2 x E$ as the  
local formation physics  parameter, two simple analytic limits of  
Eq.(\ref{fz}) can be obtained. For $x \gg x_c =   
\mu(\tau_0)^2 \tau_0^{ \frac{2\alpha}{3}}   
L^{1- \frac{2\alpha}{3} }/2E  = L\mu^2(L)/2 E$, in which case   
the formation length is large compared to the size of the medium,   
the small $Z(x,\tau)$ limit applies, leading to  
$f(Z) \approx \pi Z / 2 $. The interference pattern along the  
gluon path becomes important and accounts for the the  
non-trivial dependence of the energy loss on $L$.    
When $x \ll x_c$, i.e. the formation length   
is small compared to the plasma thickness,   
one gets $f(Z) \approx \log Z$. In the $x \gg x_c$ 
limit\cite{Gyulassy:2000gk}, 
the radiative spectrum  Eq.(\ref{de11}) becomes  
\begin{eqnarray} 
\frac{d \Delta E^{(1)}_{x\gg x_c}}{dx} \approx   
\frac{C_R \alpha_s}{2(2-\alpha)}   
\frac{\mu(\tau_0)^2\tau_0^\alpha L^{2-\alpha}}{\lambda(\tau_0)}   
\frac{1}{x} \;\;.   
\label{dIdxxbig}   
\end{eqnarray} 
In the opposite $x \ll x_c$ limit we have
\begin{eqnarray} 
\frac{d \Delta E^{(1)}_{x\ll x_c}}{dx} \approx   
\frac{6 C_R \alpha_s}{\pi (3-\alpha)}  E 
\frac{\tau_0^{\frac{\alpha}{3}} L^{1-\frac{\alpha}{3}}}{\lambda(\tau_0)}  
 \log \frac{\mu(\tau_0)^2 \tau_0^{ \frac{2\alpha}{3}}   
L^{1- \frac{2\alpha}{3} }}{2xE}   
\;\;
\label{dIdxxsml}   
\end{eqnarray} 
and the intensity spectrum has an integrable divergence 
at $x=0$. Higher orders in opacity do change the $x$-dependence of the 
radiative spectrum relative to Eqs.(\ref{dIdxxbig}) and (\ref{dIdxxsml}),  
but this change is small\cite{Gyulassy:2000er,Gyulassy:2000fs}
for $L \sim R_A \simeq 5$~fm. 
   
The mean  energy loss (to first order in $\chi$) integrates to  
\begin{eqnarray} 
\Delta E^{(1)} &=&  \frac{C_R \alpha_s}{2(2-\alpha)}   
\frac{\mu(\tau_0)^2\tau_0^\alpha L^{2-\alpha} }{\lambda(\tau_0)} 
%\nonumber   \\[.5ex]  
%&&   \times \;  
\left(  
 \log \frac{2 E}{\mu(\tau_0)^2 \tau_0^{ \frac{2\alpha}{3}}   
L^{1- \frac{2\alpha}{3} }}    
+  \cdots \right)  \;\;. \qquad    
\label{totde}  
\end{eqnarray} 
The logarithmic enhancement with energy comes from the   
$x_c<x<1$ region\cite{Gyulassy:2000er,Gyulassy:2000fs}.  
In the case of sufficiently  large jet energies   
($E\rightarrow \infty$) this term dominates. For parton energies    
$<20$~GeV, however, corrections to this leading $\log 1/x_c$  expression  
that can be exactly evaluated numerically from the GLV expression  
and are found to be  
comparable in size. The  effective $\Delta E/E$ in this energy range  is  
approximately constant. 
 
In order to study non-central collisions and azimuthal anisotropy 
we also solved the analytically tractable case of a sharp  
expanding elliptic cylinder. 
We approximate the assumed $\phi_0$ independent screening   
$\mu(\tau)\approx g T(\tau) = 2 (\rho(\tau)/2)^{1/3}$ since   
$g \simeq 2$ and $\rho = (16 \zeta(3)/ \pi^2) T^3 \simeq 2 T^3 $ for   
a gluon plasma. We define   $\tau(\phi_0)$ as the   
escape time for the jet to reach the expanding elliptic surface from   
an initial point $\vec{\bf x}_0=(x_0,y_0)$ in the azimuthal direction 
$\phi_0$:   
 \beq   
\frac{(x_0+\tau \cos \phi_0)^2}{(R_x+v_x \tau)^2}    
+\frac{(y_0+\tau \sin \phi_0)^2}{(R_y+v_y \tau)^2}=1   
\;\; . 
\eeq{escape}   
We take $\omega(\phi_0)=2\, \tau(\phi_0) \,( \rho(\tau(\phi_0))/2 )^{2/3}$   
to estimate an upper bound on the logarithmic enhancement factor.   
Performing the remaining integrals one gets:  
\begin{eqnarray}   
\Delta E^{(1)}(\phi_0)  
  &\approx& \frac{9 C_R\alpha_s^3}{4}\frac{dN^g}{dy}   
\; \log \frac{E}{\omega(\phi_0)} \,  \int^\infty_{0} d \tau \,   
\frac{1}{R_x+v_x \tau} \, 
 \frac{1}{R_y+v_y \tau } \nonumber \\[.5ex] 
&\;& \hspace{0.2in}\times 
\theta\left( 1 - \frac{(x_0+\tau \cos \phi_0)^2}{(R_x+v_x \tau)^2}    
+ \frac{(y_0+\tau \sin \phi_0)^2}{(R_y+v_y \tau)^2}\right) 
\nonumber \\[.5ex] 
&=& \frac{9}{4} \, \frac{ C_R\alpha_s^3}{R_x R_y}\frac{dN^g}{dy}   
\; \frac{\log   
\frac{1 + a_x \tau(\phi_0)}{1+ a_y \tau(\phi_0)}}{a_x-a_y}   
\; \log \frac{E}{\omega(\phi_0)}  
\;\; ,  
\label{deaz}\end{eqnarray}    
where $a_x=v_x/R_x, a_y=v_y/R_y$.  
This expression provides a simple analytic  generalization  
that interpolates between a pure Bjorken 1+1D expansion for small  
$a_{x,y} \tau$, and a 3+1D expansion at large $a_{x,y} \tau$. 

In the special case of a pure Bjorken (longitudinal) expansion  
with $v_x=v_y=0$,  
\begin{eqnarray}   
\Delta E^{(1)}_{Bj}(\phi_0)&=&    
\frac{9C_R\alpha_s^3}{4 R_x R_y}    
\frac{dN^g}{dy}  \, \tau(\phi) \,\log \frac{E}{\omega(\phi_0)}   
\; .  
\label{debj}\end{eqnarray}  
In this case, the energy loss depends  {\em linearly} on  $\tau(\phi)$.

\subsection{Centrality and Rapidity Dependence of the  Cronin
Effect at RHIC}

In the subsections that follow we will describe some of the  
important applications/predictions of the GLV reaction operator 
formalism to initial and final state multiparton interactions. We emphasize 
that these calculations can provide complementary information
about the properties of cold and hot nuclear matter such as the 
transport coefficients $\mu^2/\lambda_q$, $\mu^2/\lambda_g $ and
the effective initial gluon number density and energy density  created
in relativistic heavy ion reactions.

In Refs.\cite{Vitev:2003xu,Gyulassy:2002yv} the explicit solution 
for the transverse
momentum distribution of partons that have traversed cold nuclear
matter has been found using the GLV approach:
\begin{eqnarray}
\frac{d^3N^{f}(k_\parallel,{\bf k}_\perp)}{d k_\parallel 
d^2 {\bf k}_\perp }
& = &  \sum_{n=0}^{\infty} \frac{\chi^n}{n!} 
\int  \prod_{i=1}^n   d^2 {\bf q}_{i\,\perp} 
 \left[  \frac{1}{\sigma_{el}} 
\frac{d\sigma_{el}(R,T)}{d^2{\bf q}_{i\,\perp}} \, \right.  
\nonumber  \\[1ex]
&&  \times \left.  \left(   e^{-{\bf q}_{i\,\perp} \cdot 
\stackrel{\rightarrow}{\nabla}_{{\bf k}\;\perp }} \otimes 
 e^{- q_{i\,\parallel}\partial_{k_\parallel} }
 - 1  \right) \, \right] \times  
\, \frac{d^3N^{i}(k_\parallel,{\bf k}_\perp)}
{d k_\parallel d^2 {\bf k}_\perp } \;\;. \qquad   
\label{ropit} 
\end{eqnarray}
This leads to nuclear induced broadening  
$ \langle \, \Delta k_\perp^2 \, \rangle_{pA} = {\mu^2 \chi  \xi}$ 
for  opacity $\chi = \langle L \rangle /\lambda$ and
typical transverse momentum squared $\mu^2$. 
Beyond the naive Gaussian approximation harder fluctuations along 
the projectile path lead to a logarithmic enhancement 
$ \langle \, \Delta k_\perp^2 \, \rangle_{pA}$, i.e. 
$\xi = \ln (1 +c p_T^2 )$.  
The question of the effective longitudinal 
momentum shift associated with  this  broadening  has also been 
addressed\cite{Vitev:2003xu,Gyulassy:2002yv}:
\beq
-{ \Delta k_\parallel} =  {\mu^2 \chi  \xi} \;
 \frac{1}{2 k_\parallel} \; .  
\eeq{parshift} 
To implement {\em initial state} elastic and radiative  energy loss we 
focus on the large $Q^2 \simeq p_T^2$  partonic subprocess 
$a+b \rightarrow c+d$, where 
$k_a,k_b$ are the initial momenta involved in the 
hard part $d \sigma / dt$ of Eq.(\ref{hcrossec}).  
If partons $a$ and $b$  have lost fractions $\epsilon_\alpha$ ( 
$\alpha=a,b$)  of their 
longitudinal  momenta  according to a probability distributions 
$P_\alpha(\epsilon)$, then  
$\tilde{k}_\alpha =  k_\alpha /{1-\epsilon_\alpha}$ and
\beq
f_{\alpha/p}(x_\alpha,Q^2) \rightarrow \int d \epsilon_\alpha \, 
P_\alpha(\epsilon_\alpha) f_{\alpha/p}\left( \tilde{x}_\alpha= 
\frac{x}{1-\epsilon_\alpha},Q^2  \right) 
\theta( \tilde{x}_\alpha \leq 1)\;\;,
\eeq{initloss} 
at  asymptotic   $t= - \infty$ .
Eq.(\ref{initloss}) provides a simple modification to the 
factorized pQCD hadron production formalism. 
For bremsstrahlung processes, $P_\alpha(\epsilon)$ are sensitive  
to multiple gluon emission\cite{Gyulassy:2001nm}. 
For the simpler case where one considers only the
mean energy loss,  $P(\epsilon) =  
\delta(\epsilon  - \langle \Delta k_0 \rangle / k_0 )$. More 
specifically, for the elastic longitudinal shift that 
we consider here,  $P_\alpha(\epsilon) =  \delta\left( \epsilon  - 
{\mu^2\chi_\alpha \xi} / {2 k^2_{\alpha\; \parallel} } \right)$  and
\beq
   f_{\alpha/p}(x_\alpha,Q^2) \rightarrow  
f_{\alpha/p} \left( x_\alpha + \frac{\mu^2 \chi_\alpha \xi }{x_\alpha}
\, \frac{2}{s} , Q^2 \right)  \theta( \tilde{x}_\alpha \leq 1)\;\;.
\eeq{elsh}
The observable effects of Eqs.(\ref{initloss}) and (\ref{elsh}) 
can be very different for valence quarks, sea quarks, and gluons  
due to the different $x$-dependence of the PDFs.

\begin{figure}[htb!]
\vspace*{.5cm}
\centerline{\epsfig{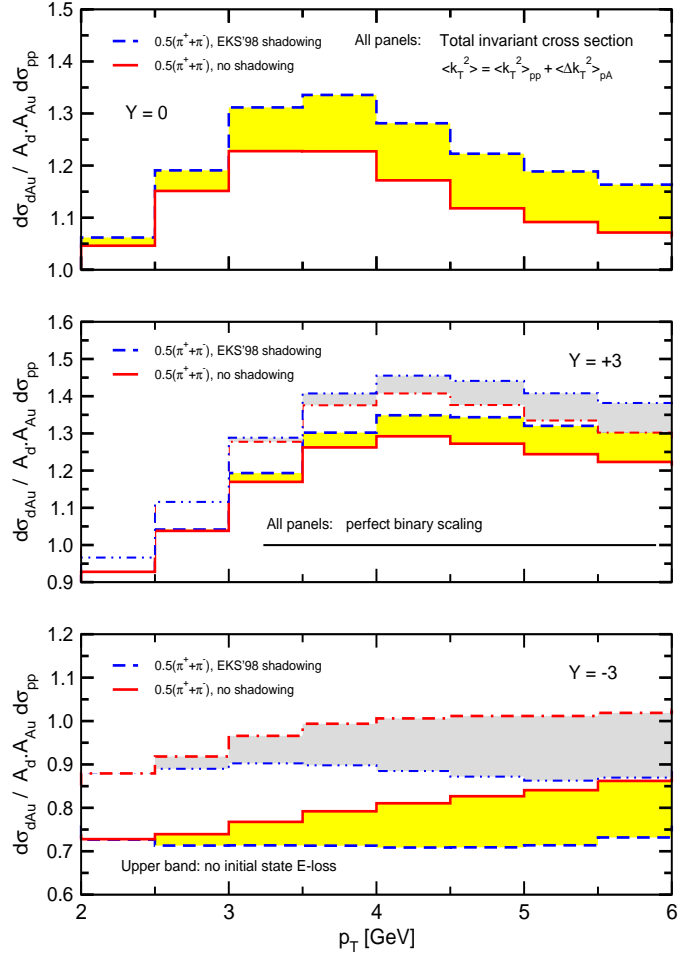}}
\vspace*{.7cm}
\caption{ Rapidity dependence of the Cronin effect in $d+Au$ 
reactions  at $\sqrt{s}_{NN}=200$~GeV with and without 
antishadowing or EMC effect from Ref.\protect\cite{Vitev:2003xu}. 
The result of switching 
off elastic energy loss is also shown via the upper bands for $y=\pm 3$.}
\label{fig_cr} 
\end{figure}

A natural  application of these results
is in the quantitative description of the Cronin effect.  
Perturbative calculations that include initial state parton 
broadening remain its only successful explanation 
to date\cite{Wang:1998ww,Vitev:2003xu,Vitev:2002pf,Accardi:2002ik,Zhang:2001ce,Kopeliovich:2002yh}. 
Comparison to low energy data on proton-nucleus 
collisions\cite{Vitev:2003xu,Vitev:2002pf} allows us to  extract the
transport coefficients of cold nuclear matter, 
$\mu^2/\lambda_q=0.06$~GeV$^2$/fm and $\mu^2/\lambda_g=0.14$~GeV$^2$/fm. 
These estimates include the effective elastic 
initial state energy loss and are  $\sim 25\%$  
bigger relative to the case when the longitudinal momentum 
shifts  have been neglected\cite{Vitev:2002pf,Arleo:2002ph}.
The transport coefficients  of cold nuclear matter change only 
slightly\cite{Vitev:2003xu}  with $\sqrt{s}$  and can
be used to predict the centrality and rapidity 
dependence of the Cronin effect in  $d+Au$ reactions at RHIC.

Fig.~\ref{fig_cr} shows the computed rapidity dependence of the 
Cronin effect in $d+Au$ reactions at RHIC. At midrapidity, $y=0$, in the 
computed transverse momentum range, we find slight Cronin enhancement
$R_{dAu\, \max}=1.2$
relative to the binary collision scaled $p+p$ result even with initial 
state elastic energy loss. Including strong antishadowing (shown with
dashed lines) leads to $R_{dAu\, \max}=1.3$. The effect of antishadowing 
becomes increasingly important at higher $p_T$. Experiments at RHIC 
can thus help constrain the poorly known nuclear modification to the PDFs 
for gluons. At forward  $y=+3$ rapidity (in the direction of the deuteron beam)
the Cronin intercept ($R_{dAu}=1$)  and maximum $R_{dAu\, \max}$ 
   are both shifted to  slightly higher $p_T$.  The most distinct prediction 
at forward rapidities  is the significantly larger Cronin enhancement 
region extending to high $p_T$.
This is understood in terms of  the softening of  the hadron  
spectra away from  midrapidity, as predicted by perturbative QCD. 
Steeper spectra tend to enhance the effect of otherwise identical 
transverse momentum kicks.  At backward $y=-3$ rapidity 
there is no enhancement since the nucleus does
not scatter multiply on the deuteron. This region is shown to be sensitive 
to the initial state energy loss\cite{Vitev:2003xu}. For further details on
the systematic  understanding of the Cronin effect versus 
$\sqrt{s}$ the reader is referred to Refs.\cite{Vitev:2002pf,Accardi:2002ik}. 
Recently it has been 
suggested by models based on gluon saturation, final state hadron
absorption, and coherent nucleon scattering, that  hadron 
cross sections  in $d+Au$ reactions will be suppressed 
%by a factor of  2    
by about 30\% for charged hadrons
relative to the 
binary collision scaled $p+p$ result. Experimental  data  at  RHIC  will  
make possible a critical test of the validity of 
different theoretical models.

\subsection{Jet Tomography of $d+Au$ and $Au+Au$ at SPS, RHIC,
and LHC}

Some of the most important applications\cite{Gyulassy:2001nm,Vitev:2002pf}  
 of the GLV results  on  medium induced  gluon 
bremsstrahlung are related to  jet tomography, 
the study of the properties of  matter through the attenuation pattern 
of fast particles that propagate and lose energy as a result of multiple 
elastic and  inelastic scatterings. Under the assumption of 
local thermal equilibrium  all relevant scales in the  problem, 
such as the Debye screening scale $\mu$, can be related to $dN^g/dy$ --
the initial effective gluon rapidity  density of the bulk soft background
matter (the quark-gluon plasma).

\begin{figure} 
\centerline{\epsfig{file=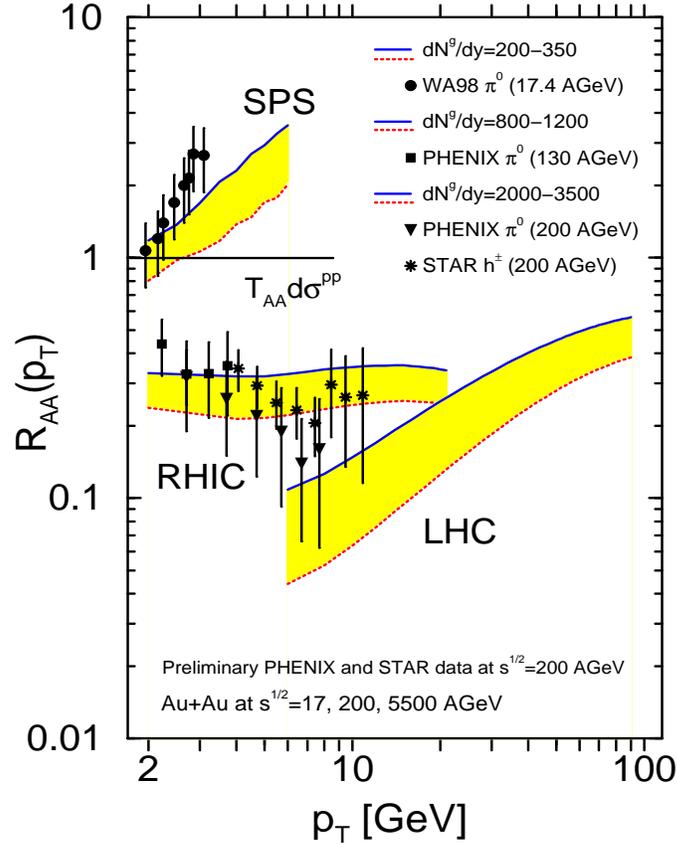,height=3.5in,width=4.5in,angle=-90}}  
\caption{The suppression/enhancement ratio $R_{AA}(p_T)$
         ($A=B=Au$)   for neutral 
         pions at $\sqrt{s}_{NN}=17$, $200$, $5500$~GeV 
         from Ref.\protect\cite{Vitev:2002pf}.   Solid (dashed) 
         lines correspond to the smaller (larger)  effective initial  
         gluon rapidity densities at given $\sqrt{s}$ that drive parton 
         energy loss. Data on
         $\pi^0$ production in central $Pb+Pb$ at $\sqrt{s}_{NN}=17.4$~GeV 
         from WA98{\protect\cite{Aggarwal:1998vh}} and in central $Au+Au$ at
         $\sqrt{s}_{NN}=130$~GeV, as well as  
         {\em preliminary} data at 
         $200$~GeV from 
         PHENIX{\protect\cite{Mioduszewski:2002wt,Adcox:2002pe}}  and 
         $h^\pm$ central/peripheral data from 
         STAR {\protect\cite{Kunde:2002pb}} are shown. The sum of 
         estimated statistical and systematic errors is indicated.}
\label{fg_tomo} 
\end{figure} 

Our main results for central $Au+Au$  reactions, which includes three 
important 
nuclear  effects, Cronin+Shadowing+Quenching, are presented  in 
Fig.~\ref{fg_tomo}. 
Jet  tomography consists of determining the  effective initial gluon 
rapidity  density $dN^g/dy$ that best reproduces the quenching pattern
of the 
data\cite{Adcox:2002pe,Mioduszewski:2002wt,Kunde:2002pb,Aggarwal:1998vh}. 
At SPS the large Cronin enhancement\cite{Vitev:2002pf}   
is reduced by a factor of two with $dN^g/dy = 350$ but the data
are more consistent with a smaller gluon density $ \leq 200$.
Unfortunately at this low energy the results are 
very sensitive to  the details of the model. At RHIC, for $p_T > 2$~GeV jet 
quenching  dominates, but surprisingly  the rate of variation with $p_T$ of 
the Cronin enhancement and jet quenching conspire to 
yield an approximately constant suppression pattern with magnitude dependent 
only on the initial $dN^g/dy$. At higher $p_T > 20$ GeV the softening
of the initial jet spectra due to the EMC modification of the   
PDFs compensates for the reduced energy loss. 
This unexpected interplay among the three nuclear effects at RHIC 
was our main prediction in Ref.\cite{Vitev:2002pf}. Preliminary 
$\sqrt{s}=200$~AGeV PHENIX and STAR results  included in 
Fig.~\ref{fg_tomo} support the  predicted magnitude and approximate  
$p_T$ dependence  of the suppression.  At LHC energies the much 
larger gluon densities $dN^g/dy  \sim 2000-3500$ are expected to 
lead to a  dramatic variation  of  quenching with $p_T$, as shown. 
In nuclear media of high opacity the mean fractional energy loss 
$\langle \Delta E \rangle / E$ of moderately hard partons can become of 
the order of unity. For LHC this may be reflected in the 
$p_T \leq 10$~GeV region through deviations from the extrapolated 
high-$p_T$  suppression trend. Hadronic fragments coming from energetic 
jets  would tend to compensate the rapidly increasing quenching 
with decreasing transverse momentum (seen in Fig.~\ref{fg_tomo}) 
and may restore the 
hydrodynamic-like  participant scaling in the soft regime.

\subsection{Enhanced Baryon/Meson Ratios}

One of the unexpected results reported  during the
first year RHIC run at $\sqrt{s}_{NN}=130$~GeV was that, in contrast 
to the strong $\pi^0$ quenching for $2\; {\rm GeV} < p_T < 5\;{\rm GeV}$,  
the corresponding charged hadrons  were found to be  
suppressed by only a  factor $\sim 2-2.5$. Even more surprisingly, 
the identified particle spectra analysis 
suggests that $R_B(p_T)=\bar{p}/\pi^-, \, p/\pi^+  \geq  1$ for
$p_T > 2$~GeV.  Thus, baryon and antibaryon production may in 
fact  dominate the  moderate to high $p_T$  hadron flavor 
yields\cite{Velkovska:2001xz,Adcox:2001mf,Xu:2001zj,Adler:2001aq}.
These and other data point to a possible novel baryon transport 
dynamics nucleus-nucleus reactions.
More recent results 
corroborate the non-perturbative baryon production hypothesis 
through equally abundant  $\Lambda$ and $\bar{\Lambda}$ 
production\cite{Adcox:2002au,Adler:2002pba}. It has also been 
observed that the mean transverse momentum $\langle p_T \rangle_B$ for 
various baryon
and antibaryon species is approximately constant and deviates from the 
common hydrodynamic flow systematics of soft hadron production in 
$A+A$ collisions. Identified particle analyses from the 
recent $\sqrt{s}_{NN}=200$~GeV RHIC run find similar puzzling 
features of moderate-$p_T$ baryon  
spectra\cite{Mioduszewski:2002wt,d'Enterria:2002bw,Kunde:2002pb}.

In Refs.\cite{Vitev:2001zn,Vitev:2002wh,Vitev:2001td} the 
GLV jet energy loss\cite{Gyulassy:2000er,Gyulassy:2000fs} was combined  
with a topological non-perturbative baryon production and transport 
mechanism\cite{Rossi:1977cy}$^-$\cite{Vance:1998vh}  to gain 
insight into the  anomalous anti+baryon behavior at RHIC. 
Phenomenological applications are currently based on  Regge theory where
a Regge trajectory $J=\alpha(0)+\alpha^\prime(0) M^2$ is specified by its
intercept $\alpha(0)$ and slope $\alpha^\prime(0)$. It has been 
argued\cite{Kharzeev:1996sq,Vance:1999pr}
 that $\alpha_J(0) \simeq 0.5$ and  $\alpha^\prime_J(0) 
\simeq 1/3 \,  \alpha^\prime_R(0)$. Regge theory gives exponential 
rapidity correlations, which in the presence of two sources  
(at $\pm Y_{\max}$) lead to net baryon rapidity density in central 
$A+A$ collisions of the form: 
\begin{equation} 
\frac{dN^{B-\bar{B}}}{dy} = (Z+N)(1-\alpha_J(0)) 
\frac{\cosh (1-\alpha_J(0)) y} {\sinh (1-\alpha_J(0))Y_{\max}} \;\; .
\label{btrans}
\end{equation}
It is evident from Eq.(\ref{btrans}) that the net baryon distribution 
integrates to $2A$ and in going to peripheral reactions scales as $N_{part}$.
At RHIC energies of $\sqrt{s}=130(200)$~GeV,  
corresponding to $Y_{\max} = 4.8(5.4)$,  in central reactions  
$dN^{B-\bar{B}}/{dy} = d(p-\bar{p})/dy+ d(n-\bar{n})/dy + 
d(\Lambda -\bar{\Lambda})/dy + \cdots \simeq 18(13.5) $. The relative 
contribution of each baryon species can be evaluated from isospin 
symmetry and strangeness conservation via comparison to midrapidity kaon 
production.

The high $p_T$ part of the hadron spectra is computed from 
Eq.(\ref{fullaa}). 
Hadronic transport in small-to-moderate $p_T$ is 
effectively controlled 
by the slope of the Regge trajectory. This would suggest that the 
baryon/meson mean inverse slopes in a phenomenological $p_T$-exponential 
($\sim e^{-p_T/T}$) soft particle production model are related as 
$T_B:T_M \simeq \sqrt{3}:\sqrt{2}$. Soft pion production, however, is  
largely  dominated by resonance decays, where the cumulative effect 
from the random walk in $p_T$ due to string breaking is destroyed. This 
leads to the relation 
$\langle p_T \rangle_\pi : \langle p_T \rangle_K : \langle p_T \rangle_B
\simeq 1: (1 \div \sqrt{2}): \sqrt{3} $ (220~MeV : 275 MeV : 
380 MeV)\cite{Vitev:2001zn,Vitev:2002wh}. One also notes 
that in the limit of pair production dominated by junction-antijunction 
loops (which we consider here)  the transverse momentum distribution of
antibaryons closely resembles that of baryons 
(with $\langle p_T \rangle_{\bar{B}} = \langle p_T \rangle_B $).

\begin{figure}[htb!] 
\centerline{\psfig{file=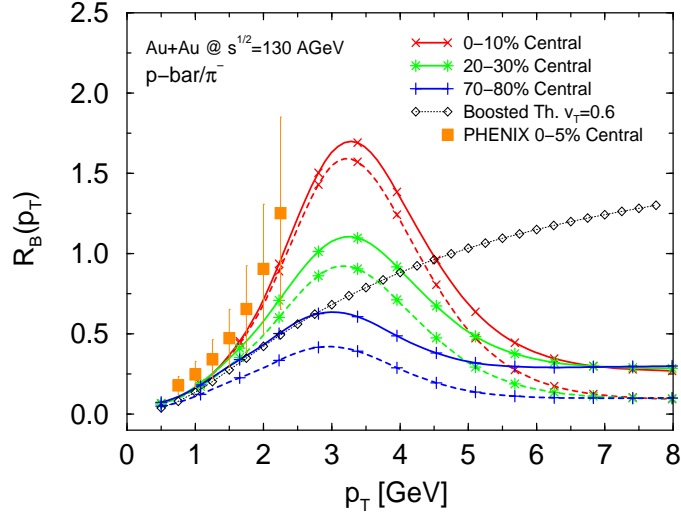,height=3.5in,width=2.750in,angle=-90}}
\caption{ The $\bar{p}/\pi^-$ ratio versus $p_T$ for 3 centrality classes 
at $\sqrt{s}_{NN}=130$~GeV from 
Refs.\protect\cite{Vitev:2001zn,Vitev:2002wh}. 
$N_{part}$(solid) versus $N_{bin}$(dashed)
geometry is shown. Boosted thermal source computation and 
ratio of {\em fits} to PHENIX data are shown  for comparison. }
\label{fg_ptopi} 
\end{figure} 

\begin{figure} 
\centerline{\psfig{file=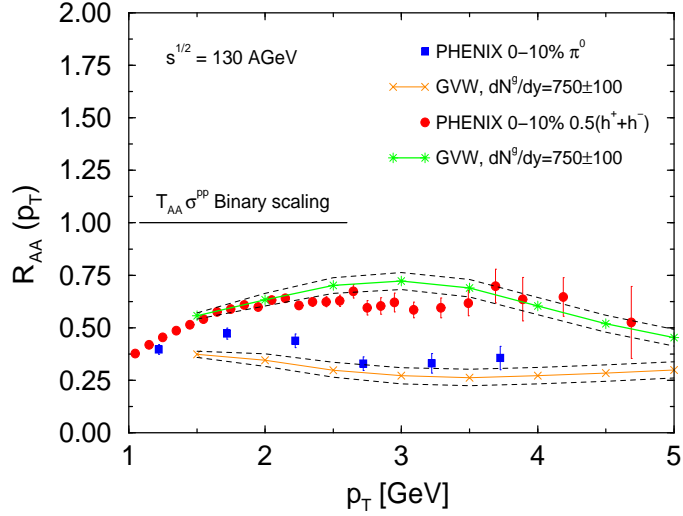,height=3.5in,width=2.750in,angle=-90}}  
\caption{ Suppression  of neutral pions and inclusive 
charged hadrons relative to the binary collision scaled $p+p$ result 
from  Refs.\protect\cite{Vitev:2001zn,Vitev:2002wh}.
Central $Au+Au$ at  $\sqrt{s}_{NN}=130$ GeV.}
\label{fg_pivsch} 
\end{figure} 

We find an enhanced  baryon/meson ratio $R_B(p_T) \geq 1$ that 
decreases with centrality  in a finite $p_T$ window as illustrated 
in Fig.~\ref{fg_ptopi}. For 
$p_T > 5$~GeV the ratio is reduced below unity  and approaches  
the perturbative calculation. The centrality dependence of  
$R_B(p_T)$ is largely 
driven by the non-Abelian energy  loss  that  suppresses  
perturbative pion production  for $p_T \geq 2$~GeV. It is important to
note that an enhanced  $p/\pi$ ratio is also observed in peripheral 
reactions and is consistent with measurements in $p+p$. 
In Fig.~\ref{fg_pivsch} baryon enhancement is reflected in the different
suppression factor $R_{AA}(p_T)$ for neutral pions and inclusive charged 
hadrons. At  high transverse momenta the suppression ratio is expected 
to be similar and driven by fragmentation of quenched jets. The rate 
at which this common $R_{AA}$ is approached depends on the 
fragmentation functions into baryons that are currently 
poorly constrained  or unknown.

\subsection{High-$p_T$ Azimuthal Asymmetry}

A new way to probe the energy loss $\Delta E$ in variable geometries 
was recently  proposed  in Ref.\cite{Wang:2000fq,Gyulassy:2000gk}.  
The idea was to exploit the  spatial azimuthal asymmetry of non-central  
nuclear collisions. The dependence of $\Delta E$ on the path 
length $L(\phi)$  naturally results in a pattern of azimuthal asymmetry 
of  high   $p_{\rm T}$ hadrons which  can be measured  via the 
differential elliptic  flow parameter (second Fourier coefficient),  
$v_2(p_{\rm T})$:
\begin{equation}
v_2(p_T) =  \frac{\langle p_x^2 \rangle - \langle p_y^2 \rangle}
{\langle p_x^2 \rangle + \langle p_y^2 \rangle }  = 
\langle \cos 2 \phi \rangle =  
\frac{\int_{0}^{2 \pi}  d\phi\,\cos 2\phi\,
\frac{dN^h}{ dy\,p_{ T}\,dp_{T}\, d\phi }  }   
{ \int_{0}^{2 \pi}  d\phi\, \frac{dN^h}{ dy\,p_{ T}\,dp_{ T}\, d\phi  } }  
\; \; .
\label{anisrelat}  
\end{equation}

In Ref.\cite{Gyulassy:2000gk} we predicted   
$v_2(p_{T})$ for two  models of  initial  
conditions\cite{Wang:2001bf} (specified by $dN^g/dy$) which differ 
by almost an  order of magnitude. 
The longitudinal expansion of the plasma and 
the corresponding mean energy loss were given  by 
Eqs.(\ref{bjork}),(\ref{totde}) with $\alpha=1$. 
A novel element of the analysis was the  discussion of the  
interplay between the azimuthally asymmetric soft  
(hydrodynamic) and hard (quenched 
jet) components of the final hadron distributions. 
In non-central $A+B$ reactions 
the low $p_{T}$ hadrons are also expected to exhibit azimuthal
asymmetry caused by the hydrodynamic 
flow\cite{Kolb:2000fh,Ollitrault:bk}. We  therefore 
modeled the soft component  with  the following Ansatz:
\begin{equation}
\frac{dN_s ({b})}{dyd^2{\bf p}_{T}} =
\frac{dn_s}{dy}\frac{e^{-p_{\rm T}/T_0}}{2\pi T^2_0}
\left(1+2 v_{2s}(p_{\rm T})\cos(2\phi) + \cdots\right) \;\;  .
\end{equation}
Here we took $T_0\approx 0.25$~GeV
and incorporate the hydrodynamic elliptic flow predicted 
in Ref.\cite{Kolb:2000fh} and  found to grow monotonically 
with $p_{T}$ as  
\begin{equation}
v_{2s}(p_{\rm T}) \approx {\rm tanh}(p_{\rm T}/(10\pm 2\;{\rm GeV}))
\;\; .
\end{equation}
The hydrodynamic elliptic flow was found\cite{Kolb:2000fh}
to be less sensitive to the initial conditions
than the high $p_T$  jet quenching studied in Ref.\cite{Gyulassy:2000gk}.

We showed that  
the combined pattern of jet quenching  in the single inclusive 
spectra and the differential elliptic flow  at high $p_{T}$ provide  
complementary tools\cite{Gyulassy:2000gk} that can determine the effective 
density of gluons created in the early stages of relativistic heavy 
ion reactions.

\begin{figure}
\centerline{\psfig{file=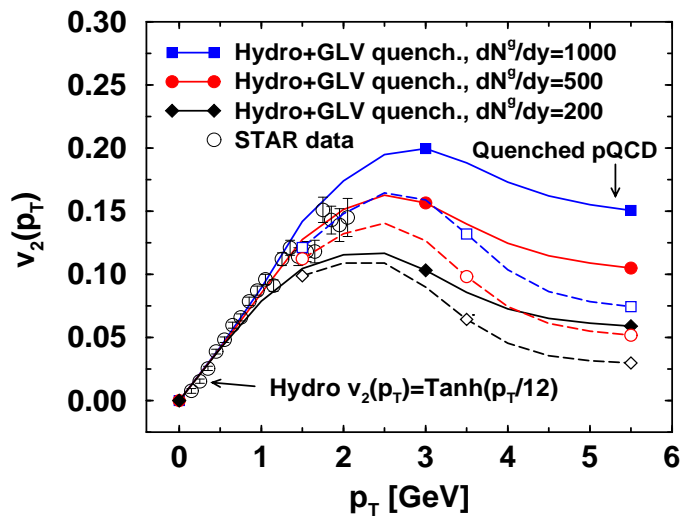,height=3.5in,width=2.750in,angle=-90}}  
\caption{ The interpolation of $v_2(p_T)$ between the soft 
hydrodynamic\protect\cite{Kolb:2000fh,Ollitrault:bk} and hard pQCD regimes 
is shown for $b=7$ fm adapted from Ref.\protect\cite{Gyulassy:2000gk}.
Solid (dashed) curves  correspond to sharp cylindrical 
(diffuse Woods-Saxon) geometries.}
\label{fg_v2} 
\end{figure}

Fig.~\ref{fg_v2} shows the predicted pattern of high-$p_T$ anisotropy.
Note the difference between sharp cylinder and diffuse Woods-Saxon geometries
at $b=7$~fm  approximating roughly  20-30\% central events.  
While  the central ($b = 0$)  inclusive quenching 
is insensitive to the density profile, non-central events clearly exhibit  
large sensitivity to the actual distribution. In particular, the sharp 
elliptic  geometry also simulates one of the possible scenarios of 
large energy loss and predominant surface emission\cite{Shuryak:2001me}.  
We conclude that $v_2(p_T>2\;{\rm GeV},b)$ 
provides essential complementary information about the geometry and impact
parameter dependence of the initial conditions in $A+A$ reactions and 
the magnitude of jet quenching.  In particular, the  rate at which 
the $v_2$  coefficient  decreases at high $p_T$ is an  indicator of 
the diffuseness of that geometry.  Minimum bias STAR data at 
RHIC\cite{Filimonov:2002xk} now confirm the predicted 
saturation  of $v_2(p_T)$.   For $p_T \geq 6$~GeV  data is 
still inconclusive due to the large error bars and can still accommodate 
different scenarios: constant versus slightly decreasing $v_2(p_T)$.   
High statistics measurements in the future RHIC runs can help to 
experimentally resolve this question and
further test the predicted\cite{Wang:2000fq,Gyulassy:2000gk} 
slow decrease of $v_2$ at large transverse momenta.

In Refs.\cite{Hirano:2003hq,Hirano:2002bm}, in an independent 
analysis the GLV radiative  energy 
loss\cite{Gyulassy:2000er,Gyulassy:2000fs} of fast 
partons  was coupled to a dynamically evolving soft hydrodynamic 
background to obtain a quantitative  description of the disappearance 
of the back-to-back jet correlations\cite{Adler:2002tq}.

\section{The WW Approach: Parton Energy Loss with Detailed Balance}

The above studies of parton energy loss have concentrated on 
gluon radiation induced by multiple scattering in a medium.
Since gluons are bosons, there should also
be stimulated gluon emission and absorption by the propagating parton
because of the presence of thermal gluons in the hot medium.  Such detailed
balance is crucial for parton thermalization and should also be important
for calculating the energy loss of an energetic parton in a hot 
medium\cite{ww01}.

\subsection{Final-state Absorption}

Let us assume that the hot medium is in thermal equilibrium 
shortly after the production of the hard parton. One should then
take into account of both stimulated emission and thermal absorption
in the final state radiation of a jet in a thermal medium with finite 
temperature $T$. One has then the
probability of gluon radiation with energy $\omega$,
 \begin{eqnarray}
 \label{prob0}
   {{dP^{(0)}}\over d\omega}&&={{\alpha_s C_F}\over 2\pi}
     \int {{dz}\over z} \int
     {{d{\bf k}_{\perp}^2}\over {\bf k}_{\perp}^2}
     \Big[ N_g(zE)\delta(\omega+zE)
 \nonumber\\
     &&+\left(1+N_g(zE)\right)\delta(\omega-zE)\theta(1-z)\Big]
     P({\omega\over E})\, ,
 \end{eqnarray}
where, $N_g(|{\bf k}|)=1/[\exp(|{\bf k}|/T)-1]$ is the thermal gluon
distribution. We have define the splitting
function $P_{gq}(z)\equiv P(z)/z=[1+(1-z)^2]/z$ for $q\rightarrow gq$.
The first term is from thermal absorption and the second term
from gluon emission with the Bose-Einstein enhancement factor.
For $E\gg T$, one can neglect the quantum statistical effect for the
leading parton. Note that the
vacuum part has a logarithmic infrared divergence
while the finite-temperature part has a linear divergence, since
$N_g(|{\bf k}|)\sim T/|{\bf k}|$ as $|{\bf k}|\rightarrow 0$.
These infrared divergences will be canceled by the virtual
corrections which also contain a zero-temperature and a 
finite-temperature part.
In addition, the virtual corrections are also essential to ensure
unitarity and momentum conservation in the QCD evolution of the
fragmentation functions. However, they do not contribute to
the effective parton energy loss. The remaining collinear divergence
in the above spectrum can be absorbed into a renormalized fragmentation
function that follows the QCD evolution equations.

Subtracting the gluon radiation spectrum in the vacuum, one then
obtains the energy loss due to final-state absorption and stimulated
emission,
 \begin{eqnarray}
 \label{eloss0}
   \Delta E^{(0)}_{abs} &=& \int d\omega \,\omega
   \left({{dP^{(0)}}\over d\omega}-
     {{dP^{(0)}}\over d\omega}\Big|_{T=0}\right)
 \nonumber\\
   &=&  {{\alpha_s C_F}\over 2\pi}E
   \int dz   \int_{\mu^2}^{{\bf k}_{\perp max}^2}
 {{d{\bf k}_{\perp}^2}\over {\bf k}_{\perp}^2}
 \Big[ -P(-z)N_g(zE)
 \nonumber\\
   && +P(z) N_g(zE) \theta(1-z) \Big].
\end{eqnarray}
Even though the stimulated emission cancels part of the contribution
from absorption, the net medium effect without rescattering is
still dominated by the final-state thermal absorption,
resulting in a net energy gain, {\it i.e.} a {\it negative} energy loss.
For asymptotically large parton energy, $E\gg T$, one can
complete the above integration approximately and have,
 \begin{eqnarray}
 \label{elossab0}
   {\Delta E^{(0)}_{abs}\over E}&\approx& -
   {{\pi\alpha_s C_F}\over 3}
   {T^2\over E^2}\left[
     \ln{4ET\over \mu^2}+2-\gamma_{\rm E}
     +{{6\zeta^\prime(2)}\over \pi^2}\right],
\end{eqnarray}
where, $\gamma_{\rm E}\approx 0.5772$ and $\zeta^\prime(2)\approx -0.9376$.
The quadratic temperature dependence of the leading contribution
is a direct consequence of the partial cancellation between stimulated
emission and thermal absorption, each having a leading contribution
linear in $T$.

\subsection{Induced Absorption}

As in the case of final-state absorption, one can also
include stimulated emission and thermal absorption when calculating
the induced radiation probability at the first order in opacity,
 \begin{eqnarray}
 \label{prob1}
   {{dP^{(1)}}\over d\omega}=&&{{C_2}\over {8\pi d_A d_R}}
     \int {{dz}\over z}  \int
      {{d^2{\bf k}_{\perp}}\over {(2\pi)^2}}
     \int {{d^2{\bf q}_{\perp}}\over {(2\pi)^2}}
      v^2({\bf q}_{\perp})P({\omega\over E}) \nonumber\\
      &\times& {N\over A_{\perp}}
      \left\langle Tr\left[|R^{(S)}|^2
     +2 Re\left(R^{(0)\dagger}
     R^{(D)}\right)\right]\right\rangle
 \nonumber\\
     &\times&\Big[\left(1+N_g(zE)\right)\delta(\omega-zE)\theta(1-z)
     +N_g(zE)\delta(\omega+zE)\Big]
 \nonumber\\
     =&&{{\alpha_s C_2 C_F C_A}\over {d_A \pi}}
     \int {{dz}\over z}  \int
      {{d{\bf k}_{\perp}^2}\over {\bf k}_{\perp}^2}
     \int {{d^2{\bf q}_{\perp}}\over {(2\pi)^2}}
     v^2({\bf q}_{\perp}) P({\omega\over E}) \nonumber\\
     &\times&
      {{{\bf k}_{\perp}\cdot {\bf q}_{\perp}}\over
      {\left({\bf k}_{\perp} - {\bf q}_{\perp}\right)^2}}
      {N\over A_{\perp}}
      \left\langle Re(1-e^{i\omega_1 y_{10}})\right\rangle
      \Big[ N_g(zE)\delta(\omega+zE)
 \nonumber\\
     &+&\left(1+N_g(zE)\right)\delta(\omega-zE)\theta(1-z)\Big]\;.
 \end{eqnarray}
The factor $1-\exp(i\omega_1 y_{10})$ reflects the destructive
interference arising from the non-Abelian LPM effect. 
The target distribution
is assumed to be an exponential form $\rho(y)=2 \exp(-2y/L)/L$.

As in the final-state absorption, the contribution from
thermal absorption associated with rescattering is larger than that of
stimulated emission, resulting in a net energy gain. However, the
zero-temperature contribution corresponds to the radiation induced
by rescattering which will lead to an effective energy loss by the
leading parton. This is the energy loss obtained the previous sections and
we denote this part as $\Delta E^{(1)}_{rad}$. The
remainder or temperature-dependent part of energy loss induced by
rescattering at the first order in opacity is then defined as
\begin{equation}
 \label{eloss1}
   \Delta E^{(1)}_{abs}=\int d\omega \,\omega
     \left({{dP^{(1)}}\over d\omega} -{{dP^{(1)}}\over d\omega}
       \Big|_{T=0}\right)\;,
 \end{equation}
which mainly comes from thermal absorption with partial cancellation
by stimulated emission in the medium. According to Eq.(\ref{prob1}),
 \begin{eqnarray}
 \label{elossem1}
   \Delta E_{rad}^{(1)} &&=
   {{\alpha_s C_F}\over \pi}{L\over \lambda_g}E
   \int dz  \int {d{\bf k}_{\perp}^2\over {\bf k}_{\perp}^2}
   \int d^2{\bf q}_{\perp}
   |{\bar v}({\bf q}_{\perp})|^2
      {{{\bf k}_{\perp}\cdot {\bf q}_{\perp}}\over
      {\left({\bf k}_{\perp} - {\bf q}_{\perp}\right)^2}}
    \nonumber \\
    &\times& P(z)
     \left\langle Re(1-e^{i\omega_1 y_{10}})\right\rangle
      \theta(1-z)\, ;
 \\
 \label{elossab1}
   \Delta E_{abs}^{(1)}&&=
   {{\alpha_s C_F}\over \pi}{L\over \lambda_g}E
   \int dz \int {d{\bf k}_{\perp}^2\over {\bf k}_{\perp}^2}
   \int d^2{\bf q}_{\perp}
    |{\bar v}({\bf q}_{\perp})|^2
      {{{\bf k}_{\perp}\cdot {\bf q}_{\perp}}\over
      {\left({\bf k}_{\perp} - {\bf q}_{\perp}\right)^2}}
 \nonumber\\
     &\times& N_g(zE)\Bigl[P(z)\left\langle
     Re(1-e^{i\omega_1 y_{10}})\right\rangle\theta(1-z)
\nonumber\\
     &-&P(-z)\left\langle
     Re(1-e^{i\omega_1 y_{10}})\right\rangle\Bigr],
\end{eqnarray}
In the limit $q_{\perp max}\rightarrow\infty$, the angular integral
can be carried out by partial integration. These
contributions to the energy loss become
 \begin{eqnarray}
 \label{elossem2}
   \Delta E_{rad}^{(1)}\approx &&
   {{\alpha_s C_F}\over 2\pi}{L\over \lambda_g}E
   \int dz P(z) h(\gamma)\theta(1-z)\, ;
 \\
 \label{elossab2}
   \Delta E_{abs}^{(1)}\approx &&
   {{\alpha_s C_F}\over 2\pi}{L\over \lambda_g}E
   \int dz N_g(zE)  h(\gamma)
   \Big[P(z) \theta(1-z) -P(-z)\Big],
 \end{eqnarray}
where, $\gamma=\mu^2 L/(4zE)$ and
\begin{equation}
h(\gamma)=\left\{
\begin{array}{ll}
{2\gamma \over \sqrt{1-4\gamma^2}} [{\pi \over 2}-\arcsin(2\gamma)]\, ,
& \gamma<1/2 \\
{2\gamma \over \sqrt{4\gamma^2-1}}\ln[2\gamma+\sqrt{4\gamma^2-1}]\, ,
&  \gamma>1/2\, .
\end{array}
\right.
\end{equation}
One can approximate $h(\gamma)$ with
$\pi\gamma+(11/4-2\pi)\gamma^2+(5/2)\gamma^3$ for $\gamma<1/2$ and
$\ln(4\gamma)+0.1/\gamma+0.028/\gamma^2$ for $\gamma>1/2$.
In the limit of $EL\gg 1$ and $E\gg \mu$, One can then get the
approximate asymptotic behavior of the energy loss,
 \begin{eqnarray}
 \label{elossem3}
   {\Delta E_{rad}^{(1)}\over E}\approx &&
   {{\alpha_s C_F \mu^2 L^2}\over 4\lambda_gE}
   \left[\ln{2E\over \mu^2L} -0.048\right]\, ;
 \\
 \label{elossab3}
   {\Delta E_{abs}^{(1)}\over E}\approx &&-
   {{\pi\alpha_s C_F}\over 3} {{LT^2}\over {\lambda_g E^2}}
   \left[
   \ln{{\mu^2L}\over T} -1+\gamma_{\rm E}-{{6\zeta^\prime(2)}\over\pi^2}
\right] .
 \end{eqnarray}
Again, the thermal absorption results in an energy gain ( or
negative energy loss). This result is
accurate through the order of $1/E$. In Eq.(\ref{elossab3}), we
have assumed $\mu^2L/T\gg 1$ and kept only the first two leading
terms. In this limit, the average formation time for stimulated
emission or thermal absorption is much smaller than the total
propagation length. Therefore, the energy gain, $\Delta
E_{abs}^{(1)}$, by thermal absorption (with partial cancellation
by the stimulated emission) is linear in $L$, as compared to the
quadratic dependence in the zero-temperature case. However, the
logarithmic dependence on $\mu^2L/T$, as compared to the factor
$\ln(4ET/\mu^2)$ in Eq.(\ref{elossab0}) for no rescattering, is
still a consequence of the LPM interference in medium. A quadratic
$L$-dependence of $\Delta E_{abs}^{(1)}$ will arise when
$\mu^2L/T\ll 1$.

To evaluate the effect of thermal absorption numerically, we assume
the Debye screening mass to be $\mu^2=4\pi\alpha_s T^2$ from
the perturbative QCD at finite temperature\cite{HTL}. The
mean-free-path for a gluon $\lambda_g$ in the GW 
model is\cite{Gyulassy:1994hr},
 \begin{equation}
 \label{lambda}
    \lambda_g^{-1}= \langle\sigma_{qg}\rho_q\rangle
+\langle\sigma_{gg}\rho_g\rangle
    \approx {2\pi\alpha^2_s\over \mu^2} 9\times 7\zeta(3)
    {T^3\over \pi^2}\, ,
 \end{equation}
where $\zeta(3)\approx 1.202$. With fixed values of $L/\lambda_g$
and $\alpha_s$, $\Delta E/\mu$ should be a function of $E/\mu$ only.
\begin{figure}
\centerline{\hspace*{-1cm}
\psfig{figure=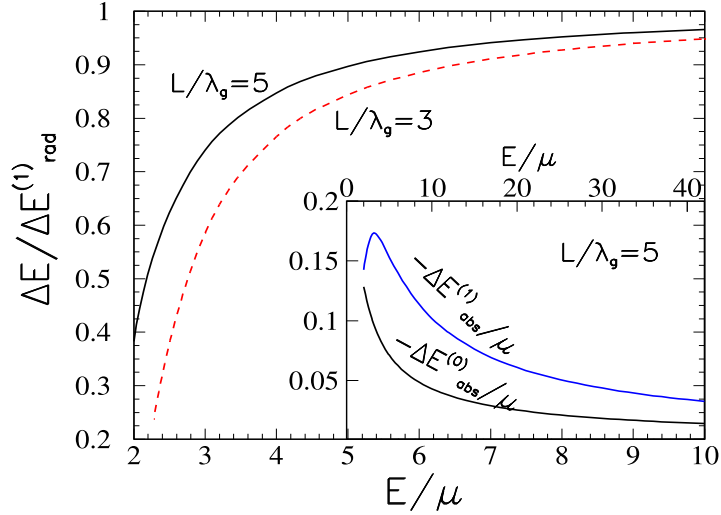,width=4.in,height=4.in}}
\vspace*{-3cm}
\caption{The ratio of effective parton energy loss
with ($\Delta E=\Delta E^{(0)}_{abs}+\Delta E^{(1)}_{abs} +\Delta
E^{(1)}_{rad}$) and without ($\Delta E^{(1)}_{rad}$) absorption as
a function of $E/\mu$. Inserted box: energy gain via absorption
with ($\Delta E^{(1)}_{abs}$) and without ($\Delta E^{(0)}_{abs}$)
rescattering.}
\label{fig:balance}
\end{figure}
Shown in Fig.~\ref{fig:balance} are ratios of the calculated radiative
energy loss with and without stimulated emission and thermal
absorption as functions of $E/\mu$ for $L/\lambda_g=3$,5 and
$\alpha_s=0.3$. Shown in the inserted
box are the energy gain via gluon absorption with ($\Delta
E^{(1)}_{abs}$) and without ($\Delta E^{(0)}_{abs}$) rescattering.
For partons with very high energy the effect of the gluon
absorption is small and can be neglected. 
However, the thermal absorption reduces the effective
parton energy loss by about 30-10\% for intermediate values of
parton energy. This will increase the energy dependence of the
effective parton energy loss in the intermediate energy region.
The observed flat or slightly $p_T$ dependence of the hadron 
spectra suppression
indicates a strong energy dependence of the
parton energy loss. This might indicate the effect of the thermal
absorption, in particular in the intermediate $p_T$ region.

\section{The WOGZ Approach: Parton Energy Loss and 
Modified Fragmentation Functions in Nuclei \label{seco2}}

So far we have reviewed the parton energy loss of an energetic
parton induced by multiple scattering in a hot QCD medium.
The effective parton energy loss is shown to be proportional 
to the gluon density. Therefore measurements of 
the parton energy loss will enable one to extract the initial 
gluon density of the produced hot medium within our assumed
screened potential model for multiple parton scattering in a
dense gluonic medium. Such an exercise will be more robust if
one can also measure the parton energy loss and the gluon
density inside a cold nuclear medium. For this purpose, one has 
to rely on other complementary experimental measurements such 
as parton energy loss in deeply inelastic scattering (DIS) off 
nuclear targets. One can then study parton energy loss and
extract the initial gluon density in high-energy heavy-ion collisions 
relative to that in a cold nucleus\cite{Wang:2002ri}.

Extending the screened potential model of multiple parton scattering
to a cold nuclear medium is somewhat problematic because it applies
mainly to a medium where colors are screened but not confined.
In this section we will study parton multiple scattering inside
a nucleus where colors are confined to the size of a nucleon within
generalized factorization theory of pQCD. In this case, 
parton propagation will be different from that in a partonic 
medium and one should expect the parton energy loss to be related to
the nucleon size or confinement scale which, as we shall show, can be
related to the gluon density inside a cold nucleus. In addition, one can
calculate directly the modification of the fragmentation functions due
to multiple scattering and induced gluon bremsstrahlung.
This section reviews developments published in 
Refs.\cite{Wang:2001if,Guo:2000nz,ow,bwzxnw},
which are referred to here as WOGZ 
and are applied in Ref.\cite{Wang:2002ri} to cold and hot nuclei.

In contrast to the situation in QED, the energy loss of a parton cannot be 
directly measured because partons are not the final experimentally observed 
particles. The total energy of a jet, traditionally defined as a cluster of 
hadrons in the phase space, will not change much due to medium induced 
radiation, because a jet so defined contains particles both from the 
leading parton and from the radiated gluons. 
This is particularly the case if multiple 
scattering and induced radiation do not {\it dramatically } change the 
energy profile of the jet in phase-space. It is also virtually impossible
to determine the jet energy event by event because of the large background 
and its fluctuation in heavy-ion collisions. One then has to resort to 
particle distributions within a jet and study the effect of parton energy loss 
by measuring the modification of the fragmentation function of 
the produced parton, $D_{a\rightarrow h}(z,\mu^2)$, where $z$ is the 
fractional energy of the parton $a$ carried by the produced particles $h$.

Since the produced quark in DIS is far off-shell, the final-state
radiation leads to the scale dependence of the fragmentation functions as
given by Dokshitzer-Gribov-Lipatov-Altarelli-Parisi (DGLAP)\cite{dglap} QCD 
evolution equations. When a parton is produced in a medium, it will 
suffer multiple scattering and induced radiation that will give rise to an
additional term in the DGLAP evolution equations. This then leads to 
the medium modification of the DGLAP evolution of the parton fragmentation 
functions. As a consequence, 
the modified fragmentation functions become softer. This can be 
directly translated into the energy loss of the leading quark.

\subsection{Generalized Factorization}
To study parton energy loss in $eA$ DIS,
we consider the semi-inclusive processes,
$e(L_1) + A(p) \longrightarrow e(L_2) + h (\ell_h) +X$,
where $L_1$ and $L_2$ are the four-momenta of the incoming and the
outgoing leptons, and $\ell_h$ is the observed hadron momentum.
The differential
cross section for the semi-inclusive process can be expressed as
\begin{equation}
E_{L_2}E_{\ell_h}\frac{d\sigma_{\rm DIS}^h}{d^3L_2d^3\ell_h}
=\frac{\alpha^2_{\rm EM}}{2\pi s}\frac{1}{Q^4} L_{\mu\nu}
E_{\ell_h}\frac{dW^{\mu\nu}}{d^3\ell_h} \; ,
\label{sigma}
\end{equation}
where $p = [p^+,0,{\bf 0}_\perp] \label{eq:frame}$
is the momentum per nucleon in the nucleus,
$q =L_2-L_1 = [-Q^2/2q^-, q^-, {\bf 0}_\perp]$ the momentum transfer,
$s=(p+L_1)^2$ and $\alpha_{\rm EM}$ is the electromagnetic (EM)
coupling constant. The leptonic tensor is given by
$L_{\mu\nu}=1/2\, {\rm Tr}(\gamma \cdot L_1 \gamma_{\mu}
\gamma \cdot L_2 \gamma_{\nu})$
while the semi-inclusive hadronic tensor is defined as,
\begin{eqnarray}
E_{\ell_h}\frac{dW_{\mu\nu}}{d^3\ell_h}&=&
\frac{1}{2}\sum_X \langle A|J_\mu(0)|X,h\rangle
\langle X,h| J_\nu(0)|A\rangle \nonumber \\
&\times &2\pi \delta^4(q+p-p_X-\ell_h)
\end{eqnarray}
where $\sum_X$ runs over all possible final states and
$J_\mu=\sum_q e_q \bar{\psi}_q \gamma_\mu\psi_q$ is the
hadronic EM current.

\begin{figure}
\centerline{\psfig{file=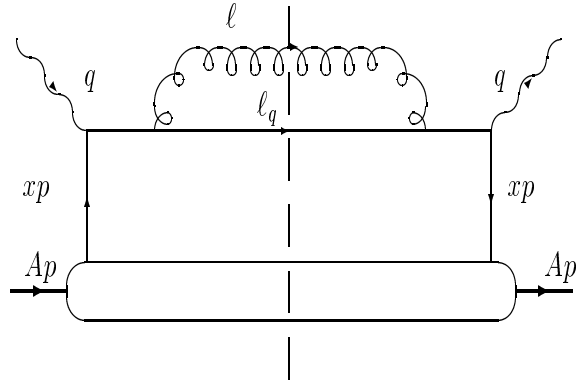,width=3in,height=2.0in}}
\caption{The hard partonic part of the first order radiative correction
to the fragmentation process.}
\label{fig:1order}
\end{figure}

In the parton model with the collinear factorization approximation,
the leading-twist contribution to the semi-inclusive cross section
can be factorized into a product of parton distributions,
parton fragmentation functions and the partonic cross section.
Including all leading log radiative corrections, the lowest order
contribution (${\cal O}(\alpha_s^0)$) from a single
hard $\gamma^*+ q$ scattering can be written as
\begin{eqnarray}
& &\frac{dW^S_{\mu\nu}}{dz_h}
= \sum_q e_q^2 \int dx f_q^A(x,\mu_I^2) H^{(0)}_{\mu\nu}(x,p,q)
D_{q\rightarrow h}(z_h,\mu^2)\, ; \label{Dq} \\
& &H^{(0)}_{\mu\nu}(x,p,q) = \frac{1}{2}\,
{\rm Tr}(\gamma \cdot p \gamma_{\mu} \gamma \cdot(q+xp) \gamma_{\nu})
\, \frac{2\pi}{2p\cdot q} \delta(x-x_B) \, , \label{H0}
\end{eqnarray}
where the momentum fraction carried by the hadron is defined as
$z_h=\ell_h^-/q^-$ and $x_B=Q^2/2p^+q^-$ is the Bjorken variable.
$\mu_I^2$ and $\mu^2$ are the factorization scales for the initial
quark distributions $f_q^A(x,\mu_I^2)$ in a nucleus and the fragmentation
functions $D_{q\rightarrow h}(z_h,\mu^2)$, respectively.
Considering the leading logarithm approximation of the radiative
correction to the fragmentation process as shown in Fig. \ref{fig:1order},
the renormalized quark fragmentation function
$D_{q\rightarrow h}(z_h,\mu^2)$ satisfies the DGLAP\cite{dglap} 
QCD evolution equations,
\begin{eqnarray}
  \frac{\partial D_{q\rightarrow h}(z_h,\mu^2)}{\partial \ln \mu^2} & = &
  \frac{\alpha_s(\mu^2)}{2\pi} \int^1_{z_h} \frac{dz}{z} 
\left[ \gamma_{q\rightarrow qg}(z)
D_{q\rightarrow h}(z_h/z,\mu^2) \right. \nonumber \\
& + & \left. \gamma_{q\rightarrow gq}(z) 
D_{g\rightarrow h}(z_h/z,\mu^2)\right]\; , \label{eq:ap1}
\end{eqnarray}
where
\begin{eqnarray}
\gamma_{q\rightarrow qg}(z) &=& C_F\left[\frac{1+z^2}{(1-z)_+} 
  + \frac{3}{2}\delta(1-z)\right] \; , \label{eq:split1}\\
\gamma_{q\rightarrow gq}(z) &=& \gamma_{q\rightarrow qg}(1-z) \; 
 \label{eq:split2}
\end{eqnarray}
are the splitting functions. The `$+$' function is defined as
\begin{equation}
\int_0^1 dz \frac{F(z)}{(1-z)_+} \equiv \int_0^1 dz \frac{F(z)-F(1)}{1-z}
\label{eq:plus}
\end{equation}
with $F(z)$ being any function which is sufficiently smooth at $z=1$.

In a nuclear medium, the propagating quark in DIS will experience additional
scatterings with other partons from the nucleus. The rescatterings may
induce additional gluon radiation and cause the leading quark to lose
energy. Such induced gluon radiations will effectively give rise to
additional terms in the evolution equation leading to the modification of the
fragmentation functions in a medium. These are the so-called higher-twist
corrections since they involve higher-twist parton matrix elements and
are power-suppressed. We will consider those contributions that
involve two-parton correlations from two different nucleons inside
the nucleus. They are proportional to the size of the nucleus\cite{ow} 
and thus are enhanced by a nuclear factor $A^{1/3}$ as 
compared to two-parton correlations
in a nucleon. As in previous studies\cite{Wang:2001if,Guo:2000nz}, 
we will neglect
those contributions that are not enhanced by the nuclear medium.

\begin{figure}
\centerline{\psfig{file=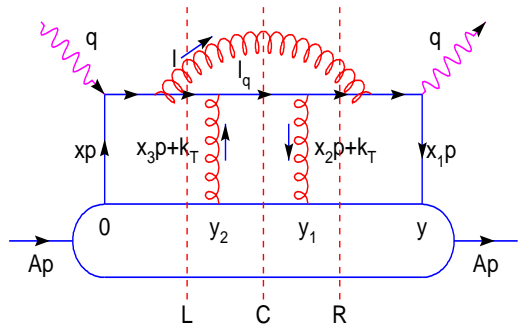,width=3in,height=2.0in}}
\caption{A typical diagram for quark-gluon re-scattering processes with three
possible cuts, central(C), left(L) and right(R).}
\label{fig:tw4-1}
\end{figure}

We will employ the generalized factorization of multiple scattering
processes\cite{LQS}. In this approximation, the double scattering
contribution to radiative correction from processes like the one
illustrated in Fig.~\ref{fig:tw4-1} can be written in the following form,
\begin{eqnarray}
\frac{dW_{\mu\nu}^D}{dz_h} &=& 
\sum_q \,\int_{z_h}^1\frac{dz}{z}D_{q\rightarrow h}(z_h/z)
\int \frac{dy^-}{2\pi}\, dy_1^-\, dy_2^-\,\frac{d^2y_T}{(2\pi)^2}
d^2k_T \nonumber \\
&\times& e^{- i\vec{k}_T \cdot\vec{y}_T}
\overline{H}^D_{\mu\nu}(y^-,y_1^-,y^-_2,k_T,p,q,z); \nonumber \\
&\ \times & \frac{1}{2} \,\langle A |
 \bar{\psi}_q(0)\, \gamma^+ \,A^+(y_{2}^{-},0_{T})\,
          A^+(y_{1}^{-},y_{T})\, \psi_q(y^{-}) |A\rangle \ .
\label{WD1}
\end{eqnarray}
Here $\overline{H}^D_{\mu\nu}(y^-,y_1^-,y^-_2,k_T,p,q,z)$ is the
Fourier transform of the partonic hard part 
$\widetilde{H}_{\mu\nu}(x,x_1,x_2,k_T,p,q,z)$ in momentum space,
\begin{eqnarray}
\overline{H}^D_{\mu\nu}(y^-,y_1^-,y^-_2,k_T,p,q,z) 
&=& \int dx\, \frac{dx_1}{2\pi}\, \frac{dx_2}{2\pi}\, 
       e^{ix_1p^+y^- + ix_2p^+y_1^-} \nonumber \\
&\ & \hspace{-1.0in}\times e^{i(x-x_1-x_2)p^+y_2^-}
\widetilde{H}^D_{\mu\nu}(x,x_1,x_2,k_T,p,q,z)\ ,
\label{eq-mH}
\end{eqnarray}
where $k_T$ is the relative transverse momentum carried by the second parton
in the double scattering. Values of the momentum fractions $x, x_1$ and $x_2$ 
are fixed by $\delta$-functions and poles in the partonic hard part. 
They normally depend on $k_T$.

In order to pick up the next-leading-twist contribution, we 
expand the partonic hard part around $k_T=0$,
\begin{eqnarray}
\overline{H}^D_{\mu\nu}(y^-,y_1^-,y^-_2,k_T,p,q,z)
&=& \overline{H}^D_{\mu\nu}(y^-,y_1^-,y^-_2,k_T=0,p,q,z) \nonumber \\
&\ & \hspace{-1.in}
+\left. \frac{\partial \overline{H}^D_{\mu\nu}}{\partial k_{T}^{\alpha}}
    \right|_{k_{T}=0}\ k_{T}^{\alpha}\, 
+\, \left. \frac{1}{2}\, \frac{\partial^{2}\overline{H}^D_{\mu\nu}}
                {\partial k_{T}^{\alpha} \partial  k_{T}^{\beta}} 
    \right|_{k_{T}=0}\ k_{T}^{\alpha}\, k_{T}^{\beta} + \ldots\ .
\label{eq-expand1}
\end{eqnarray}
This is known as the collinear expansion\cite{qiu}.
On the right-hand-side of Eq.(\ref{eq-expand1}), the first term gives
the eikonal contribution to the leading-twist results. It does not 
correspond to the physical double scattering, but simply makes the 
matrix element in a single scattering gauge invariant. The second term for 
unpolarized initial and final states vanishes after being integrated 
over $k_T$.  The third term will give a finite contribution to the double 
scattering process. Substituting Eq.(\ref{eq-expand1}) into
Eq.(\ref{WD1}), integrating over $d^2k_T$ and $d^2y_T$, we obtain
\begin{eqnarray}
\frac{dW_{\mu\nu}^D}{dz_h}
&=& \sum_q \,\int_{z_h}^1\frac{dz}{z}D_{q\rightarrow h}(z_h/z)
   \int \frac{dy^{-}}{2\pi}\, dy_1^-dy_2^- \nonumber \\
&\times& \frac{1}{2}\, \langle A | \bar{\psi}_q(0)\,
           \gamma^+\, F_{\sigma}^{\ +}(y_{2}^{-})\, 
 F^{+\sigma}(y_1^{-})\,\psi_q(y^{-})| A\rangle \nonumber \\
&\times &
    \left(-\frac{1}{2}g^{\alpha\beta}\right)
\left[\frac{1}{2}\, \frac{\partial^{2}}
{\partial k_{T}^{\alpha} \partial k_{T}^{\beta}}\,
\overline{H}^D_{\mu\nu}(y^-,y_1^-,y^-_2,k_T,p,q,z)\right]_{k_T=0} ,
\label{eq-expand2}
\end{eqnarray}
where $k_T^{\alpha}A^+k_T^{\beta}A^+$ are converted into field strength 
$F^{\alpha+}F^{\beta+}$ by partial integrations.

\subsection{Double Parton Scattering}

There are all together 23 cut-diagrams that contribute to the
leading twist-four corrections to the quark fragmentation
function in $eA$ DIS. For simplification of the calculation,
one can use the helicity amplitude approximation by Guo and 
Wang\cite{Wang:2001if,Guo:2000nz}
in the limit of soft gluon radiation. Such an approximation 
enables one to simplify the calculation of the radiation amplitudes.
However, a complete calculation of the all cut-diagrams were recently 
done by Zhang and Wang\cite{bwzxnw}. Before we list the results, we first 
consider the contribution from Fig.~\ref{fig:tw4-1} in detail. 

Using the conventional Feynman rule, one can write down the hard partonic
part of the central cut-diagram of 
Fig.~\ref{fig:tw4-1}\cite{Wang:2001if,Guo:2000nz},
\begin{eqnarray}
\overline{H}^D_{C\,\mu\nu}(y^-,y_1^-,y_2^-,k_T,p,q,z)&=&
\int dx\frac{dx_1}{2\pi}\frac{dx_2}{2\pi}
e^{ix_1p^+y^- + ix_2p^+y_1^-} \nonumber \\
 &\ & \hspace{-1.0in} \times
e^{i(x-x_1-x_2)p^+y_2^-} \int \frac{d^4\ell}{(2\pi)^4}
\frac{1}{2}{\rm Tr}\left[p\cdot\gamma\gamma_\mu p^\sigma p^\rho
\widehat{H}_{\sigma\rho}\gamma_\nu \right]  \nonumber \\
&\ & \hspace{-1.0in}\times 2\pi\delta_+(\ell^2)\,
\delta(1-z-\frac{\ell^-}{q^-}) \; . \label{eq:fig1-1}
\end{eqnarray}
\begin{eqnarray}
\widehat{H}_{\sigma\rho} &=&
\frac{C_F}{2N_c}g^4\frac{\gamma\cdot(q+x_1 p)}{(q+x_1p)^2-i\epsilon}
\,\gamma_\alpha\,\frac{\gamma\cdot(q + x_1 p
-\ell)}{(q+x_1p-\ell)^2-i\epsilon}
\,\gamma_\sigma \gamma\cdot\ell_q\,\gamma_\rho
\nonumber \\
&\times &\varepsilon^{\alpha\beta}(\ell)\frac{\gamma\cdot(q+xp -\ell)}
{(q+xp-\ell)^2+i\epsilon} \,\gamma_\beta\,
\frac{\gamma\cdot(q + xp)}{(q+xp)^2+i\epsilon}
\,\,2\pi \delta_+(\ell_q^2)\, , \label{eq:fig1-2}
\end{eqnarray}
where $\varepsilon^{\alpha\beta}(\ell)$ is the polarization tensor of a
gluon propagator in an axial gauge, $n\cdot A=0$ with
$n=[1,0^-,\vec{0}_\perp]$, and
$\ell$, $\ell_q=q+(x_1+x_2)p+k_T-\ell$
are the 4-momenta carried by the gluon and the final quark, respectively.
$z=\ell_q^-/q^-$ is the fraction of longitudinal momentum
(the large minus component) carried by the final quark.

To simplify the calculation, we apply the collinear
approximation to complete the trace of the product of
$\gamma$-matrices,
\begin{equation}
p^\sigma\widehat{H}_{\sigma\rho}p^\rho
\approx \gamma\cdot\ell_q \,\frac{1}{4\ell_q^-}
{\rm Tr} \left[\gamma^- p^\sigma\widehat{H}_{\sigma\rho}p^\rho\right] \; .
\label{coll}
\end{equation}
After carrying out momentum integrations in $x$, $x_1$, $x_2$
and $\ell^{\pm}$ with the help of contour integration
and $\delta$-functions, the partonic hard part
can be factorized into the product of $\gamma$-quark scattering
matrix $H^{(0)}_{\mu\nu}(x,p,q)$ [Eq.(\ref{H0})]
and the quark-gluon rescattering part $\overline{H}^D$,
\begin{eqnarray}
\overline{H}^D_{\mu\nu}(y^-,y_1^-,y_2^-,k_T,p,q,z)& = &
\int dx H^{(0)}_{\mu\nu}(x,p,q)\ \nonumber \\
&\times&\overline{H}^D(y^-,y_1^-,y_2^-,k_T,x,p,q,z)\, . \label{eq:hc0}
\end{eqnarray}
Contributions from all the diagrams have this factorized from.
Therefore, we will only list the rescattering part
$\overline{H}^D$ for different diagrams in the following. For the
central-cut diagram in Fig.~\ref{fig:tw4-1} it 
reads\cite{Wang:2001if,Guo:2000nz},
\begin{eqnarray}
\overline{H}^D_{C(Fig.\ref{fig:tw4-1}) }(y^-,y_1^-,y_2^-,k_T,x,p,q,z)&=&
\int \frac{d\ell_T^2}{\ell_T^2}\, \frac{\alpha_s}{2\pi}\,
 C_F\frac{1+z^2}{1-z} \nonumber \\
&\ &\hspace{-1.5in}\times\frac{2\pi\alpha_s}{N_c}
\overline{I}_{C(Fig.\ref{fig:tw4-1}) }(y^-,y_1^-,y_2^-,\ell_T,k_T,x,p,q,z)
 \, , \label{eq:hc1}
\end{eqnarray}
\begin{eqnarray}
\overline{I}_{C(Fig.\ref{fig:tw4-1}) }(y^-,y_1^-,y_2^-,\ell_T,k_T,x,p,q,z)
&=&e^{i(x+x_L)p^+y^- + ix_Dp^+(y_1^- - y_2^-)}
\theta(-y_2^-)\nonumber \\
&\ & \hspace{-1.5in} \times \theta(y^- - y_1^-) 
(1-e^{-ix_Lp^+y_2^-})(1-e^{-ix_Lp^+(y^- - y_1^-)}) \; .
\label{eq:Ic1}
\end{eqnarray}
Here, the fractional momentum is defined as
\begin{eqnarray}
  x_L&=&\frac{\ell_T^2}{2p^+q^-z(1-z)} \,\, ,\,\,
  x_D=\frac{k_T^2-2\vec{k}_T\cdot \vec{\ell}_T}{2p^+q^-z} \, ,
\label{eq:xld}
\end{eqnarray}
and $x=x_B=Q^2/2p^+q^-$ is the Bjorken variable.

The above contribution resembles the cross section of a dipole scattering
and contains essentially four terms. The first diagonal term
corresponds to the so-called hard-soft process where the
gluon radiation is induced by the hard scattering between the virtual photon
and an initial quark with momentum fraction $x$. The quark is
knocked off-shell by the virtual photon and becomes on-shell again after
radiating a gluon. Afterwards the on-shell 
quark (or the radiated gluon) will have a
secondary scattering with another soft gluon from the nucleus.
The second diagonal term is due to the double hard process
where the quark is on-shell after the first hard scattering with the
virtual photon. The gluon radiation is then induced by the scattering of
the quark with another gluon that carries finite momentum fraction $x_L+x_D$.
The other two off-diagonal terms are interferences between hard-soft
and double hard processes. In the limit of collinear
radiation ($x_L\rightarrow 0$) or when the formation time of the
gluon radiation, $\tau_f\equiv 1/x_Lp^+$, is much larger
than the nuclear size, the two processes have destructive interference,
leading to the LPM interference effect.

\begin{figure}
\centerline{\psfig{file=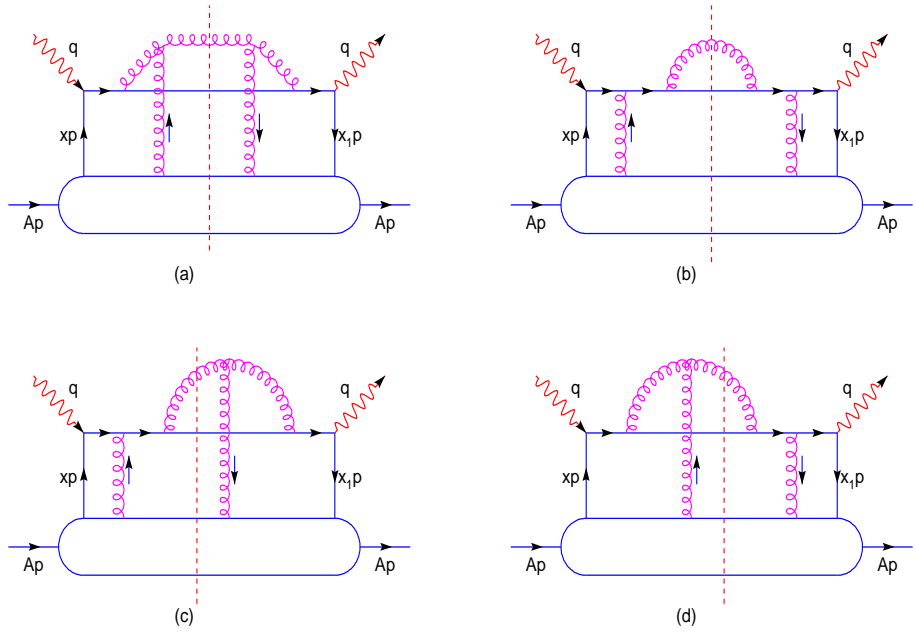,width=5.in,height=3.5in}}
\caption{Four central-cut diagrams that contribute to the final results.}
\label{fig:tw4-2}
\end{figure}

One can similarly obtain the  rescattering part $\overline{H}^D$
of other central-cut diagrams (a-d) in Fig.~\ref{fig:tw4-2}:
\begin{eqnarray}
\overline{H}^D_{C(a)}(y^-,y_1^-,y_2^-,k_T,x,p,q,z)&=&
\int \frac{d\ell_T^2}{(\vec{\ell_T}-\vec{k_T})^2}\, \frac{\alpha_s}{2\pi}\,
 C_A\frac{1+z^2}{1-z} \nonumber \\
&\ & \hspace{-1.0in} \times \frac{2\pi\alpha_s}{N_c}
\overline{I}_{C(a)}(y^-,y_1^-,y_2^-,\ell_T,k_T,x,p,q,z)
 \, , \nonumber \\
\overline{I}_{C(a)}(y^-,y_1^-,y_2^-,\ell_T,k_T,x,p,q,z)&=&
e^{i(x+x_L)p^+y^-+ix_Dp^+(y_1^--y_2^-)}
\theta(-y_2^-)\nonumber \\
&\ & \hspace{-1.0in}\times \theta(y^- - y_1^-) 
[e^{ix_Dp^+y_2^-/(1-z)}-e^{-ix_Lp^+y_2^-}] \nonumber \\
&\ & \hspace{-1.0in}
\times[e^{ix_Dp^+(y^- - y_1^-)/(1-z)}-e^{-ix_Lp^+(y^- - y_1^-)}] \, ,
\label{eq:hc(a)}
\end{eqnarray}
\begin{eqnarray}
\overline{H}^D_{C(b)}(y^-,y_1^-,y_2^-,k_T,x,p,q,z)&=&
\int \frac{d\ell_T^2}{(\vec{\ell_T}-(1-z)\vec{k_T})^2}\, 
\frac{\alpha_s}{2\pi}\,
 C_F\frac{1+z^2}{1-z} \nonumber \\
&\ &\hspace{-1.0in}\times\frac{2\pi\alpha_s}{N_c}
\overline{I}_{C(b)}(y^-,y_1^-,y_2^-,\ell_T,k_T,x,p,q,z)
 \, , \nonumber \\
\overline{I}_{C(b)}(y^-,y_1^-,y_2^-,\ell_T,k_T,x,p,q,z)&=&
e^{i(x+x_L)p^+y^-+ix_Dp^+(y_1^--y_2^-)}
\theta(-y_2^-) \nonumber \\
&\ & \hspace{-1.0in} \times\theta(y^- - y_1^-)
e^{-ix_Lp^+(y^- - y_1^-)}e^{-ix_Lp^+y_2^-} \, ,
\label{eq:hc(b)}
\end{eqnarray}
\begin{eqnarray}
\overline{H}^D_{C(c)}(y^-,y_1^-,y_2^-,k_T,x,p,q,z)&=&
\int d\ell_T^2\frac{ (\vec{\ell_T}-\vec{k_T})\cdot
(\vec{\ell_T}-(1-z)\vec{k_T}) }
{(\vec{\ell_T}-\vec{k_T})^2 (\vec{\ell_T}-(1-z)\vec{k_T})^2}\,
\nonumber \\
&\ &\hspace{-1.1in} \times\frac{\alpha_s}{2\pi}\,
\frac{C_A}{2}\frac{1+z^2}{1-z} 
\frac{2\pi\alpha_s}{N_c}
\overline{I}_{C(c)}(y^-,y_1^-,y_2^-,\ell_T,k_T,x,p,q,z), \nonumber \\
\overline{I}_{C(c)}(y^-,y_1^-,y_2^-,\ell_T,k_T,x,p,q,z)&=&
e^{i(x+x_L)p^+y^-+ix_Dp^+(y_1^--y_2^-)} \,
\theta(-y_2^-) \nonumber \\
&\ & \hspace{-1.0in}\times \theta(y^- - y_1^-) e^{-ix_Lp^+y_2^-} \nonumber \\
&\ &\hspace{-1.0in}\times
[e^{ix_Dp^+(y^- - y_1^-)/(1-z)}-e^{-ix_Lp^+(y^- - y_1^-)}] \, ,
\label{eq:hc(c)}
\end{eqnarray}
\begin{eqnarray}
\overline{H}^D_{C(d)}(y^-,y_1^-,y_2^-,k_T,x,p,q,z)&=&
\int d\ell_T^2\frac{ (\vec{\ell_T}-\vec{k_T})\cdot
(\vec{\ell_T}-(1-z)\vec{k_T}) }
{(\vec{\ell_T}-\vec{k_T})^2 (\vec{\ell_T}-(1-z)\vec{k_T})^2}\,
 \nonumber \\
&\ &\hspace{-1.1in} \times\frac{\alpha_s}{2\pi}\,
\frac{C_A}{2}\frac{1+z^2}{1-z}
\frac{2\pi\alpha_s}{N_c}
\overline{I}_{C(d)}(y^-,y_1^-,y_2^-,\ell_T,k_T,x,p,q,z), \nonumber \\
\overline{I}_{C(d)}(y^-,y_1^-,y_2^-,\ell_T,k_T,x,p,q,z)&=&
e^{i(x+x_L)p^+y^-+ix_Dp^+(y_1^--y_2^-)} \,
\theta(-y_2^-) \nonumber \\
&\ & \hspace{-1.0in}\times
\theta(y^- - y_1^-)e^{-ix_Lp^+(y^- - y_1^-)} \nonumber \\
&\ & \hspace{-1.0in}\times
[e^{ix_Dp^+y_2^-/(1-z)}-e^{-ix_Lp^+y_2^-}] \, .
\label{eq:hc(d)}
\end{eqnarray}

To complete the calculation we also have to consider
the asymmetrical-cut diagrams(left-cut and right-cut) that represent
interferences between single and triple scatterings.
They can be obtained with similar procedures. We refer readers
to Ref.\cite{bwzxnw} for a list of the the rescattering 
part $\overline{H}^D$ of all those asymmetrical-cut diagrams.

To obtain the double scattering contribution
to the semi-inclusive processes of hadron production in
Eq.(\ref{eq-expand2}), one will
then have to calculate the second derivatives of the
rescattering part $\overline{H}^D$.
After a closer examination of these rescattering parts,
one can find that all contributions from the
asymmetrical-cut diagrams have the form
\begin{eqnarray}
\overline{H}^D_{asym}=\frac{ \vec{\ell_T} \cdot
(\vec{\ell_T}-f(z)\vec{k_T}) }
{\ell_T^2(\vec{\ell_T}-f(z)\vec{k_T})^2}
e^{iXp^+Y^-}, \label{eq:asym}
\end{eqnarray}
where $ f(z)=0,\,1,\,1-z,\,z$,
 $X$ is the longitudinal momentum fraction and $Y^-$ represents the
spatial coordinates. One can prove that the second
derivative of the above expression vanishes at $k_T=0$,
\begin{eqnarray}
\nabla^2_{k_T}\frac{ \vec{\ell_T} \cdot (\vec{\ell_T}-f(z)\vec{k_T}) }
{\ell_T^2(\vec{\ell_T}-f(z)\vec{k_T})^2}  =0 \,.
\label{derivative}
\end{eqnarray}
Therefore, all contributions from the asymmetrical-cut(right-cut
and left-cut) diagrams will vanish after we take the second
partial derivative with respect to $k_T$ when we keep only the
leading terms up to ${\cal O}(x_B/Q^2\ell_T^2)$,
\begin{eqnarray}
\nabla^2_{k_T}\overline{H}^D_{asym}|_{k_T=0}
&=&0 +{\cal O}(x_B/Q^2\ell_T^2).
\label{ep:asym-deri}
\end{eqnarray}
In the same way we find that some of the central-cut diagrams will
not contribute to the final results, either.
In fact, after taking the second partial derivative
with respect to $k_T$ only four central-cut
diagrams shown in Fig.~\ref{fig:tw4-2} will contribute to the final result.

Including only those contributions that do not vanish after
the second derivative with respect to $k_T$, we have
\begin{eqnarray}
\nabla^2_{k_T}\overline{H}^D|_{k_T=0}
&=&\int d\ell_T^2 \frac{\alpha_s}{2\pi} \frac{1+z^2}{1-z}
e^{i(x+x_L)p^+y^-} \frac{2\pi\alpha_s}{N_c}
\theta(-y_2^-)\theta(y^--y_1^-)
\nonumber \\
&\times&\left[
\frac{4C_A}{\ell_T^4}(1-e^{-ix_Lp^+y_2^-})
(1-e^{-ix_Lp^+(y^--y_1^-)}) \right. \nonumber  \\
&+&\frac{4C_F(1-z)^2}{\ell_T^4}
e^{-ix_Lp^+(y^--y_1^-)}e^{-ix_Lp^+y_2^-} \nonumber \\
&+&\frac{2C_A(1-z)}{\ell_T^4}e^{-ix_Lp^+y_2^-}
(1-e^{-ix_Lp^+(y^--y_1^-)})  \nonumber  \\
&+& \frac{2C_A(1-z)}{\ell_T^4}e^{-ix_Lp^+(y^--y_1^-)}
(1-e^{-ix_Lp^+y_2^-}) \nonumber  \\
&+& \left. {\cal O}(x_B/Q^2\ell_T^2)\right] \,. \label{eq:expand1}
\end{eqnarray}
The first term at the right-hand side 
in Eq.(\ref{eq:expand1}) comes from the
contribution of $ \overline{H}^D_{C(a)} $ which is the main 
contribution in the helicity amplitude 
approximation\cite{Wang:2001if,Guo:2000nz}.
It contains hard-soft, double hard processes and their interferences.
The other three terms come from diagram (b),(c),(d) of
Fig.~\ref{fig:tw4-2} respectively. They constitute corrections
to the first term in powers of $1-z$. The second term that is
proportional to $(1-z)^2$ is from the
final state radiation from the quark in the double hard
process in Fig.~\ref{fig:tw4-2}(b).
The third and fourth terms are the results of the interference of
the final state radiation from the quark and other radiation processes
(initial state radiation and radiation from the gluon line).
They contain both double
hard processes and interferences between hard-soft and double hard
processes in Fig.~\ref{fig:tw4-2}(c) and (d).

Substituting Eq.(\ref{eq:expand1}) into Eqs.(\ref{eq:hc0}) and
(\ref{eq-expand2}), we have the semi-inclusive tensor from double
quark-gluon scattering including the contribution beyond the helicity
amplitude approximation,
\begin{eqnarray}
\frac{W_{\mu\nu}^{D,q}}{dz_h}
&=&\sum_q \,\int dx H^{(0)}_{\mu\nu}(xp,q)
\int_{z_h}^1\frac{dz}{z}D_{q\rightarrow h}(z_h/z)
\frac{\alpha_s}{2\pi} C_A \frac{1+z^2}{1-z} \nonumber \\
&\times&\int \frac{d\ell_T^2}{\ell_T^4} \frac{2\pi\alpha_s}{N_c}
\left[T^{A}_{qg}(x,x_L) +(1-z) T^{A(1)}_{qg}(x,x_L) \right.
 \nonumber \\
 &\ & \hspace{1.in}
+\left. \frac{C_F}{C_A} (1-z)^2 T^{A(2)}_{qg}(x,x_L)\right]
\, , \label{wd1}
\end{eqnarray}
where
\begin{eqnarray}
T^{A}_{qg}(x,x_L)&=& \int \frac{dy^{-}}{2\pi}\, dy_1^-dy_2^-
(1-e^{-ix_Lp^+y_2^-})(1-e^{-ix_Lp^+(y^--y_1^-)})
\nonumber  \\
&\times& e^{i(x+x_L)p^+y^-}\theta(-y_2^-)\theta(y^- -y_1^-)\nonumber  \\
&\times &\frac{1}{2}\langle A | \bar{\psi}_q(0)\,
\gamma^+\, F_{\sigma}^{\ +}(y_{2}^{-})\, F^{+\sigma}(y_1^{-})\,\psi_q(y^{-})
| A\rangle  \;\; , \label{Tqg} \\
T^{A(1)}_{qg}(x,x_L)&=& \int \frac{dy^{-}}{2\pi}\, dy_1^-dy_2^-
e^{i(x+x_L)p^+y^-}
\left[e^{-ix_Lp^+(y^- - y^-_1)}+e^{-ix_Lp^+y^-_2} \right. \nonumber  \\
&-&\left. 2e^{-ix_Lp^+(y^--y^-_1+y_2^-)}\right]
\theta(-y_2^-)\theta(y^- -y_1^-) 
\nonumber  \\
&\times& \frac{1}{4}\langle A | \bar{\psi}_q(0)\,
\gamma^+\, F_{\sigma}^{\ +}(y_{2}^{-})\, F^{+\sigma}(y_1^{-})\,\psi_q(y^{-})
| A\rangle \;\; , \label{Tqg1} \\
T^{A(2)}_{qg}(x,x_L)&=& \int \frac{dy^{-}}{2\pi}\, dy_1^-dy_2^-
e^{ixp^+y^- +ix_Lp^+(y^-_1-y^-_2)}\theta(-y_2^-)\theta(y^- -y_1^-)
\nonumber  \\
& & \frac{1}{2}\langle A | \bar{\psi}_q(0)\,
\gamma^+\, F_{\sigma}^{\ +}(y_{2}^{-})\, F^{+\sigma}(y_1^{-})\,\psi_q(y^{-})
| A\rangle 
\label{Tqg2}
\end{eqnarray}
are twist-four parton matrix elements of the nucleus. Evidently these
parton matrix elements are not independent of each other. $T^A_{qg}(x,x_L)$
has the complete four terms of soft-hard, double hard processes
and their interferences. Therefore it contains essentially four
independent parton matrix elements. $T^{A(1)}_{qg}(x,x_L)$ and
$T^{A(2)}_{gq}(x,x_L)$ are the results of the corrections beyond
the helicity amplitude approximation. But these two matrix elements are
already contained in $T^A_{qg}(x,x_L)$.

During the collinear expansion, we have kept $\ell_T$ finite and
took the limit $k_T\rightarrow 0$. As a consequence, the gluon field
in one of the twist-four parton matrix elements in
Eqs.(\ref{Tqg})-(\ref{Tqg2}) carries zero momentum in
the soft-hard process. However, the
gluon distribution $xf_g(x)$ at $x=0$ is not defined in QCD.
As argued in Refs.\cite{Wang:2001if,Guo:2000nz}, this is due to the omission
of higher order terms in the collinear expansion.
As a remedy to the problem, a subset of the higher-twist
terms in the collinear expansion can be resummed to
restore the phase factors such as $\exp(ix_Tp^+y^-)$,
where $x_T\equiv \langle k_T^2\rangle/2p^+q^-z$ is related to
the intrinsic transverse momentum of the initial partons.
As a result, soft gluon fields in the parton matrix elements
will carry a fractional momentum $x_T$.

Using the factorization 
approximation\cite{Wang:2001if,Guo:2000nz,LQS,ow}, we can
relate the twist-four parton matrix elements of the nucleus
to the twist-two parton distributions of nucleons and the
nucleus,
\begin{eqnarray}
T^A_{qg}(x,x_L)&=&\frac{C}{x_A}
(1-e^{-x_L^2/x_A^2}) [f_q^A(x+x_L)\, x_Tf_g^N(x_T)
\nonumber \\
&+&f_q^A(x)(x_L+x_T)f_g^N(x_L+x_T)] \, ,
\end{eqnarray}
where C is a constant, $x_A=1/MR_A$, $f_q^A(x)$ is the quark
distribution inside a nucleus, and $f_g^N(x)$ is the gluon
distribution inside a nucleon. A Gaussian distribution in the
light-cone coordinates was assumed for the nuclear distribution,
$\rho(y^-)=\rho_0 \exp({y^-}^2/2{R^-_A}^2)$, where
$R^-_A=\sqrt{2}R_AM/p^+$ and $M$ is the nucleon mass. We should
emphasize that the twist-four matrix element is proportional to
$1/x_A=R_AM$, or the nuclear size\cite{ow}.

Notice that the off-diagonal matrix elements that correspond to
the interferences between hard-soft and double hard processes
are suppressed by a factor of $\exp(-x_L^2/x_A^2)$.
This is because in the interferences between double-hard and
hard-soft processes, there is actually
momentum flow of $x_Lp^+$ between the two nucleons that
the initial quark and gluon come from.
Without strong long range two-nucleon correlation inside a
nucleus, the amount of momentum flow $x_Lp^+$ should then
be restricted to the amount allowed by the uncertainty
principle, $1/R^-_A\sim p^+/R_AM$.
Similarly, the other two-parton matrix elements in
Eqs.(\ref{Tqg1}) and (\ref{Tqg2}) can be approximated as
\begin{eqnarray}
T^{A(1)}_{qg}(x,x_L)
&=&\frac{C}{2x_A}
\left\{ \left[f_q^A(x+x_L)\, x_Tf_g^N(x_T)\right.\right. \nonumber\\
&+&\left. f_q^A(x)(x_L+x_T)f_g^N(x_L+x_T)\right]e^{-x_L^2/x_A^2} \nonumber\\
&-&\left. 2f_q^A(x)(x_L+x_T)f_g^N(x_L+x_T) \right\} \, , \\
T^{A(2)}_{qg}(x,x_L)
&=&\frac{C}{x_A}f_q^A(x)(x_L+x_T)f_g^N(x_L+x_T) \, .
\end{eqnarray}

>From the above estimate of the matrix elements, both $T^A_{qg}(x,x_L)$
and $T^{A(1)}_{qg}(x,x_L)$ contain a factor $1-e^{-x_L^2/x_A^2}$
because of the LPM interference effect. Such an interference
factor will effectively cut off the integration over the
transverse momentum at $x_L\sim x_A$ in Eq.(\ref{wd1}). As we
will show later in the calculation of the effective energy loss,
the integration with such a restriction in the transverse momentum
due to LPM interference effect will give rise to a factor $1/x_A$
in addition to the coefficient $f^A_q(x)/x_A$. Consequently,
contributions from double scattering in Eq.(\ref{wd1}) that are
associated with $T^A_{qg}(x,x_L)$ and $T^{A(1)}_{qg}(x,x_L)$ will
be proportional to $R_A^2f^A_q(x)$. These are the leading double
scattering contributions in the limit of a large nuclear size. On
the other hand, the third term $T^{A(2)}_{qg}(x,x_L)$ in
Eq.(\ref{wd1}), which does not contain any interference effect,
will only contribute to a correction that is proportional to
$R_Af^A_q(x)$. In the limit of a large nucleus, $A^{1/3}\gg 1$, we
will neglect this term in our study of the double scattering processes.

\subsection{Virtual Corrections}

So far we have not considered virtual corrections which will ensure the
final result to be infrared safe. The calculation of the virtual 
corrections can be calculated similarly as the radiative corrections.
On the other hand, they can also be obtained via unitarity
requirement.  When cast into the DGLAP evolution equation as 
Eq.(\ref{eq:ap1}), the real corrections can be interpreted as
the probability for the quark to radiate a gluon with momentum 
fraction $1-z$.
Then one must also take into account the probability of no
gluon radiation in the evolution to ensure unitarity. Such unitarity
requirement gives rise to the same virtual correction as calculated 
from the virtual diagrams. 
The virtual contribution to the quark fragmentation in double
scattering processes is, for example,
\begin{eqnarray}
\frac{W_{\mu\nu}^{D(v),q}}{dz_h}
&=&-\sum_q \,\int dx H^{(0)}_{\mu\nu}(xp,q)
D_{q\rightarrow h}(z_h) \nonumber\\
&\times& \frac{\alpha_s}{2\pi} 
C_A\int_{0}^1dz\,\frac{1+z^2}{1-z}
\int \frac{d\ell_T^2}{\ell_T^4} \frac{2\pi\alpha_s}{N_c} 
T^{A(m)}_{qg}(x,x_L) \, , \label{eq:WD-fqv1}
\end{eqnarray}
where
\begin{equation}
T^{A(m)}_{qg}(x,x_L) \equiv T^{A}_{qg}(x,x_L)
+(1-z)T^{A(1)}_{qg}(x,x_L) \, . \label{modT}
\end{equation}
Assuming a function $F(z)$ which is sufficiently smooth at $z=1$,
One can single out the infrared divergent part of the following
integral,
\begin{eqnarray}
\int_{0}^1dz\,\frac{1+z^2}{1-z}F(z)
&=&F(1) \int_{0}^1dz\frac{2}{1-z}
-\Delta F \; ; \nonumber \\
\Delta F &\equiv &
\int_0^1 dz\frac{1}{1-z}\left[ 2 F(1)
-(1+z^2) F(z)\right] \, . \label{eq:vsplit0}
\end{eqnarray}
The second term $\Delta F$ is finite since $F(z)$ is a 
smooth function of $z$.

We can apply this procedure to the integral in the modified
fragmentation function.
The divergent term can be combined with the radiative contribution 
in Eq.(\ref{wd1}) to cancel the infrared divergency. 
With the help of the `$+$'function\cite{dglap},
the final result can be expressed as
\begin{eqnarray}
\frac{W_{\mu\nu}^{D,q}}{dz_h}
&=&\sum_q \,\int dx H^{(0)}_{\mu\nu}(xp,q)
\frac{2\pi\alpha_s}{N_c}\int \frac{d\ell_T^2}{\ell_T^4}  
\int_{z_h}^1\frac{dz}{z}D_{q\rightarrow h}(z_h/z) 
 \frac{\alpha_s}{2\pi} C_A \nonumber \\
&\times& \left[
\frac{1+z^2}{(1-z)_+}T^{A(m)}_{qg}(x,x_L)+\delta(z-1)
\Delta T^{A(m)}_{qg}(x,\ell_T^2) \right] \, ;\label{eq:WD-fq2}
\end{eqnarray}
\begin{eqnarray}
\Delta T^{A(m)}_{qg}(x,\ell_T^2) &\equiv&
\int_0^1 dz\frac{1}{1-z}\left[ 2 T^{A(m)}_{qg}(x,x_L)|_{z=1} \right.
\nonumber\\
&\ & \hspace{1.in}
-\left. (1+z^2) T^{A(m)}_{qg}(x,x_L)\right] \, . \label{eq:vsplit}
\end{eqnarray}
Here the implicit $z$-dependence of $T^{A(m)}_{qg}(x,x_L)$ 
plays an important role in the
final result. The above integrand will be proportional
to the splitting function for a single scattering if
one ignores the $z$ dependence of $T^A_{qg}(x,x_L)$.
Similarly, the final result for contributions from gluon fragmentation is
\begin{eqnarray}
\frac{W_{\mu\nu}^{D,g}}{dz_h}
&=&\sum_q \,\int dx H^{(0)}_{\mu\nu}(xp,q)
\frac{2\pi\alpha_s}{N_c}\int \frac{d\ell_T^2}{\ell_T^4}  
\int_{z_h}^1\frac{dz}{z}D_{g\rightarrow h}(z_h/z) \nonumber \\
&\ &\hspace{-0.8cm}\times\frac{\alpha_s}{2\pi} C_A\left[
\frac{1+(1-z)^2}{z_+}T^{A(m)}_{qg}(x,x_L)+\delta(z)
\Delta T^{A(m)}_{qg}(x,\ell_T^2) \right], 
\label{eq:WD-fg2}
\end{eqnarray}
where we have used the fact that $x_L$ in Eq.(\ref{eq:xld}) is invariant
under the transform $z\rightarrow 1-z$ and so is $T^A_{qg}(x,x_L)$.

\subsection{Modified Fragmentation Function and Parton Energy Loss}

Including these virtual corrections and the single scattering
contribution, we can rewrite the semi-inclusive tensor in
terms of a modified fragmentation function
$\widetilde{D}_{q\rightarrow h}(z_h,\mu^2)$,
\begin{equation}
\frac{dW_{\mu\nu}}{dz_h}=\sum_q \int dx \widetilde{f}_q^A(x,\mu_I^2)
H^{(0)}_{\mu\nu}(x,p,q)
\widetilde{D}_{q\rightarrow h}(z_h,\mu^2) \label{eq:Wtot}
\end{equation}
where $\widetilde{f}_q^A(x,\mu_I^2)$ is the quark distribution function
which in principle should also include the
higher-twist contribution\cite{MQiu} of the
initial state scattering. The modified effective quark
fragmentation function  is defined as
\begin{eqnarray}
\widetilde{D}_{q\rightarrow h}(z_h,\mu^2)&\equiv&
D_{q\rightarrow h}(z_h,\mu^2)\nonumber \\
&+&\int_0^{\mu^2} \frac{d\ell_T^2}{\ell_T^2}
\frac{\alpha_s}{2\pi} \int_{z_h}^1 \frac{dz}{z}
\left[ \Delta\gamma_{q\rightarrow qg}(z,x,x_L,\ell_T^2)
D_{q\rightarrow h}(z_h/z) \right. \nonumber \\
&+& \left. \Delta\gamma_{q\rightarrow gq}(z,x,x_L,\ell_T^2)
D_{g\rightarrow h}(z_h/z)\right] \, , \label{eq:MDq}
\end{eqnarray}
where $D_{q\rightarrow h}(z_h,\mu^2)$ and
$D_{g\rightarrow h}(z_h,\mu^2)$ are the leading-twist
fragmentation functions. The modified splitting functions are
given as
\begin{eqnarray}
\Delta\gamma_{q\rightarrow qg}(z,x,x_L,\ell_T^2)&=&
\left[\frac{1+z^2}{(1-z)_+}T^{A(m)}_{qg}(x,x_L) +
\delta(1-z)\Delta T^{A(m)}_{qg}(x,\ell_T^2) \right]\nonumber \\
&\times& \frac{2\pi\alpha_s C_A}
{\ell_T^2 N_c\widetilde{f}_q^A(x,\mu_I^2)}\, ,
\label{eq:r1}\\
\Delta\gamma_{q\rightarrow gq}(z,x,x_L,\ell_T^2)
&=& \Delta\gamma_{q\rightarrow qg}(1-z,x,x_L,\ell_T^2). \label{eq:r2}
\end{eqnarray}

To further simplify the calculation, we assume
$x_T\ll x_L \ll x$. The modified parton matrix elements can be
approximated by
\begin{equation}
T^{A(m)}_{qg}(x,x_L)\approx \frac{\widetilde{C}}{x_A}
(1-e^{-x_L^2/x_A^2}) f_q^A(x) \left[1-\frac{1-z}{2}\right],
\label{modT2}
\end{equation}
where $\widetilde{C}\equiv 2C x_Tf^N_g(x_T)$ is a coefficient which
should in principle depend on $Q^2$ and $x_T$. Here we will simply take
it as a constant.

In the above matrix element, one can identify $1/x_Lp^+=2q^-z(1-z)/\mu^2$ 
as the formation time of the emitted gluons. For large formation time 
as compared to the nuclear size, the above matrix element vanishes,
demonstrating a typical LPM interference effect.
This is because the emitted gluon (with long formation time) and the leading 
quark are still a coherent system when they propagate through the nucleus.
Additional scattering will not induce more gluon radiation, 
thus limiting the energy loss of the leading quark. 

The reduction of phase space available for gluon radiation due to the 
LPM interference effect is critical for applying the LQS formalism to 
the problem in this paper. In the original LQS approach\cite{LQS}, the 
generalized factorization for processes with a large final transverse
momentum $\ell_T^2\sim Q^2$ allows one to consider the leading 
contribution in $1/Q^2$, which is enhanced by the nuclear 
size $R_A \sim A^{1/3}$. For large $Q^2$ and $A$, the higher-twist contribution
from double parton rescattering that is proportional to $\alpha_s R_A/Q^2$
will then be the leading nuclear correction. One can neglect contributions 
from more than two parton rescattering. In deriving the modified fragmentation 
functions, we however have to take the leading logarithmic approximation 
in the limit $\ell_T^2\ll Q^2$, where $\ell_T$ is the transverse 
momentum of the radiated gluon. Since the LPM interference suppresses 
gluon radiation whose formation time ($\tau_f \sim Q^2/\ell_T^2p^+$)
is larger than the nuclear 
size $MR_A/p^+$ in our chosen frame, $\ell_T^2$ should then have a 
minimum value of $\ell_T^2\sim Q^2/MR_A\sim Q^2/A^{1/3}$. 
Here $M$ is the nucleon mass.
Therefore, the logarithmic approximation is still valid for
large nuclei ($MR_A\gg 1$). In the meantime, the leading higher-twist
contribution proportional to $\alpha_s R_A/\ell_T^2 \sim \alpha_s R_A^2/Q^2$
can still be small for large $Q^2$ so that one can neglect
processes with more than two parton rescattering. The parameter for
twist expansion of the fragmentation processes inside a nucleus is 
thus $\alpha_sA^{2/3}/Q^2$ as compared to $\alpha_sA^{1/3}/Q^2$ for
processes with large final transverse momentum as studied by LQS\cite{LQS}.
This is why the nuclear modification to the fragmentation function
as derived in this paper depends quadratically on the nuclear size $R_A$.

\begin{figure}
\centerline{\psfig{figure=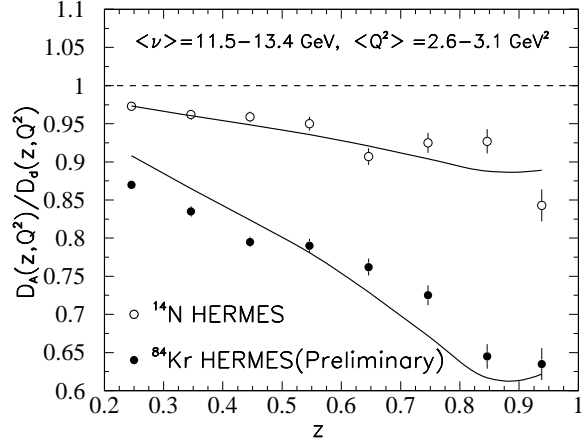,width=3.0in,height=2.3in}}
\caption{Predicted nuclear modification of jet fragmentation function
is compared to the HERMES data \protect\cite{hermes} on ratios of
hadron distributions between $A$ and $d$ targets in DIS.}
\label{fig:hermes1}
\end{figure}

\begin{figure}
\centerline{\psfig{figure=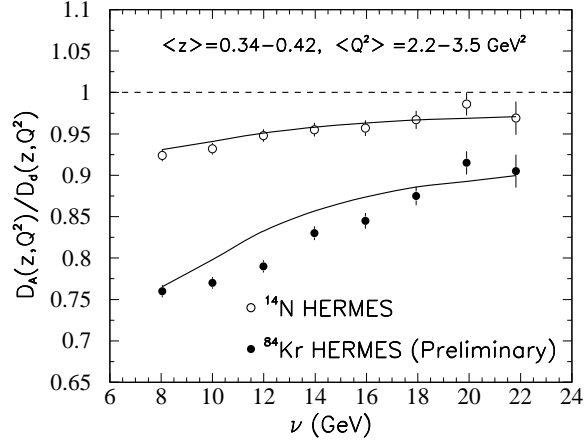,width=3.0in,height=2.3in}}
\caption{Energy dependence of the nuclear modification compared
with the HERMES data \protect\cite{hermes}.}
\label{fig:hermes2}
\end{figure}

Because of momentum conservation, the fractional momentum in a
nucleon is limited to $x_L<1$. Though the Fermi motion effect in a
nucleus can allow $x_L>1$, the parton distribution in this region
is still significantly suppressed. It therefore provides a natural
cut-off for $x_L$ in the integration over $z$ and $\ell_T$ in 
Eq.(\ref{eq:MDq}). With the assumption of the factorized form 
of the twist-4 nuclear parton matrices, there is only one free 
parameter $\widetilde{C}(Q^2)$
which represents quark-gluon correlation strength inside nuclei.
Once it is fixed, one can predict the $z$, energy and
nuclear dependence of the medium modification of the fragmentation
function. Shown in Figs.~\ref{fig:hermes1} 
and \ref{fig:hermes2} are the calculated 
nuclear modification factor of the fragmentation functions for $^{14}N$ 
and $^{84}Kr$ targets as compared to the recent HERMES data\cite{hermes}.
There are strong correlations among values of $Q^2$, $\nu$ and $z$ in the
HERMES data which are also taken in account in our calculation.
The predicted shape of the $z$- and $\nu$-dependence agrees well 
with the experimental data.  A remarkable feature of the prediction
is the quadratic $A^{2/3}$ nuclear size dependence, which is verified 
for the first time by an experiment. This quadratic dependence comes 
from the combination of the QCD radiation spectrum and the modification 
of the available phase space in $\ell_T$ or $x_L$ due to the LPM 
interferences. Note that the numerical results shown here are
obtained with the original helicity amplitude 
approximation\cite{Wang:2001if,Guo:2000nz}.
The numerical calculation beyond the helicity approximation differs only
about a few percent\cite{bwzxnw}.

By fitting the overall suppression for one nuclear target, 
we obtain the only parameter in our calculation,
$\widetilde{C}(Q^2)=0.0060$ GeV$^2$ 
with $\alpha_{\rm s}(Q^2)=0.33$ at $Q^2\approx 3$ GeV$^2$.
This parameter is also related to nuclear broadening of 
the transverse momentum of the Drell-Yan dilepton in 
$pA$ collisions\cite{guo},
$\langle\Delta q_\perp^2\rangle\approx\widetilde{C}\pi\alpha_{\rm s}/N_cx_A$.
With an experimental\cite{peng} value of 
$\langle\Delta q_\perp^2\rangle=0.016A^{1/3}$ GeV$^2$ and
$\alpha_{\rm s}(M^2_{l\bar{l}})=0.21$ ($<M_{l\bar{l}}^2>\approx 40$ GeV$^2$),
one finds $\widetilde{C}(M^2_{l\bar{l}})=0.013$ GeV$^2$, 
which is about a factor 2
larger than the value obtained in our fit to the HERMES data. The value of 
$\widetilde{C}$ determined from nuclear broadening in photo-production
of a di-jet is even larger\cite{LQS} at $Q^2=4p_T^2\approx 64$ GeV$^2$.
Such a strong scale dependence of $\widetilde{C}(Q^2)$ 
is in line with one's expectation,
since it is related to gluon distribution $xg(x,Q^2)$ at 
small $x$ in nuclei\cite{ow}.

In principle the modification of the fragmentation functions would be the only
experimental effect of induced gluon radiation via multiple scattering. 
One can never directly measure the energy loss of the leading
quark. The net effect of the energy loss is the suppression of leading 
particles on one hand and the enhancement of soft particles on the other, 
leading to the modification of the fragmentation functions. One can then
experimentally characterize the parton energy loss via the momentum transfer 
from large to small momentum regions of the fragmentation functions.

Upon a close examination of Eq.(\ref{eq:MDq}), we see that 
the first term is the renormalized fragmentation
function in vacuum. The rest is particle production induced by the 
rescattering of the quark through the nuclear medium. In particular,
the last term is particle production from the fragmentation of the gluon
which is induced by the secondary scattering. Such particle production
is at the expense of the energy loss of the leading quark. We can thus
quantify the quark energy loss by the momentum fraction carried by 
the radiated gluon,
\begin{eqnarray}
\langle\Delta z_g\rangle(x_B,\mu^2)
&=& \int_0^{\mu^2}\frac{d\ell_T^2}{\ell_T^2}
\int_0^1 dz \frac{\alpha_s}{2\pi}
 z\,\Delta\gamma_{q\rightarrow gq}(z,x_B,x_L,\ell_T^2) \nonumber \\
&=&\frac{C_A\alpha_s^2}{N_c} \int_0^{\mu^2}\!\!\!\!\frac{d\ell_T^2}{\ell_T^4}
\int_0^1 \!\!dz [1+(1-z)^2]
\frac{T^{A(m)}_{qg}(x_B,x_L)}{\widetilde{f}_q^A(x_B,\mu_I^2)}\, .\nonumber \\
\label{eq:loss1}
\end{eqnarray}
Using the approximation for the modified twist-four parton matrix elements
in Eq.(\ref{modT2}), we have
\begin{eqnarray}
\langle\Delta z_g\rangle(x_B,\mu^2)
&=&\widetilde{C}\frac{C_A\alpha_s^2}{N_c}
\frac{x_B}{x_AQ^2} \int_0^1 dz \frac{1+(1-z)^2}{z(1-z)}\nonumber \\
&\times&
\int_0^{x_\mu} \frac{dx_L}{x_L^2}(1-\frac{z}{2})\,(1-e^{-x_L^2/x_A^2}),
\label{eq:heli-loss}
\end{eqnarray}
where $x_\mu=\mu^2/2p^+q^-z(1-z)=x_B/z(1-z)$ if we choose the
factorization scale as $\mu^2=Q^2$.
When $x_A\ll x_B\ll 1$ we can estimate the leading quark energy
loss roughly as
\begin{eqnarray}
\langle \Delta z_g\rangle(x_B,\mu^2)& \approx &
\widetilde{C}\frac{C_A\alpha_s^2}{N_c}\frac{x_B}{Q^2
x_A^2}5\sqrt{\pi}[\ln\frac{1-2x_B}{x_B}-\frac{1}{2}] \, .
\label{eq:appr1-loss}
\end{eqnarray}
Since $x_A=1/MR_A$, the energy loss $\langle \Delta
z_g\rangle$ thus depends quadratically on the nuclear size.

In the rest frame of the nucleus, $p^+=m_N$, $q^-=\nu$, and
$x_B\equiv Q^2/2p^+q^-=Q^2/2m_N\nu$. One can
get the averaged total energy loss as
$ \Delta E=\nu\langle\Delta z_g\rangle
\approx  \widetilde{C}(Q^2)\alpha_{\rm s}^2(Q^2)
m_NR_A^2(C_A/N_c) 3\ln(1/2x_B)$.
With the determined value of $\widetilde{C}$, 
$\langle x_B\rangle \approx 0.124$ in the HERMES experiment\cite{hermes}
and the average distance $\langle L_A\rangle=R_A\sqrt{2/\pi}$
for the assumed Gaussian nuclear distribution,
one gets the quark energy 
loss $dE/dL\approx 0.5$ GeV/fm inside a $Au$ nucleus.

\subsection{Energy Loss in Hot Medium at RHIC}

To extend our study of modified fragmentation functions to 
jets in heavy-ion collisions and to relate to results obtained
in the opacity expansion approach, we can
assume $\langle k_T^2\rangle\approx \mu^2$ (the Debye screening mass)
and a gluon density profile
$\rho(y)=(\tau_0/\tau)\theta(R_A-y)\rho_0$ for a 1-dimensional 
expanding system. Since the initial jet production 
rate is independent of the final gluon density which can be 
related to the parton-gluon scattering cross 
section\cite{Baier:1996sk} 
[$\alpha_s x_TG(x_T)\sim \mu^2\sigma_g$], one has then
\begin{equation}
\frac{\alpha_s T_{qg}^A(x_B,x_L)}{f_q^A(x_B)} \sim
\mu^2\int dy \sigma _g \rho(y)
[1-\cos(y/\tau_f)],
\end{equation}
where $\tau_f=2Ez(1-z)/\ell_T^2$ is the gluon formation time. One
can recover the form of energy loss in a thin plasma obtained 
in the opacity expansion approach\cite{Gyulassy:2000gk},
\begin{eqnarray}
\langle\Delta z_g\rangle &=&\frac{C_A\alpha_s}{\pi}
\int_0^1 dz \int_0^{\frac{Q^2}{\mu^2}}du \frac{1+(1-z)^2}{u(1+u)} \nonumber \\
&\times&\int_{\tau_0}^{R_A} d\tau\sigma_g\rho(\tau) 
\left[1-\cos\left(\frac{(\tau-\tau_0)\,u\,\mu^2}{2Ez(1-z)}\right)\right].
\end{eqnarray}
Keeping only the dominant contribution and assuming 
$\sigma_g\approx C_a 2\pi\alpha_s^2/\mu^2$ ($C_a$=1 for $qg$ and 9/4 for
$gg$ scattering), one obtains the averaged energy loss,
\begin{equation}
\langle \frac{dE}{dL}\rangle \approx \frac{\pi C_aC_A\alpha_s^3}{R_A}
\int_{\tau_0}^{R_A} d\tau \rho(\tau) (\tau-\tau_0)\ln\frac{2E}{\tau\mu^2}.
\end{equation}
Neglecting the logarithmic dependence on $\tau$, the averaged energy loss
in a 1-dimensional expanding system can be expressed as
\begin{equation}
\langle\frac{dE}{dL}\rangle_{1d} \approx \frac{dE_0}{dL} \frac{2\tau_0}{R_A},
\end{equation}
where $dE_0/dL\propto \rho_0R_A$
is the energy loss in a static medium with the same gluon density $\rho_0$ 
as in a 1-d expanding system at time $\tau_0$.
Because of the expansion, the averaged energy loss $\langle dE/dL\rangle_{1d}$
is suppressed as compared to the static case and does not depend linearly
on the system size. This could be one of the reasons why the effect of
parton energy loss is found to be negligible in $AA$ collisions at
$\sqrt{s}=17.3$ GeV\cite{wangsps}.

An effective model of modified fragmentation functions was
proposed in Ref.\cite{Wang:1996yh,Wang:1996pe}:
\begin{equation}
\widetilde{D}_{a\rightarrow h} (z)\approx \frac{1}{1-\Delta z} 
D_{a\rightarrow h}\left(\frac{z}{1-\Delta z}\right), \label{dbar0}
\end{equation}
with $\Delta z$ to account for the fractional parton energy loss. 
This effective model is found to
reproduce the pQCD result 
from Eq.(\ref{eq:MDq}) very well, but only when
$\Delta z$ is set to be $\Delta z\approx 0.6 \langle z_g\rangle$.
Therefore the actual averaged parton energy loss should be
$\Delta E/E=1.6\Delta z$ with $\Delta z$ extracted from the 
effective model. The factor 1.6 is mainly
caused by the unitarity correction effect in 
the pQCD calculation. A similar effect is also found in the 
opacity expansion approach\cite{Gyulassy:2001nm}.

\begin{figure}
\centerline{\psfig{figure=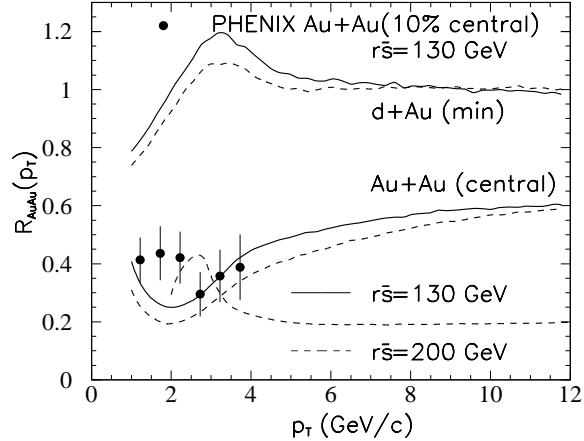,width=3.0in,height=2.3in}}
\caption{Calculated nuclear modification factor of $\pi^0$ $p_T$
spectra for $d+Au$ and central $Au+Au$ collisions at $\sqrt{s}=130$ (solid)
and 200 GeV (dashed) as compared to PHENIX data \protect\cite{Adcox:2002pe}
The lower dashed line uses the energy loss as given 
in Eq.(\protect\ref{para}).}
\label{fig:phenix}
\end{figure}

Both PHENIX  and STAR experiments have 
reported\cite{Adcox:2002pe,Kunde:2002pb} strong suppression 
of high $p_T$ hadrons in central $Au+Au$ collisions at 
$\sqrt{s}=130$ and 200 GeV, indicating for the first 
time a large parton energy loss in heavy-ion collisions. 
To extract the parton energy loss, we compare the data
with the calculated hadron $p_T$ spectra in heavy-ion collisions 
using the above effective model for medium modified jet fragmentation 
functions\cite{Wang:1998ww}.
Shown in Fig.~\ref{fig:phenix} are the nuclear modification 
factors $R_{AA}(p_T)$ as the ratios of hadron spectra in $AA$ ($pA$) 
and $pp$ collisions normalized by the number of binary 
collisions\cite{Wang:2001bf}.
Parton shadowing and nuclear broadening of the intrinsic $k_T$ 
are also taken into account in the calculation which decrsibes $pA$
data for energies up to $\sqrt{s}=40$ GeV\cite{Wang:1998ww}.
The nuclear $k_T$-broadening gives the Cronin 
enhancement at large $p_T$ in $pA$ collisions, where there is no
parton energy loss induced by a hot medium. 
Fitting the PHENIX data 
yields  $\langle dE/dL\rangle_{1d} \approx 0.34(\ln E/\ln 5)$ GeV/fm, 
including the 
factor of 1.6 from the unitarity correction effect. We consider only
$\pi^0$ data here, since at large $p_T$ the charged hadrons are dominated by
baryons, which could be influenced mainly by non-perturbative
dynamics\cite{Gyulassy:2000gk}.

Taking into account the expansion, the averaged parton
energy loss extracted from the PHENIX data 
would be equivalent to $(dE/dL)_0=0.34 (R_A/2\tau_0)\ln E/\ln 5$ 
in a static system 
with the same gluon density as the initial value of the
expanding system at $\tau_0$. With $R_A\sim 6$ fm and $\tau_0\sim 0.2$ fm,
this would give $(dE/dL)_0\approx 7.3$ GeV/fm for a 10-GeV parton, 
which is about 15 times of that in a cold $Au$ nucleus, as extracted 
from the HERMES data. Since the parton energy loss is directly 
proportional to gluon density of the medium, this
implies that the gluon density in the initial stage of $Au+Au$
collisions at $\tau_0=0.2$ fm/$c$ is about 15 times 
higher than that inside a cold nucleus. 
We can predict the $\pi^0$ spectra at $\sqrt{s}=200$ GeV as given by
the dashed lines in Fig.~\ref{fig:phenix}, assuming that the initial parton 
density in central $Au+Au$ collisions at $\sqrt{s}=200$ GeV is 
about 10\% higher than at 130 GeV.

In Fig.~\ref{fig:phenix}, we have shown the predicted suppression factor at
large $p_T$ with the same logarithmic energy-dependence of the 
energy loss as in
the cold nuclear matter which gives a suppression factor that increases
with $p_T$. However, new RHIC data\cite{Mioduszewski:2002wt} at 200 GeV show
a constant suppression factor at larger $p_T$. This might be
the indication of the importance of the detailed balance\cite{ww01} 
which gives much stronger energy dependence of the energy loss. 
We show as the lower dashed line in Fig.~\ref{fig:phenix} the result
for an effective parton energy loss
\begin{equation}
\Delta E\propto (E/\mu-1.6)^{1.20}/(7.5+E/\mu)
\label{para}
\end{equation}
which is parameterized according to the result from Ref.\cite{ww01}
where both stimulated gluon emission and thermal absorption are
included in the calculation of the total energy loss. The detailed
balance between emission and absorption both reduces the effective parton
energy loss and increases the energy dependence. 
The threshold is the consequence of gluon absorption that competes
with radiation that effectively shuts off the energy loss. The
parameter $\mu$ is set to be 1 GeV in the calculation 
shown in Fig.~\ref{fig:phenix}.

\section{Summary} 

In this report we reviewed two recent approaches to the
problem of non-Abelian radiative energy energy loss in dense but finite
QCD matter. In the first section, we highlighted some of the striking
new high $p_T$ phenomena observed  for the first time in $A+A$
reactions at RHIC with $\sqrt{s}> 100$ AGeV. An interesting pattern of hadron
suppression, already beginning at moderate $p_T> 3$ GeV, was seen in
the hadron flavor dependence of single inclusive spectra, in the large
azimuthal asymmetry,  
and in the preliminary two-hadron correlations.
We interpret these phenomena as manifestations of jet quenching in 
ultra-dense matter produced in such reactions.   Our predictions for 
these phenomena in both approaches are reviewed in later sections 
and depend  on the energy loss, $\Delta E = \int dx \;
dE/dx$, of fast quarks and gluons propagating through rapidly
expanding QCD matter. The two approaches, GLV and WW/WOGZ reviewed here,
provide a systematic way to compute $\Delta E$  via an opacity or
higher twist expansion in finite nuclear matter.  Elsewhere 
reviewed asymptotic approaches such as the BDMS/Z/SW\cite{Baier:2000mf,salgado}
are  designed for applications to ``thick'' or macroscopic 
media at asymptotic energies. An analytic approximation to the sum of 
the infinite multiple
collision series is obtained through an approximate color dipole
quantum diffusion analogous to the  Moliere series in electrodynamics. 
The complications due to finite kinematic bounds are neglected.
As shown in\cite{Arleo:2002kh}, the phenomenological
applications of the asymptotic expressions tend to overpredict
quenching at RHIC and lead  to a too rapid 
variation of the suppression factor with $p_T$,
inconsistent  with the RHIC data. In our approach, on the other hand,
the opacity series is computed to arbitrary order in opacity
for applications to finite opacity  systems  where the 
non-Gaussian  (Rutherford) tails of distributions are not yet eclipsed by 
the approximate Gaussian small $p_T$ component. In addition, our 
expressions can be applied to arbitrary 
3D expanding and time dependent media such as created 
in nuclear collisions.

The GLV reaction operation approach discussed in Section 2 is based on
a general algebraic recursive method that describes the propagation and
interaction of systems through dense nuclear matter, taking into
account the severe destructive LPM interference effects due to the long
formation times of ultra-relativistic partons. It provides a means to
compute the multiple collision amplitudes to any order $\chi^n$ in the
opacity $\chi=\int dz \, \sigma \rho$.  So far it has been successfully
applied to obtain solutions for the nuclear broadening\cite{Gyulassy:2002yv}
and the final state medium induced 
radiation\cite{Gyulassy:2000er,Gyulassy:2000fs} resulting from
multiple elastic and inelastic projectile scatterings. 
The range of applicability of the calculations can be
significantly extended by careful treatment of kinematic bounds.

The final state double differential distributions of jets and gluons are
presented as an infinite series in powers of the mean number of
scatterings $\chi$.  For elastic scatterings this series can be summed to
all orders (or equivalently all twist parton-parton correlations)   
to reproduce the standard Glauber theory result, but more
general than the simple color dipole approximation used in 
asymptotic analyses.  For inelastic processes, a closed sum to all
orders can unfortunately only  be  carried  out in  the  dipole
approximation as in BDMS/Z/SW (see\cite{salgado} for most 
recent developments). 
However, our analytic\cite{Gyulassy:2000er,Gyulassy:2002yv}
expressions  at  any  finite order in opacity can be evaluated
numerically\cite{Gyulassy:2000er,Gyulassy:2000fs}.  
Numerical evaluation of the expressions through
the first three orders (up to twist 8) has been carried out
for phenomenological applications.  
Each power in opacity adds a twist 2 parton-parton correlation.  
Future work via  this  approach includes computation of multiple 
gluon emission beyond the Poisson  approximation and the broadening and 
radiation of dipole-like,  possibly heavy, $\bar{q} q$ systems. 
Applications to heavy quark energy loss  including both the 
Ter-Mikayelian gluon dispersion effects as well
as the ``dead cone'' effect are also underway\cite{magda}.

The WW approach reviewed in Section 3  
extends the calculation of energy loss to include
the positive feedback (jet acceleration)
due to absorption of thermal gluons in the medium.
Absorption counteracts the induced energy loss 
for jet momenta less than the typical thermal energy scale $\sim 3 T$.
In an expanding hydrodynamic medium with transverse boost rapidity
$\eta_T$ we expect that this absorption feedback contribution 
is blue shifted to higher momenta $p_T\sim 3T e^{\eta_T}$.
Thus jet quenching cannot suppress the spectrum below the 
local equilibrium hydrodynamic limit.

In Section 4, several recent works were combined
(referred to as the WOGZ approach\cite{Wang:2001if,Guo:2000nz,ow,bwzxnw})
to elaborate on a general twist-expansion of multiple parton scattering 
beyond the GW model of the medium. In that approach
one can calculate explicitly the modified parton fragmentation function
up to twist 4 thus far.
The LPM interference effect is then embedded in the twist-four
parton matrix elements of the nucleus which also contains the
fundamental properties of the nuclear medium -- parton density
correlations inside a nucleus. These matrix elements
replace the Debye screened interactions used in the GLV approach.
One can demonstrate explicitly that the 
quadratic dependence of the modification of fragmentation functions
and the effective parton energy loss on the nuclear size $R_A$
is caused both by the LPM interference and the specific form
of gluon radiation spectra in QCD. The predicted nuclear
modification of the fragmentation function, both the energy and
nuclear dependence, is found to agree well with the recent experimental 
data\cite{hermes} on jet fragmentation in $e+A$. 
This is an important test of both the twist or opacity expansions
since it shows the dominance of the twist 4 (first order in opacity)
contribution to the energy loss in finite nuclear systems
as found numerically in the GLV expansion up to twist 8.
Extending the results to a parton
propagating in a hot QCD medium, we have shown that the twist expansion
to order 4 is equivalent to 
the first order opacity GLV result under certain simplifying
assumptions about the form of the twist 4 matrix elements. 
The phenomenological application of WOGZ to both DIS on 
cold nuclear targets and
high-energy heavy-ion collisions suggests that the
parton energy loss in an expanding system at RHIC would be 
equivalent to $(dE/dx)_0\approx 7.3$ GeV/fm in a static medium,
which is almost 15 times higher than that in a cold $Au$ nucleus.

If the jet quenching pattern is  confirmed by further measurements 
and theoretical  refinements, current RHIC data may have already 
provided  the first tomographic evidence  that initial parton densities  
on the order of 100 times nuclear matter density were achieved in  
$Au+Au$ collisions. The full analysis of the flavor composition, 
shape, and azimuthal moments of the high $p_T$ spectra appears to be  
a promising diagnostic probe of the  evolution of the produced
quark-gluon plasma.

However, in spite of the consistency of our jet tomography
analysis with current RHIC data, it is still too early to 
draw definitive conclusions.  The main uncertainty is the magnitude 
of gluon shadowing\cite{Armesto:2003bq}
in the initial nuclear wavefunction.
While estimates of shadowing vary wildly, unfortunately
nothing is yet known experimentally on this important question.
QCD  analysis\cite{Mueller:wy}
suggests that non-linear corrections to DGLAP evolution
may cause the gluon density to saturate at small $x$.
Saturation is found in classical Yang-Mills  
models\cite{McLerran:1994vd,McLerran:1993ka,Iancu:2002xk,Mueller:2001fv} 
as well.
In Ref.\cite{Kharzeev:2002pc} it was proposed
that gluon saturation or shadowing phenomena alone may  account for 
a significant part of the observed  high $p_T$ hadron  suppression 
pattern. Our estimates of shadowing based on ESK'98\cite{Eskola:1998df}
do not support this idea and neither do recent 
calculations\cite{Eskola:2002yc,Eskola:2003gc}
in the MQ approach\cite{Mueller:wy}
of the upper limit of the saturation scale and the rate of 
disappearance of its effects on the small and moderate 
$p_T$ spectra.  Fortunately, a decisive 
experimental test is now under way at RHIC via
$d+ Au$ reactions. If our theory is correct, 
and the observed suppression pattern
in $Au+Au$ is due to jet quenching via final state interactions
in the dense QCD matter produced, then instead of suppression, 
the Cronin effect is predicted to dominate over shadowing
in the $x  >  0.01$ range accessible at RHIC. We therefore
predict as an upper limit an {\em enhancement} of the moderate 
$p_T$ hadrons by  $\sim 10-30 \%$ in $d+A$ relative to binary scaled $p+p$.
On the other hand, if the saturation model is correct, 
and the suppression pattern in $Au+Au$
is a consequence of deep gluon shadowing in the initial state,
then a suppression about 30\% of moderate and high 
$p_T$ hadrons is predicted to occur in $d+Au$\cite{Kharzeev:2002pc}.

The experimental answer to this decisive question may
be known by the publication time of
this review. In either case, the answer will be exciting.

\vspace*{.4cm}
\noindent{\bf Acknowledgments} 
\vspace*{.3cm}
 
M.G. and I.V. gratefully acknowledge extensive collaboration 
with Peter Levai  on the GLV approach. X.N.W. and B.W.Z. would
like to thank Enke Wang on extensive collaboration on the WW
approach with detailed balance and many fruitful discussions.
X.N.W. would also like to acknowledge collaborations with
X. F. Guo and J. Osborne on the twist expansion approach.
This work was supported by  the Director, Office of Energy  
Research, Office of High Energy and Nuclear Physics,  
Division of Nuclear Physics, and by the Office of Basic Energy 
Science,  Division of Nuclear Science, of   
the U.S. Department of Energy  under Contracts No.  
DE-FG02-87ER40371,  DE-FG02-93ER40764, and DE-AC03-76SF00098
and by National Natural Science Foundation of China under project 
No. 19928511 and No. 10135030.

\end{document}